\documentclass[a4paper,11pt]{article}
\pdfoutput=1 
             
\usepackage{jheppub}

\usepackage[T1]{fontenc} 
\usepackage{comment}
\usepackage{tikz}
\usetikzlibrary{decorations.markings}

\def\eps{\epsilon}

\newcommand{\p}[1]{(\ref{#1})}

\newcommand \vev [1] {\langle{#1}\rangle}

\newcommand{\pa}{\partial}
\newcommand{\ep}{\epsilon}

\title{Anomalous Ward identities for on-shell amplitudes at the conformal fixed point}
\author[a]{Dmitry Chicherin}
\author[b]{Johannes Henn}
\author[c]{Simone Zoia}

\affiliation[a]{LAPTh, Universit\'e Savoie Mont Blanc, CNRS, B.P.\ 110, F-74941 Annecy-le-Vieux, France}
\affiliation[b]{Max-Planck-Institut f\"{u}r Physik, Werner-Heisenberg-Institut, D-80805 M\"{u}nchen, Germany}
\affiliation[c]{Dipartimento di Fisica and Arnold-Regge Center, Universit\`a di Torino, and INFN, Sezione di Torino, Via P.\ Giuria 1, I-10125 Torino, Italy}

\emailAdd{chicherin@lapth.cnrs.fr}
\emailAdd{henn@mpp.mpg.de}
\emailAdd{simone.zoia@unito.it}

\preprint{LAPTH-036/22, MPP-2022-83}

\abstract{
Conformal symmetry underlies many massless quantum field theories, but little is known about the consequences of this powerful symmetry for on-shell scattering amplitudes. 
Working in a dimensionally-regularised $\phi^3$ model at the conformal fixed point,
we show that the on-shell renormalised amplitudes satisfy anomalous conformal Ward identities. 
Each external on-shell state contributes two terms to the anomaly.
The first term is proportional to the elementary field anomalous dimension, and thus involves only lower-loop information. We show that the second term can be given as the convolution of a universal collinear function and lower-order amplitudes. The computation of the conformal anomaly is therefore simpler than that of the amplitude at the same perturbative order, which gives our anomalous conformal Ward identities a strong predictive power in perturbation theory.
Finally, we show that our result is also of practical importance for dimensionally-regularised amplitudes away from the conformal fixed point.
}

\begin{document}

\setcounter{tocdepth}{2}
\maketitle
\setcounter{page}{1}

\section{Introduction}
Four-dimensional quantum field theories such as Yang-Mills, massless quantum electrodynamics or quantum chromodynamics, 
Yukawa and $\phi^4$ theory have conformal symmetry at the level of the Lagrangian. 
To date, the consequences of this powerful symmetry have been widely studied and exploited 
for off-shell correlation functions, and mostly in position space~\cite{Todorov:1978rf,DiFrancesco:1997nk,Fradkin:1997df,Braun:2003rp}.
Motivated by problems in cosmology, several groups have studied conformal symmetry for off-shell momentum-space correlation functions~\cite{Coriano:2013jba,Bzowski:2013sza,Bzowski:2019kwd,Bzowski:2020kfw}. 
However, surprisingly little is known about the implications of conformal symmetry for on-shell scattering amplitudes.
{The latter are fundamental ingredients for theoretical predictions at particle colliders.}
The goal of the present paper is to contribute to our understanding of the implications of this powerful symmetry for such objects. 

At tree level, scattering amplitudes are annihilated by the conformal generators, as was discussed in \cite{Witten:2003nn} for gluon scattering amplitudes. The resulting differential equations are second-order in momentum space.
Very interestingly, the tree-level gluon amplitudes possess a bigger symmetry group when they are embedded into the supermultiplet of ${\mathcal N}=4$ super Yang-Mills.
It was shown that the conformal symmetry, together with enhanced symmetries of ${\mathcal N}=4$ super Yang-Mills, is enough to uniquely fix the form of the tree-level scattering amplitudes~\cite{Korchemsky:2009hm, Bargheer:2009qu}.

At loop level, to our knowledge, conformal symmetry predictions are mostly restricted to certain finite objects in scattering amplitudes
(see however \cite{Bargheer:2009qu,Beisert:2010gn} for interesting work in the context of maximally supersymmetric Yang-Mills theory).
Firstly, it is known that certain coefficients of transcendental functions at loop level, such as for example the leading singularities~\cite{Arkani-Hamed:2010pyv}, are conformally invariant. The reason is that the latter are obtained via generalised unitarity~\cite{Bern:1994zx} from tree-level amplitudes in a way that preserves conformal symmetry. 
For example, conformally-invariant expressions have been found to play an important role in all-plus scattering amplitudes~\cite{Badger:2019djh,Henn:2019mvc,Chicherin:2022bov}, but they also appear in more general helicity configurations.
Secondly, it was shown that conformal symmetry constrains infrared- and ultraviolet-finite loop integrals \cite{Chicherin:2017bxc}.
The authors found anomalous conformal Ward identities for an $L$-loop integral, where the anomaly term is expressed as an integral over the collinear region of a simpler, $(L-1)$-loop integral. This insight can therefore be useful for practical calculations, as was demonstrated in references \cite{Zoia:2018sin,Chicherin:2018ubl,Chicherin:2018rpz}.

A key difficulty in generalizing these results to full scattering amplitudes (in classically conformally-invariant theories)
are divergences, which na\"ively break the symmetry. 
In this paper we wish to generalise the results of \cite{Chicherin:2017bxc} to ultraviolet-divergent (but infrared-finite) scattering amplitudes, which makes dimensional regularisation necessary. 
We study the amplitudes 
at a conformal fixed point where the $\beta$ function vanishes \cite{Wilson:1971dc}.

Although our ultimate aim is to apply conformal methods to Yang-Mills scattering amplitudes at loop level, we find it convenient for this first study to use a $\phi^3$ toy model in $d=6-2\eps$ dimensions, where $\eps$ is the dimensional regularisation parameter~\cite{Gracey:2020tkk}. 
For our purposes we find it insightful to have a further parameter, and therefore we take the scalar field to be matrix-valued, in the fundamental representation of $su(\textsc{n})$.

\newpage

We show that on-shell renormalised amplitudes satisfy anomalous conformal Ward identities (CWIs) at the conformal fixed point.
The anomaly has a local nature, with each external on-shell leg giving a separate additive contribution made of two terms. The first term is proportional to the anomalous dimension of the elementary field. The second term is given by the convolution of a universal collinear function with lower-loop amplitudes. The anomaly for an $\ell$-loop amplitude is therefore determined by $(\ell-1)$ information. Because of this, our result is not only interesting from a theoretical point of view, but also useful for computing amplitudes perturbatively.
The amplitude away from the fixed point can then be recovered from the conformal result by supplying lower-loop information, which makes our anomalous CWIs a new valuable tool for computing amplitudes in perturbation theory.

\smallskip

The starting point of our derivation are the conformal properties of the off-shell correlators, which we review in section~\ref{sec:offshell}.
After setting the notation in section~\ref{sec:conformal_tree}, in section~\ref{sec:CWI_corr_funcs} we recall that the UV-renormalised correlators satisfy conformal Ward identities with an anomaly term which is proportional to the $\beta$ function. At the conformal fixed point, thus, the renormalised correlators are conformally invariant. From a perturbative point of view, this is an all-order statement which requires a very intricate conspiracy among infinitely many Feynman diagrams. We elaborate on this in section~\ref{sec:conformalVSperturbative}.
Section~\ref{sec:CWI_amplitudes} is devoted to the on-shell amplitudes. We define them in section~\ref{sec:LSZ} by applying the Lehmann--Symanzik--Zimmermann (LSZ) reduction procedure to the off-shell correlators. The non-commutativity of the generator of conformal boosts with the on-shell limit in the LSZ reduction generates an anomaly in the CWIs for the on-shell amplitudes.
We pinpoint the origin of the anomaly to certain terms of the asymptotic expansion of the amputated correlators in the on-shell limit. 
We discuss this thoroughly in section~\ref{sec:WardId3pt} for the three-point correlator with one leg put on-shell, and then generalise to the $n$-particle case in section~\ref{sec:multi-point}. The anomaly receives contributions of two types. One is proportional to the elementary field anomalous dimensions, and thus needs to be computed at a lower loop order. In section~\ref{sec:collinear} we show that the second type of contribution to the conformal anomaly can be expressed as the convolution of a universal function with lower-loop amplitudes, so that the whole anomaly is entirely determined by lower-loop information.
In section~\ref{sec:checks} we present the calculation of the conformal anomaly in a few examples. We conclude in section~\ref{sec:conclusions} with a discussion of our results, and an outlook on the future studies.

We provide useful technical details in several appendices. In appendix~\ref{app:CWIren} we review the proof of the Ward identities for the renormalised correlation functions. In appendix~\ref{app:implications_away} we show how the knowledge of a correlator/amplitude at the conformal fixed point constrains it away from the fixed point. In appendix~\ref{app:momcons} we prove that the generator of conformal boosts commutes with the momentum-conservation delta function.
Appendix~\ref{sec:correlators} is devoted to an instructive comparison between the perturbative results and the expression fixed by conformal symmetry for the three-point correlator. The latter is well known and very simple in position space. Its Fourier transform to momentum space is given by a one-loop triangle Feynman integral with non-integer powers of the propagators. 
We show in appendix~\ref{app:calculation_3pt} how to efficiently use the differential equations method in this novel setting.


\section{Correlation functions at the conformal fixed point}
\label{sec:offshell}

In this section we study the conformal properties of off-shell correlators in momentum space as a preparatory step before considering on-shell amplitudes.
We define the $su(\textsc{n})$-matrix $\phi^3$ theory and recall its conformal properties at tree level in section~\ref{sec:conformal_tree}. In section~\ref{sec:CWI_corr_funcs} we review that the renormalised correlators satisfy CWIs with an anomalous term proportional to the $\beta$ function. At the conformal fixed point the $\beta$ function vanishes, and the correlators are annihilated by the conformal generators. We discuss this explicitly in  section~\ref{sec:conformalVSperturbative} for the three-point correlator. The latter gives a spectacular example of the constraining power of conformal symmetry, and is the perfect toy example for the derivation of our anomalous CWIs for on-shell amplitudes in section~\ref{sec:CWI_amplitudes}. 

\subsection{Conformal invariance at the classical level}
\label{sec:conformal_tree}

In this section we review the conformal invariance of the $su(\textsc{n})$-matrix $\phi^3$ theory at the tree level. The bare action in $d=6$ dimensions is given by
\begin{align} \label{eq:action}
S_0 = \int \mathrm{d}^d x \left[\frac{1}{2} (\partial \phi_0^a)^2 + \frac{g_0}{6} d^{abc} \phi_0^a \phi_0^b \phi_0^c \right]\,,
\end{align}
where $\phi_0 = \sum_a \phi_0^a t^a$, the matrices $t^a$ are the $su(\textsc{n})$ generators in the fundamental representation normalised as $\text{tr} \, t^a t^b = \delta^{ab}/2$, $d^{abc} = 2 \text{tr} \, t^a \{ t^b, t^c \}$ with the curly brackets denoting anti-commutation, i.e.\ $\{A,B\}:=A B + B A$, and $a,b,c=1,\ldots,\textsc{n}^2-1$. The $0$ subscript denotes bare quantities.

The action in eq.~\eqref{eq:action} is invariant under conformal transformations. On top of the usual Poincar\'e group, in fact, it is invariant also under the following infinitesimal variation of the fields,
\begin{align}
\label{eq:confvarphi}
\phi_0^a(x) \to \phi_0^a(x) + \textup{i} \omega D_{\Delta_{\phi}} \phi_0^a(x) + \textup{i} \varepsilon_\mu K^\mu_{\Delta_{\phi}} \phi_0^a(x)  \,,
\end{align}
where $\omega$ and $\varepsilon^{\mu}$ are infinitesimal parameters.
Here, $D_\Delta$ is the generator of infinitesimal \emph{dilatations}, 
\begin{align} \label{eq:Dill}
D_\Delta = - \textup{i} \left( x^\mu \pa_\mu + \Delta\right) \,,
\end{align}
which are a rescaling of the spacetime coordinates,
\begin{align}
    x^{\mu} \to e^{ \omega} \, x^{\mu} \,.
\end{align}
$\Delta$ is the conformal weight (or scaling dimension) of the field $\phi_0$, and its canonical value ---~namely the value for which the action in eq.~\eqref{eq:action} is invariant with $d=6$~--- is $\Delta_{\phi}=2$.
$K^\mu_{\Delta}$ in eq.~\eqref{eq:confvarphi} is instead the generator of \emph{conformal boosts} (or special conformal transformations),
\begin{align} \label{eq:KX} 
    K^\mu_\Delta = \textup{i} \left(x^2 \pa^\mu - 2 x^\mu x^\nu \pa_\nu - 2 \Delta x^\mu \right) \,,
\end{align}
which at the level of the spacetime coordinates corresponds to
\begin{align}
    x^{\mu} \to \frac{x^{\mu} - \varepsilon^{\mu} x^2}{1 - 2 \varepsilon\cdot x + \varepsilon^2 x^2}\,.
\end{align}
The latter can be more intuitively viewed as the composition of an inversion ($x^{\mu} \to x^{\mu}/x^2$), followed by a translation $ x^\mu\to x^\mu - \varepsilon^{\mu}$, and finally another inversion. 

The conformal invariance of the bare action implies that the correlation functions are conformally invariant at tree level, namely that the tree-level correlators of $n$ fields are annihilated by the generators of conformal symmetry,\footnote{We consider only correlators of elementary fields of the theory with action~\p{eq:action}, which are (renormalised) $n$-point Green functions. We always tacitly imply the time ordering in the notation $\vev{\ldots}$.}
\begin{align} \label{eq:treeCWI}
    K^{\mu}_{\Delta_{\phi}} \, \langle \phi_0(x_1) \ldots \phi_0(x_n) \rangle \biggl|_{\text{tree}} = 0 \,,
\end{align}
and similarly for the other generators.

The representation of the generators acting on a function of the scalar field $\phi_0$ at $n$ points $x_i$ in position space is obtained by summing over all points the single-point generators in eqs.~\eqref{eq:Dill} and~\eqref{eq:KX},
\begin{align} \label{eq:Dx}
 & D_{\Delta} = -\textup{i} \sum_{j=1}^n \left(x_j^{\mu} \frac{\partial}{\partial x_j^{\mu}} + \Delta \right) \,, \\
 \label{eq:Kx}
 & K^{\mu}_{\Delta} = \textup{i} \sum_{j=1}^n \left( x_j^2 \frac{\partial}{\partial x_{j\,\mu}} - 2 x_j^{\mu} x_j^{\nu} \frac{\partial}{\partial x_{j}^{\nu}} - 2 \Delta \, x_j^{\mu} \right) \,.
\end{align}
For example, it is straightforward to show that the two- and three-point tree-level correlators are annihilated by the generators of dilatations and conformal boosts,
\begin{align}\label{positionspace2pttree}
 & \langle  \phi^a_0(x_1) \phi^b_0(x_2) \rangle \biggl|_{\text{tree}} = \frac{1}{4\pi^3} \frac{\delta^{ab}}{\left(x_{12}^2\right)^2} \,, \\
 & \langle  \phi^a_0(x_1) \phi^b_0(x_2) \phi^c_0(x_3) \rangle \biggl|_{\text{tree}} = -\frac{g}{(2\pi)^6} \frac{d^{abc}}{x_{12}^2 x_{23}^2 x_{31}^2} \,,\label{positionspace3pttree}
\end{align}
where $x_{ij} = x_i - x_j$.

We are interested in the implications of conformal symmetry for scattering amplitudes, which are defined in on-shell momentum space. {We work in Minkowski spacetime to describe light-like momenta.} The (off-shell) momentum-space realisation of the conformal generators can be obtained by Fourier transforming the position-space one, obtaining\footnote{To avoid the proliferation of symbols, we denote with the same symbol the conformal boost generator in both momentum and position spaces.}
\begin{align}\label{eq:Dmomentum}
    D_{\Delta} = \textup{i} \sum_{j=1}^n \left( p_j^{\mu}   \frac{\partial}{\partial p_{j}^{\mu}} + d - \Delta \right) \,,
\end{align}
and
\begin{align} \label{eq:Kmu}
  K^{\mu}_{\Delta} = \sum_{j=1}^n \left( -p_j^{\mu } \frac{\partial}{\partial p_{j\,\nu}} \frac{\partial}{\partial p_{j}^{\nu}} + 2 p_j^{\nu} \frac{\partial}{\partial p_{j}^{\nu}} \frac{\partial}{\partial p_{j\,\mu}} + 2 (d-\Delta) \frac{\partial}{\partial p_{j\,\mu}} \right) \,.
\end{align}
While in position space all conformal generators are first-order differential operators, the conformal boost generator is realised in momentum space by a second-order operator. 
This is one of the reasons why studying conformal symmetry in momentum space is more challenging than in position space. Indeed, in section~\ref{sec:correlators} we will review how conformal symmetry allows us to draw sweeping, all-order conclusions about the correlation functions in position space, whereas the corresponding results in momentum space are substantially more involved beyond the tree level. 

The Fourier transform of the tree-level two- and three-point functions~\eqref{positionspace2pttree} and~\eqref{positionspace3pttree} gives the following momentum-space correlators,
\begin{align}
 & \langle  \phi^a_0(p_1) \phi^b_0(p_2) \rangle \biggl|_{\text{tree}} = \frac{\textup{i} \delta^{ab}}{p_1^2} (2 \pi)^6 \delta^{(6)}(p_1+p_2) \,, \\
 & \langle  \phi^a_0(p_1) \phi^b_0(p_2) \phi^c_0(p_3) \rangle \biggl|_{\text{tree}} = \frac{g \, d^{abc}}{p_1^2 p_2^2 p_3^2} (2 \pi)^6 \delta^{(6)}(p_1+p_2+p_3)  \,.
\end{align}
One can show that the conformal generators~\eqref{eq:Dmomentum} and~\eqref{eq:Kmu} annihilate these expressions.

The dilatation and Lorentz symmetry generators are first-order in momentum space, and hence their constraints on the correlators are easy to take into account. For this reason we will focus in this paper on the conformal boost generator.

In addition to this technical complication, two conceptual issues need to be addressed: the appearance of divergences in the correlators beyond the tree level, and the definition of an on-shell scattering amplitude starting from the off-shell correlator. We tackle the latter problem in section~\ref{sec:CWI_amplitudes}, and devote the next section to the former.

\subsection{Conformal invariance for renormalised correlation functions}
\label{sec:CWI_corr_funcs}

Beyond tree level the theory is ultraviolet divergent and needs to be renormalised. This obscures the conformal properties of the correlators. In this section we review that the renormalised correlators satisfy CWIs with a different conformal weight and an anomalous term proportional to the $\beta$ function.

We work in dimensional regularisation, namely we analytically continue the spacetime dimension to $d=6-2 \eps$ with $\eps>0$. The canonical conformal weight of $\phi$ in $d$ dimensions is
\begin{align}
    \Delta_{\phi} = \frac{d}{2}-1 \,.
\end{align}
The theory is multiplicatively renormalisable.
The renormalised field $\phi$ and coupling $g$ are defined as
\begin{align} \label{eq:renormalisation}
\phi_0 = Z_1^{\frac{1}{2}} \phi \, , \qquad \qquad g_0 = \mu^{\eps} Z_g \, g \,,
\end{align}
introducing the renormalisation scale $\mu$. The renormalisation factors $Z_1, Z_g$ are calculated in perturbation theory as series in the coupling
\begin{align} \label{eq:u}
u = \frac{g^2}{(4 \pi)^3} \, .
\end{align}
Their two-loop expressions, given in ref.~\cite{Braun:2013tva}, are reproduced in Appendix~\ref{app:CWIren}. The $\beta$ function and the anomalous dimension $\gamma$ of the elementary field are 
\begin{align}
& \beta = \frac{d u}{d\log \mu} = -2 \eps u - \frac{\textsc{n}^2-20}{2 \textsc{n}} u^2 - \frac{5360 - 496 \textsc{n}^2 + 5 \textsc{n}^4}{ 72 \textsc{n}^2} u^3 + \mathcal{O}(u^4) \, , \label{eq:beta} \\
& \gamma = \frac{1}{2} \frac{d \log Z_1}{d \log \mu} = u \frac{\textsc{n}^2-4}{12 \textsc{n}} \left(1 + u \frac{\textsc{n}^2-100}{36 \textsc{n}} \right) + \mathcal{O}(u^3) \, .  \label{eq:gamma}
\end{align}
The four-loop approximation of $\beta$ and $\gamma$ can be found in \cite{Gracey:2015tta}.

Renormalisation introduces the scale running of the coupling. As a result, conformal symmetry is obscured, but not entirely lost. The renormalised correlators in fact satisfy CWIs with the conformal dimension shifted by the anomalous dimension $\gamma$,
\begin{align} \label{eq:Deltag}
    \Delta_{\gamma} = \Delta_{\phi} + \gamma(u) \,.
\end{align}
The dilatation generator acts on the renormalised correlation function as follows,
\begin{align} 
D_{\Delta_{\gamma}} \langle \phi (x_1)\ldots \phi(x_n) \rangle = -\frac{\beta(u)}{2u} \int \mathrm{d}^d x \,\left\langle  \phi(x_1) \ldots \phi(x_n) \,  {\cal O}(x) \right\rangle \,, \label{eq:WardIdDil}
 \end{align}
where ${\cal O}(x) := g \mu^\ep\left[ d^{a b c} \phi_a \phi_b \phi_c \right]_R(x)/6$ is the renormalised local operator which is the UV-finite counterpart of the theory interaction, and we recall that $\phi$ denotes the renormalised field. The dilatation Ward identity is equivalent to the renormalisation group Callan--Symanzik equation for the correlation function, which describes the evolution of the correlator with the scale $\mu$. The conformal boost Ward identity has an analogous form,
\begin{align} \label{eq:KCx}
K^{\mu}_{\Delta_{\gamma}} \langle \phi (x_1)\ldots \phi(x_n) \rangle = -\frac{\beta(u)}{u} \int \mathrm{d}^d x \, x^{\mu} \left\langle \phi(x_1) \ldots \phi(x_n)  \, {\cal O}(x) \right\rangle \,.
\end{align}
The CWIs for multi-point renormalised correlation functions are well-known and date back to the early days of the conformal symmetry, see e.g.~\cite{Parisi:1972zy}. 
They were also studied in non-abelian gauge theories, see~\cite{Braun:2003rp} for a review. For the convenience of the readers, we outline the derivation of the CWIs \p{eq:Deltag} and \p{eq:KCx} in Appendix~\ref{app:CWIren}.

The invariance of the action~\p{eq:action} under conformal transformations generates currents which are conserved at classical level. Beyond the classical approximation dilatations and conformal boosts are not exact symmetries, and the conservation of the corresponding currents becomes anomalous. The right-hand sides of eqs.~\p{eq:WardIdDil} and~\p{eq:KCx} provide the quantum anomalies of the dilatation and conformal boost currents. In section~\ref{sec:CWI_amplitudes} we consider conformal anomalies of the on-shell amplitudes, which are not to be confused with the quantum anomalies of the currents originating from the running of the coupling.

The fact that the right-hand sides of eqs.~\eqref{eq:WardIdDil} and~\eqref{eq:KCx} are proportional to the $\beta$ function suggests to use the notion of \emph{conformal fixed point} (alias Wilson--Fisher fixed point~\cite{Wilson:1971dc}), i.e.\ a special value of the coupling constant $u^*$ such that the $\beta$ function vanishes,
\begin{align}
    \beta\left(u^*\right) = 0 \,.
\end{align}
The coupling at the conformal point is a function of the dimensional regulator $\eps$. Up to order $\eps^2$ it is given by
\begin{align}
    \label{eq:fixed_point}
u^* = \frac{4 \textsc{n} \eps}{20 - \textsc{n}^2 } + \frac{4 \textsc{n} \eps^2}{9} \frac{5 \textsc{n}^4 - 496 \textsc{n}^2 + 5360}{(20-\textsc{n}^2)^3} + \mathcal{O}(\eps^3) \,.
\end{align}
Note that the real-valuedness of the coupling $g$ implies that $u>0$ and, since $\eps>0$, this is possible only for sufficiently small values of the rank $\textsc{n}$.\footnote{We are mostly interested in the perturbative implications of conformal symmetry at the conformal fixed point, and therefore do not study the nature of the latter \cite{Giombi:2015haa,Gracey:2020tkk}.
}
The renormalised correlators are therefore conformally invariant at the conformal fixed point,
\begin{align} \label{eq:CWI_fixed_point}
    K^{\mu}_{\Delta_{\phi}+\gamma} \, \langle \phi(x_1) \ldots \phi(x_n) \rangle \biggl|_{u = u^*} = 0 \,,
\end{align}
though with a conformal dimension shifted by the anomalous dimension $\gamma$. This is the starting point for the derivation of our anomalous CWIs for the on-shell amplitudes, which we discuss in section~\ref{sec:CWI_amplitudes}. It is important to stress that the CWIs~\eqref{eq:CWI_fixed_point} give very strong constraints on the form of the renormalised correlators also away from the conformal fixed point. Indeed, in appendix~\ref{app:implications_away} we show that the finite part of a renormalised correlator at a given loop order is entirely determined by its expression at the conformal fixed point and lower loop information.


\subsection{A toy example: the three-point correlator}
\label{sec:conformalVSperturbative}

Seen from the point of view of a perturbative computation, having exact conformal invariance at the conformal fixed point implies a remarkable conspiracy of infinitely many Feynman diagrams at all orders. It is instructive to illustrate this for the three-point correlator, which will be a useful toy example for deriving the anomalous Ward identities for the on-shell amplitudes in section~\ref{sec:CWI_amplitudes}. We give here only the salient points, and refer to appendix~\ref{sec:correlators} for a thorough discussion.

\begin{figure}[t]
\centering
\includegraphics[scale=0.25]{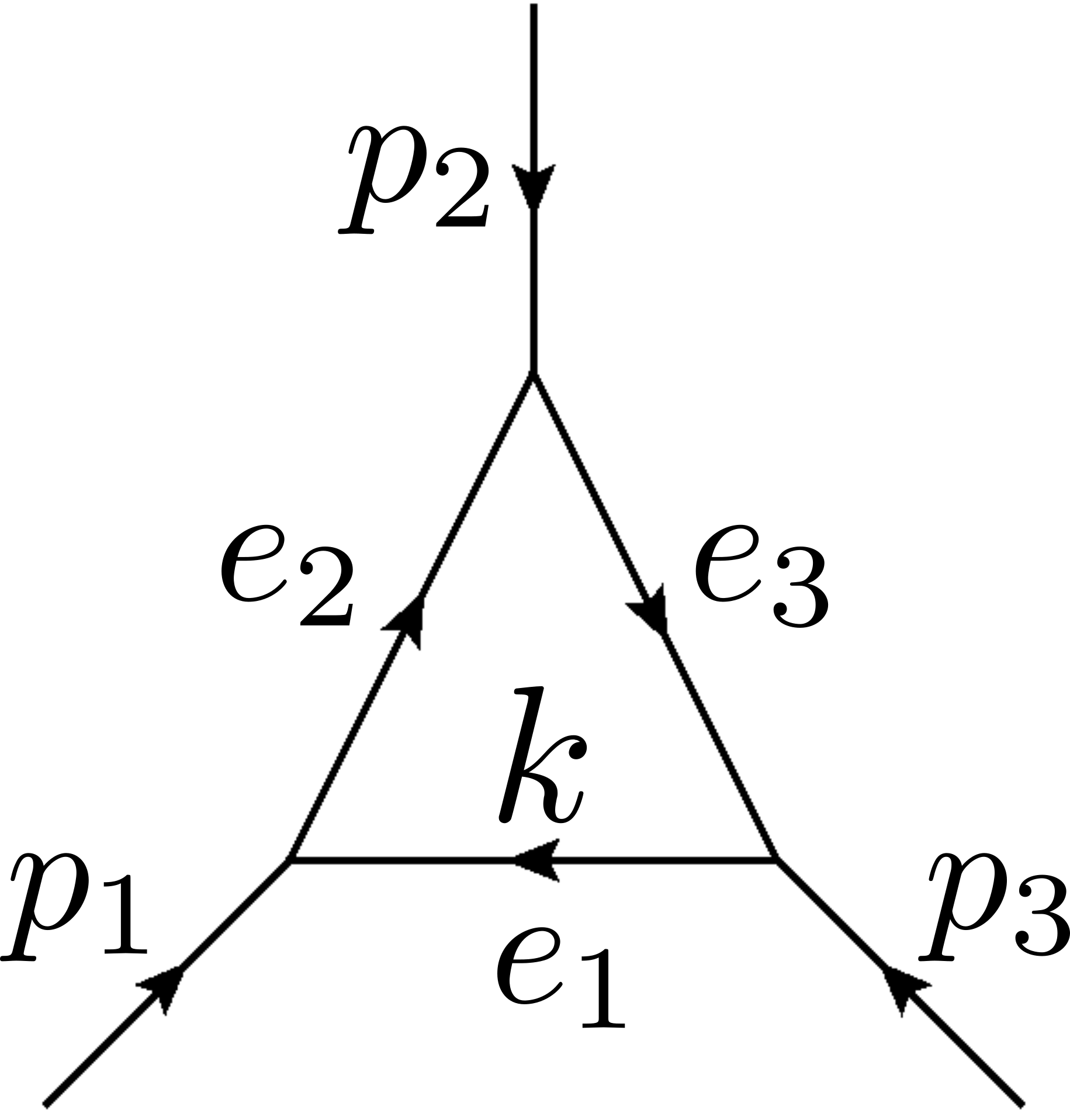}
\caption{Graph representing the one-loop triangle integral family defined by eq.~\eqref{eq:3massTriangle}. The arrows denote the momentum flow.}
\label{fig:triangle1L}
\end{figure}

It is well known that conformal symmetry fixes the three-point correlators up to an overall constant (see e.g.\ ref.~\cite{Polyakov:1970xd}). The expression is extremely simple in position space,\footnote{We recall that we are working in the Minkowski space-time signature.  The subtraction of a small positive imaginary part  $\textup{i} 0$ from each $x_{ij}^2$ corresponds to Feynman's prescription for the propagators in momentum space.}
\begin{align}
\label{eq:C3confX}
    \langle \phi^a(x_1) \phi^b(x_2) \phi^c(x_3) \rangle \biggl|_{u=u^*} = \frac{d^{abc} \, c_{123}(u^*)}{\left(x_{12}^2- \textup{i} 0\right)^{\frac{\Delta_{\gamma}}{2}} \left(x_{23}^2- \textup{i} 0\right)^{\frac{\Delta_{\gamma}}{2}} \left(x_{31}^2- \textup{i} 0\right)^{\frac{\Delta_{\gamma}}{2}}} \,,
\end{align}
where $x_{ij}=x_i - x_j$, but becomes more complicated in momentum space. We recall that we have to work directly in momentum space in order to study on-shell scattering amplitudes.
Expressions for the Fourier transform of the three-point conformal correlator have been obtained in terms of Appell's hypergeometric function $F_4$ or triple-$K$ integrals~\cite{Barnes:2010jp,Coriano:2013jba,Bzowski:2013sza,Bzowski:2015yxv,Isono:2019ihz,Bautista:2019qxj,Gillioz:2019lgs,Bzowski:2020lip}. We adopt a different approach, and use of the technology developed for computing Feynman integrals. The Fourier transform of the three-point correlator can in fact be written as
\begin{align} \label{eq:C3Imain}
 \langle \phi^a(p_1) \phi^b(p_2) \phi^c(p_3) \rangle \biggl|_{u=u^*} \propto d^{abc} \, \delta^{(d)}(p_1+p_2+p_3) \,
     I\left(2-\frac{\eps+\gamma}{2}, 2-\frac{\eps+\gamma}{2}, 2- \frac{\eps+\gamma}{2} \right) \,,
\end{align}
up to an overall kinematic-independent factor, where $I(e_1,e_2,e_3)$ is a Feynman integral of the one-loop ``three-mass triangle family,''
\begin{align}
\label{eq:3massTriangle}
I\left(e_1,e_2,e_3\right) = e^{\eps \gamma_\text{E}} \int \frac{\mathrm{d}^d k}{\textup{i} \pi^{\frac{d}{2}}} \frac{1}{\left[-k^2 -\textup{i} 0\right]^{e_1} \left[-(k+p_1)^2 -\textup{i} 0 \right]^{e_2} \left[ -(k+p_1+p_2)^2-\textup{i} 0 \right]^{e_3}  }\,,
\end{align}
whose defining graph is shown in figure~\ref{fig:triangle1L}. 
In contrast with the usual Feynman integrals, the propagators in eq.~\eqref{eq:C3Imain} are raised to non-integer, $\eps$-dependent powers.
Given this aspect of novelty, we discuss thoroughly the computation using the method of differential equations~\cite{KOTIKOV1991158,Bern:1993kr,Remiddi:1997ny,Gehrmann:1999as} in canonical form~\cite{Henn:2013pwa} in appendix~\ref{app:DEs}. As a result, we obtain the analytic expression of the three-point conformal correlator in momentum space expanded around $\eps=0$.\footnote{The expansion is truncated by the limited perturbative knowledge of the anomalous dimension $\gamma$. The latter is given up to four-loop order in~\cite{Gracey:2015tta}, which suffices to expand the correlator up to order $\eps^4$.} It is written in terms of algebraic functions and two-dimensional harmonic polylogarithms~\cite{Gehrmann:2001jv}. Remarkably, conformal symmetry captures entirely this complexity, in an expression which in position space is ridiculously compact. 

\begin{figure}
\centering
\begin{tabular}{cccccc}
  \raisebox{-.5\height}{\includegraphics[scale=0.13]{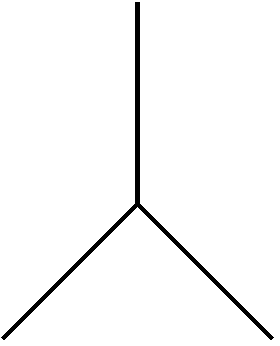}} & 
  \raisebox{-.5\height}{\includegraphics[scale=0.15]{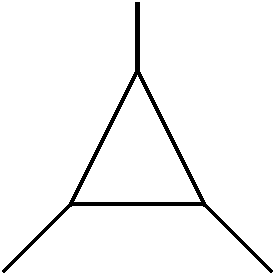}} & 
  \raisebox{-.5\height}{\includegraphics[scale=0.14]{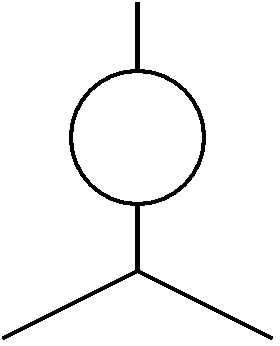}} & 
  \raisebox{-.5\height}{\includegraphics[scale=0.14]{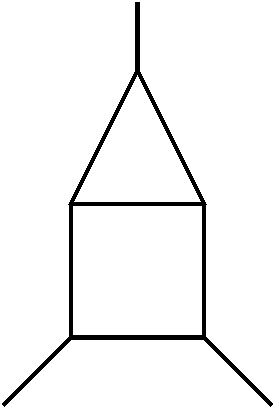}} & 
  \raisebox{-.5\height}{\includegraphics[scale=0.15]{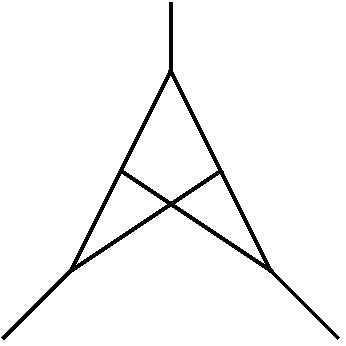}} &
  \raisebox{-.5\height}{\includegraphics[scale=0.17]{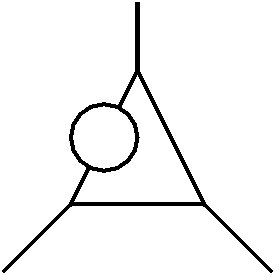}} \\ 
  \raisebox{-.5\height}{\includegraphics[scale=0.15]{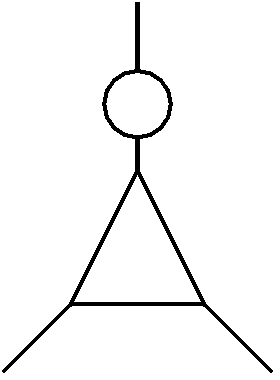}} &
  \raisebox{-.5\height}{\includegraphics[scale=0.15]{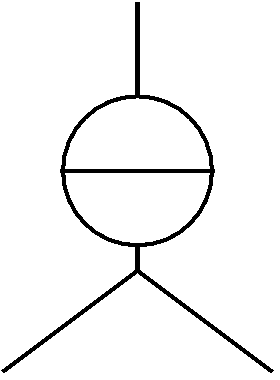}} &
  \raisebox{-.5\height}{\includegraphics[scale=0.15]{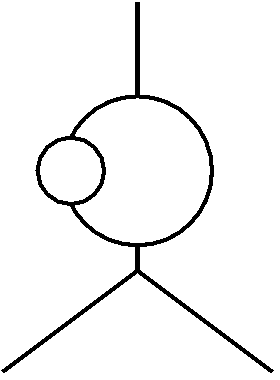}} &
  \raisebox{-.5\height}{\includegraphics[scale=0.16]{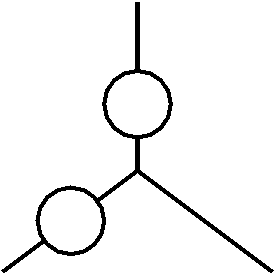}} & 
  \raisebox{-.5\height}{\includegraphics[scale=0.15]{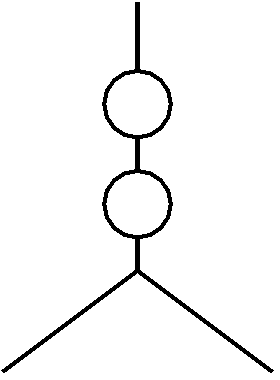}} & \\
\end{tabular}
\caption{Representative Feynman diagrams contributing to the three-point correlator $C^{(3)}$ up to two-loop order. 
}
\label{fig:diags_3pt}
\end{figure}

This is even more remarkable if contrasted with a perturbative computation of the correlator. The latter in fact involves  infinitely many Feynman diagrams (see e.g.\ figure~\ref{fig:diags_3pt}), which evaluate to complicated transcendental functions. All these pieces fit together, to all orders, to reproduce the conformally invariant expression at the conformal fixed point, fixing the normalisation factor $c_{123}$ in eq.~\eqref{eq:C3confX}. We check this explicitly with a perturbative computation of the two- and three-point correlators up to two-loop order in appendix~\ref{sec:perturbative}. 

The analysis of the three-point correlator gives two important lessons. First, it gives a paradigmatic example of the constraining power of conformal symmetry. Second, it shows that conformal symmetry underlies the perturbative results in a very non-trivial way. This is due to the second-order differential nature of the conformal constraints in momentum space. A precise understanding of how these constraints are implemented is therefore imperative in order to unveil the underlying conformal symmetry and exploit it.  

In the next section we investigate how this extends to on-shell scattering amplitudes. The latter are obtained from the correlators by amputating the external legs and taking the light-like limits. These operations have a non-trivial interplay with conformal symmetry, and generate an anomaly in the CWIs. 
The full analytic control over the three-point correlator~\eqref{eq:C3Imain} achieved through Feynman integral techniques will be instrumental to derive the conformal anomaly in the on-shell limit.

\section{Anomalous conformal Ward identities for scattering amplitudes}
\label{sec:CWI_amplitudes}

In the previous section we have shown that the renormalised correlators are annihilated by the conformal boost generator at the conformal fixed point,
\begin{align} \label{eq:KCn}
    K^{\mu}_{\Delta_{\gamma}} \, \langle \phi(x_1) \ldots \phi(x_n) \rangle \biggl|_{u=u^*} = 0 \,.
\end{align}
In this section we discuss how this translates into a constraint on the corresponding scattering amplitude. 
The latter is obtained from the correlator through the Lehmann--Symanzik--Zimmermann (LSZ) reduction formula~\cite{Lehmann:1954rq} (see ref.~\cite{Gillioz:2020mdd} for a formulation of the LSZ reduction procedure in a conformal field theory).

We begin in section~\ref{sec:LSZ} by discussing the interplay between the LSZ reduction formula and conformal symmetry. We show that the conformal boost generator does not commute with the on-shell limit of the reduction, and how we overcome this obstacle by analysing the asymptotic expansion of the correlator in the light-like limit. We do this explicitly for the three-point case in section~\ref{sec:WardId3pt}. We amputate one of the external legs and work out the anomalous CWIs satisfied by the resulting amplitude. We then extend this procedure to a multi-point correlator with several LSZ-reduced legs in section~\ref{sec:multi-point}. The anomaly in general comprises two terms. One is proportional to the elementary field anomalous dimension, and thus involves only lower-loop information. In section~\ref{sec:collinear} we argue that the second term is governed by collinear regions of loop momentum, and that it can be represented as the convolution of a universal function and lower-order amplitudes. This gives a handle on the computation of the conformal anomaly, which this way becomes a lower-order problem.
In section~\ref{sec:checks} we give a fewWe recall from examples of computation of the conformal anomaly: for the three-point case with one on-shell leg, both exactly to all orders and perturbatively up to one loop, and for the fully on-shell four-point case perturbatively up to one loop.

\subsection{Definition of scattering amplitudes via LSZ formula and conformal symmetry}
\label{sec:LSZ}

The LSZ reduction procedure takes from the correlator to the scattering amplitude by removing the bubble corrections from the external legs and setting the external momenta on shell. In this section we show that the conformal boost generator can be pulled through the amputation of the external bubbles, but does not commute with the on-shell limit.

We define the renormalised momentum-space correlators at the conformal fixed point by Fourier transforming the position-space ones as
\begin{align} \label{eq:correlators_Fourier}
    C^{(n)}\left(p_1,\ldots,p_n\right) \, (2 \pi)^d \, \delta^{(d)}\left(p_1 + \ldots + p_n \right) = \int \left( \prod_{j=1}^n \mathrm{d}^d x_j e^{\textup{i} p_j \cdot x_j} \right) \langle \phi(x_1) \ldots \phi(x_n)\rangle \biggl|_{u=u^*} \,.
\end{align}
We strip off the overall momentum-conservation $\delta$ function and a factor of $(2 \pi)^d$ to simplify the expressions. In appendix~\ref{app:momcons} we show that the conformal boost generator commutes with the momentum-conservation $\delta$ function when acting on a correlator. We will thus neglect the $\delta$ function hereinafter, and act with the conformal boost generator directly on $C^{(n)}$. 

We begin with the amputation of the bubble corrections on each external leg. The latter sum up to the renormalised two-point correlator. %
Their amputation therefore amounts to
\begin{align} \label{eq:GnDef}
    G^{(n)}(p_1,\ldots,p_n) := \left[ \prod_{j=1}^n   C^{(2)}(p_j, -p_j)\right]^{-1} C^{(n)}(p_1,\ldots,p_n) \,.
\end{align}
We call this intermediate object, $G^{(n)}$, the \textit{amputated correlator}.
The two-point renormalised correlator is given to all orders by (see appendix~\ref{sec:C2})
\begin{align}
    C^{(2)}(p_j, -p_j) = \tilde{c}_{12}\left(u^*\right) \left(-p_j^2\right)^{-1+\gamma} \,,
\end{align}
where $\tilde{c}_{12}\left(u^*\right)$ is a kinematic-independent factor given by eq.~\eqref{eq:ctilde12}.
We can pull the conformal boost generator through the inverse two-point correlation functions thanks to the intertwining relation
\begin{align}
    K^{\mu}_{\Delta} \left(p^2\right)^{\Delta-\frac{d}{2}} = \left(p^2\right)^{\Delta-\frac{d}{2}} K^{\mu}_{d-\Delta} \,.
\end{align}
Setting $\Delta = \Delta_{\gamma}$, this relation, together with the invariance of the correlator~\eqref{eq:KCn}, implies that the amputated correlator $G^{(n)}$ satisfies the CWI
\begin{align} \label{eq:KGn}
    K^{\mu}_{d-\Delta_{\gamma}} G^{(n)}(p_1,\ldots,p_n) = 0 \,.
\end{align}
Therefore, pulling the conformal boost generator through the inverse two-point correlators simply amounts to changing the conformal weight from $\Delta_{\gamma}$ to its ``shadow'' $d-\Delta_{\gamma}$. 
Equivalently, instead of the correlation function $C^{(n)}$ of elementary fields $\phi$, we could have considered a correlator of the shadow operators \cite{Ferrara:1972uq}, which carry the shadow conformal dimension. The latter coincides with $G^{(n)}$, and the amputation procedure \p{eq:GnDef} relates the two pictures. In the representation theoretic language, representations of the conformal algebra with weights $\Delta_{\gamma}$ and $d-\Delta_{\gamma}$ are equivalent \cite{Knapp1971}.

The scattering amplitude, $M^{(n)}$, is then given by the on-shell limit of the amputated correlator,\footnote{In the following, it will also be convenient to consider analogous objects where only some of the external legs are taken on-shell, which we will also denote by the letter $M$.}
\begin{align}
    M^{(n)}(p_1,\ldots,p_n) = \lim_{p_1^2\to 0} \ldots \lim_{p_n^2\to 0} G^{(n)}(p_1,\ldots,p_n) \,.
\end{align}
The conformal boost generator, however, does not commute with this limit. In order to prove this, let us consider a toy example: a generic scalar function of a single squared momentum, $f(p^2)$, which is finite at $p^2 \to 0$. The conformal boost generator acting on it is 
\begin{align} \label{eq:Kmup}
  K^{\mu}_{\Delta} = -p^{\mu } \frac{\partial}{\partial p_{\nu}} \frac{\partial}{\partial p^{\nu}} + 2 p^{\nu} \frac{\partial}{\partial p^{\nu}} \frac{\partial}{\partial p_{\mu}} + 2 (d-\Delta) \frac{\partial}{\partial p_{\mu}} \,.
\end{align}
The commutator of the conformal boost generator and the on-shell limit acting on $f$ is then given by
\begin{align} \label{eq:commutator}
    \left[K^{\mu}_{\Delta}, \lim_{p^2\to 0} \right] f\left(p^2\right) = - 4 \lim_{p^2\to 0} p^{\mu} \left[p^2 f''\left(p^2\right) + \left(\frac{d}{2}+1 -\Delta\right) \, f'\left(p^2\right) \right] \,.
\end{align}
The two terms on the right-hand side of the previous equation correspond to two different physical effects that leads to an on-shell conformal anomaly:
\begin{itemize}
    \item The first term on the right-hand side of eq.~\eqref{eq:commutator}
    also appeared in ref.~\cite{Chicherin:2017bxc} in the context of finite conformal integrals. There it was shown that the action of the conformal boost leads to an anomaly that is localised on collinear regions of the loop integration (more on this in section~\ref{sec:collinear}). 
\item The second term on the right-hand side of eq.~\p{eq:commutator} vanishes for $\Delta = d/2+1$, or equivalently $\Delta = d-\Delta_{\phi}$. However, we see that there is a mismatch with the conformal weight $\Delta = d-\Delta_{\gamma}$ that is needed to annihilate the amputated correlator, see eq.~\eqref{eq:KGn}. The discrepancy is proportional to the anomalous dimension $\gamma$. This generates a new type of conformal anomaly, which we investigate in section~\ref{sec:WardId3pt}. Since it is related to the anomalous dimension, one could say that it is of ultraviolet origin.
\end{itemize}

The amputated correlator $G^{(n)}$ is finite at $p_i^2=0$, but its derivatives do not vanish in general. To see this, consider for example the one-loop triangle integral defined by eq.~\eqref{eq:3massTriangle} with unit propagator powers. 
In a perturbative computation of $G^{(3)}$, this is the only diagram contributing at one-loop order. In the light-like limit $s_1\to 0$, the triangle integral has the following asymptotic expansion,
\begin{align} \label{eq:triangleExp}
I(1,1,1) = \sum_{m\ge 0} (-s_1)^m I^{\text{hard}}_{[m]}(1,1,1) + \sum_{m\ge 0} (-s_1)^{1-\eps+m} I^{\text{coll}}_{[m]}(1,1,1) \,.
\end{align}
The terms of the expansion can be computed using the method of the \emph{expansion by regions}~\cite{Beneke:1997zp,Smirnov:1999bza,Jantzen:2011nz}.
One splits the loop integration domain into regions in which the expansion in the small parameter ($s_1$ in this case) and the integration commute, and it is thus possible to expand in series the integrand under the integral sign. The terms with integer exponents of $s_1$ in the expansion~\eqref{eq:triangleExp} stem from what is typically called the \emph{hard region} in the literature, which corresponds to expanding the loop integrand as is under the integral sign, and integrating term by term. The terms in eq.~\eqref{eq:triangleExp} with $\eps$-dependent powers originate instead from the \emph{collinear region} of the loop integration. Computing them requires re-scaling appropriately the loop momenta so as to catch the contributions from this region, before we expand and integrate. We discuss this thoroughly for the triangle integrals in appendix~\ref{app:AnomalyComputation}.
In a perturbative expansion around $\eps=0$, the collinear region generates terms of the form $s_1 \log(-s_1)$, which are finite at $s_1=0$ but have singular derivative. In general, the terms of order $p^2$ and $p^2 \log(-p^2)$ in the asymptotic expansion of $f(p^2)$ in eq.~\eqref{eq:commutator} give non-vanishing contributions to the commutator at $p^2=0$.

These terms in the asymptotic expansion of the amputated correlator $G^{(n)}$ prevent us from simply pulling the conformal boost generator through the on-shell limit. The amplitude $M^{(n)}$ therefore does not inherit the conformal invariance from the amputated correlator $G^{(n)}$ in general. Instead, it satisfies CWIs with an anomaly term,
\begin{align} \label{eq:anCWI}
    K^{\mu}_{d-\Delta_{\gamma}} M^{(n)}(p_1,\ldots,p_n) = \mathcal{A}^{(n)\,\mu}(p_1,\ldots,p_n) \,,
\end{align}
where we recall that the external momenta are on-shell, $p_i^2=0$. The anomaly on the right-hand side originates from the non-commutativity of the conformal boost generator and the on-shell limit. Given that the terms on the right-hand side of eq.~\eqref{eq:commutator} have a specific physical origin, we aim to find a universal description for them.

Before we present our idea to reach this goal, a word of caution is in order. The non-commutativity of the conformal boost generator with the on-shell limit not only prevents us from effortlessly deriving the implications of conformal symmetry for the scattering amplitude.
It also leads to a practical inconvenience:
Regardless of how we obtain the anomalous CWI~\eqref{eq:anCWI}, the external momenta are put on-shell and the amplitude, the anomaly, and ---~through the chain rule~--- the conformal generator are written in terms of some set of independent variables. The latter are chosen so as to implement momentum conservation and the on-shellness of the momenta, but are otherwise arbitrary.
If the conformal generator and the on-shell limit commuted, we would be free to change the variables at any step. 
Since this is not the case, deriving the CWI~\eqref{eq:anCWI} with a first choice of variables and changing it to a second one gives a different result than using the second from the beginning. The difference stems from the non-vanishing commutator in eq.~\eqref{eq:commutator} with $\Delta = d-\Delta_{\gamma}$, and is hence proportional to the anomalous dimension $\gamma$. This unusual feature may be uncomfortable, but causes no harm provided we choose the independent variables once and for all at the beginning of the derivation. 

Let us now lay out our plan to derive the implications of conformal symmetry for the scattering amplitude from the invariance of the correlator. The basic idea is to study its asymptotic expansion in the light-like limits $p_i^2 \to 0$, identify which terms contribute to the anomaly, and find a convenient method to compute them. In the next section we discuss this explicitly in the $n=3$ case. The procedure and the conclusions can be straightforwardly generalised to a generic number of particles, as we show in section~\ref{sec:multi-point}.

\subsection{Anomalous Ward identity for a three-point amplitude}
\label{sec:WardId3pt}

In this section we study what happens to the conformal properties of the three-point correlator when amputating external legs and taking the on-shell limit.
As discussed above, the crucial step is the on-shell limit. 
We will study the case where one of the external legs goes on-shell.
Studying this case will allow us to understand the mechanism which generates the anomaly in the CWI, and to generalise it to the general case.

Let us start from the amputated correlator $G^{(3)}$, which satisfies the CWI
\begin{align} \label{eq:KG3}
    K^{\mu}_{d-\Delta_{\gamma}} G^{(3)}\left(s\right) = 0 \,.
\end{align}
We choose as independent variables $s=(s_1, s_2, s_3)$, and use the chain rule to express the conformal boost generator as a differential operator in the $s_i$,
\begin{align} \label{eq:Kexp}
K^{\mu}_{\Delta} = \sum_{i=1}^3 p_i^{\mu} \hat{K}^{(i)}_{\Delta} \,,
\end{align}
where 
\begin{align} \label{eq:hatKi}
\hat{K}^{(i)}_{\Delta} = 4 s_i \frac{\partial^2}{\partial s_i^2} + 2 (d+2- 2 \Delta) \frac{\partial}{\partial s_i} \,.
\end{align}
By substituting eq.~\eqref{eq:Kexp} into eq.~\eqref{eq:KG3} and using momentum conservation we obtain the following scalar constraints,
\begin{align} \label{eq:scalarCWIs1}
& \left( \hat{K}^{(2)}_{d-\Delta_{\gamma}} - \hat{K}^{(3)}_{d-\Delta_{\gamma}} \right) G^{(3)}\left(s\right) = 0 \,, \\
\label{eq:scalarCWIs2}
& \left( \hat{K}^{(2)}_{d-\Delta_{\gamma}} - \hat{K}^{(1)}_{d-\Delta_{\gamma}} \right) G^{(3)}\left(s\right) = 0 \,.
\end{align}

Next, we take the on-shell limit $s_1 \to 0$ of the amputated correlator and define an amplitude-like object,
\begin{align}
    M^{(3)}\left(s_2,s_3\right) = \lim_{s_1 \to 0} G^{(3)}\left(s\right)\,.
\end{align}
We choose $(s_2, s_3)$ as independent on-shell variables.
In order to establish a link between the amputated correlator and the amplitude in such a way that we can translate the constraints~\eqref{eq:scalarCWIs1} and~\eqref{eq:scalarCWIs2} into conditions on the latter, we consider the asymptotic expansion of the former in the $s_1 \to 0$ limit. The result of this analysis, which we discuss in appendix~\ref{app:AnomalyComputation}, is that the amputated correlator admits the following asymptotic expansion, 
\begin{align} \label{eq:G3exp1}
    G^{(3)}\left(s\right) = \sum_{m\ge 0} (-s_1)^m \, G^{(3),\text{hard}}_{[m]}\left(s_2,s_3\right) + \sum_{m\ge 0} (-s_1)^{1-\gamma+m} \, G^{(3),\text{coll}}_{[m]}\left(s_2,s_3\right) \,.
\end{align}
The series with integer powers of $s_1$ originates from the hard region of the loop integration defining the conformal correlator $C^{(3)}$ in eq.~\eqref{eq:C3confX}. It corresponds to expanding the integral around $s_1=0$ at the integrand level. The zero-th term gives the amplitude as one would obtain it from a perturbative computation, setting $s_1=0$ from the beginning,
\begin{align}
 M^{(3)}\left(s_2,s_3\right) = G^{(3),\text{hard}}_{[0]}\left(s_2,s_3\right)\,.
\end{align}
The series with powers of $(-s_1)^{1-\gamma+m}$ in the asymptotic expansion~\eqref{eq:G3exp1} stems instead from the collinear region of the loop integration.

Upon perturbative expansion around $\eps=0$, the terms from the collinear region generate logarithms of $s_1$.
For our purpose it is convenient to spell out these logarithms, by rewriting eq.~\eqref{eq:G3exp1} as
\begin{align} \label{eq:G3exp}
G^{(3)}\left(s\right) = M^{(3)}\left(s_2,s_3\right) + \sum_{m\ge 1} \sum_{k\ge 0} s_1^m \, \log^k\left(-s_1\right) G^{(3)}_{m;k}\left(s_2,s_3\right) \,.
\end{align}
We then substitute this expansion into eqs.~\eqref{eq:scalarCWIs1} and~\eqref{eq:scalarCWIs2}. Since $\hat{K}^{(2)}_{\Delta}$ and $\hat{K}^{(3)}_{\Delta}$ do not act on $s_1$, we can set $s_1 = 0$ in eq.~\eqref{eq:scalarCWIs1}, and get a first constraint on the amplitude $M^{(3)}$,
\begin{align} \label{eq:scalarCWIM3s1}
    \left(\hat{K}^{(2)}_{d-\Delta_{\gamma}} - \hat{K}^{(3)}_{d-\Delta_{\gamma}}\right) M^{(3)}\left(s_2,s_3\right) =  0 \,.
\end{align}
On the other hand, we cannot set $s_1 = 0$ na\"ively in eq.~\eqref{eq:scalarCWIs2}, because of the operator $\hat{K}^{(1)}_{\Delta}$ on the right-hand side. Instead, we act with $\hat{K}^{(1)}_{\Delta}$ on the asymptotic expansion of the amputated correlator~\eqref{eq:G3exp}, and set $s_1 = 0$ after differentiation. The result is
\begin{align} \label{eq:scalarCWIM3s2}
    \hat{K}^{(2)}_{d-\Delta_{\gamma}} M^{(3)}\left(s_2,s_3\right) = 4 \gamma G^{(3)}_{1;0}\left(s_2,s_3\right)+4  G^{(3)}_{1;1}\left(s_2,s_3\right) \,.
\end{align}
This simple computation, the details of which can be found in Appendix~\ref{app:K1G3}, also leads to an interesting bonus result. All terms $G_{m;k}^{(3)}$ in the asymptotic expansion~\eqref{eq:G3exp} are entirely determined by a sequence of conformal boost generators acting on just two terms: the amplitude $M^{(3)}$ and $G^{(3)}_{1;1}$, namely the coefficient of $s_1 \log(-s_1)$ in the asymptotic expansion of the amputated correlator in eq.~\eqref{eq:G3exp}. Let us also note that the sum of the two terms in the expression of the anomaly on the right-hand side of  eq.~\p{eq:scalarCWIM3s2} is equal to $ -4 \gamma G_{[1]}^{(3),\text{hard}}\left(s_2,s_3\right)$, so that the collinear region contributes neither to the amplitude nor to the conformal anomaly. 

Finally, eqs.~\eqref{eq:scalarCWIM3s1} and~\eqref{eq:scalarCWIM3s2} can be combined into the desired anomalous CWI,
\begin{align} \label{eq:anomCWI}
K^{\mu}_{d-\Delta_{\gamma}} M^{(3)}\left(s_2,s_3\right) = \mathcal{A}^{(3) \mu} \,,\end{align}
where the anomaly is given by
\begin{align} \label{eq:anomalyM3}
 \mathcal{A}^{(3) \mu} = -4 p_1^{\mu} \left( \gamma G^{(3)}_{1;0}\left(s_2,s_3\right) + G^{(3)}_{1;1}\left(s_2,s_3\right) \right) \,,
\end{align}
and the conformal boost generator,
\begin{align} \label{eq:K23}
K^{\mu}_{\Delta} = \sum_{i=2,3} p_i^{\mu} \hat{K}^{(i)}_{\Delta} \,,
\end{align}
is exactly the generator~\eqref{eq:Kmu} restricted to a function of $s_2$ and $s_3$, with $s_1=0$. 
Several comments are in order.
\begin{itemize}
\item What predictive power does the anomalous CWI~\eqref{eq:anomCWI} hold? In section~\ref{sec:exactsol} we write down a closed-form expression for the anomaly, and solve the anomalous CWI for the three-particle amplitude $M^{(3)}$ exactly. 

\item In section~\ref{sec:multi-point} we discuss how the anomalous CWI generalises to the $n$-point correlator. Although in this case an all-order result is out of reach, the anomalous CWI~\eqref{eq:anomCWI} retains a strong predictive power in a perturbative approach, i.e.\ for computing the amplitude up to a fixed order in $\eps$. To this end, the computation of the anomaly must be simpler than that of the amplitude at the same order.
Let us then look at the two terms of the anomaly in eq.~\eqref{eq:anomCWI} from a perturbative point of view. 

\item The term with the coefficient $G^{(3)}_{1;0}$ is proportional to the anomalous dimension. Since the latter starts at order $\eps$, $G^{(3)}_{1;0}$ is needed at a lower order in $\eps$ than the amplitude.

\item The second term of the anomaly, containing $G^{(3)}_{1;1}$, is instead needed at the same order as the amplitude on the left-hand side. In section~\ref{sec:collinear} we propose a method to compute $G^{(3)}_{1;1}$ which relies on lower-loop information only. 

\item The tree-level amplitudes cannot contribute any logarithms, and therefore $G^{(3)}_{1;1}$ starts at order $\eps$. The entire anomaly then starts at order $\eps$, which is in agreement with the well-known conformal invariance at tree level for $\eps=0$,
\begin{align}
    K^{\mu}_{d-\Delta_{\gamma}}   M^{(3)}\left(s_2,s_3\right) = \mathcal{O}\left(\eps\right) \,.
\end{align}
\end{itemize}

In this section we have proposed a procedure to extract the conformal anomaly associated with amputating and putting on-shell one of the external legs in the three-point correlator. In the next section we will iterate this procedure on multiple legs of an $n$-point correlator, and generalise the anomalous CWI to the multi-point case.

\subsection{Conformal Ward identity for \texorpdfstring{$n$}{n}-particle amplitudes} \label{sec:multi-point}

In the previous sections we considered in detail the three-point correlator at the conformal fixed point, and studied its conformal anomaly generated by putting one of its legs on-shell. Many of the observed properties hold for $n$-point correlators when putting a subset $\Lambda$ of its legs on-shell,
\begin{align}
s_i \equiv p_i^2 = 0 \; , \qquad i \in \Lambda \subseteq \{1,\ldots,n\} \,.
\end{align}
As we have already mentioned above, the precise form of the conformal anomaly equation depends on the choice of independent kinematic variables. Nevertheless, some generic features, which we develop below, hold for any choice of the variables. 

We complement $\{ s_i \}_{i \in \Lambda} \equiv s_{\Lambda} $  with a set $v$ of Mandelstam variables in order to parameterise the kinematics of the $n$-point correlator. Putting $\Lambda$ of the $n$ legs of the amputated correlator $G^{(n)}$ on-shell results in an amplitude-like object $M^{(n)}$, 
\begin{align}
M^{(n)}\left(v \right) = \left( \prod_{i\in \Lambda }\lim_{s_i \to 0} \right) G^{(n)}\left(v, s_{\Lambda}\right) \,.
\end{align}

The amputated correlator $G^{(n)}$ is conformally invariant with conformal weight $d-\Delta_\gamma$ with respect to each leg. Among the conformal algebra generators, only the conformal boost symmetry is broken by an anomaly in the on-shell limit. We rewrite the momentum-space conformal boost generator~\p{eq:Kmu} in the Mandelstam variables $v, s_{\Lambda}$ using the chain rule as 
\begin{align}
K^\mu_{\Delta} = \sum_{i=1}^{n} p_i^\mu K^{(i)}_\Delta    \,.
\label{eq:Knsum}
\end{align}
The scalar operators $K^{(i)}_\Delta$ have the following form,
\begin{align}
& K^{(i)}_\Delta  = \mathbb{K}^{(i)}_\Delta + \hat{K}^{(i)}_\Delta + \sum_{j\in \Lambda} s_j \pa_{s_j} A^{(ij)} +   \sum_{j\in \Lambda} s_j B^{(ij)} \;, \qquad i \in \Lambda \,, \label{eq:Konshell}
\end{align}
where $\hat{K}^{(i)}_\Delta$ defined in eq.~\p{eq:hatKi} acts only on the $s_i$ variables. As we will see below, only this piece of the conformal boost generates the anomaly on the light-cone. The differential operators $\mathbb{K}^{(i)}_\Delta ,  A^{(ij)}, B^{(ij)}$ act only on Mandelstam invariants $v$, and do not involve any of the $s_{\Lambda}$. Their explicit form can be easily worked out for any given choice of the Mandelstam variables $v$. Similarly, the operators referring to the remaining legs are as follows,
\footnote{The three-point correlator and the amplitude considered in section~\ref{sec:WardId3pt} correspond to $\Lambda =\{1\}$, $s_{\Lambda} = \{ s_1\}$, $v = \{ s_2 ,s_3\}$ in the notation of the current section. Also, we have $A^{(ij)} = B^{(ij)} = \mathbb{K}_{\Delta}^{(1)} = 0$, $\mathbb{K}_{\Delta}^{(2)} = \hat{K}^{(2)}_\Delta$ and $\mathbb{K}_{\Delta}^{(3)} = \hat{K}^{(3)}_\Delta$.\label{ftnt:compare3}} 
\begin{align}
& K^{(i)}_\Delta  = \mathbb{K}^{(i)}_\Delta  + \sum_{j\in \Lambda} s_j \pa_{s_j} A^{(ij)} +   \sum_{j\in \Lambda} s_j B^{(ij)} \;, \qquad i \in \{1,\ldots,n\} \backslash \Lambda \,.
\label{eq:Koffshell}
\end{align}

In the asymptotic expansion of the amputated correlator in the on-shell limit $s_i \to 0$ for all $i\in \Lambda$, each of the legs $\Lambda$ contributes either $s_i^{m_i}$ with $m_i \geq 0$, or $s_i^{m_i} \log^{k_i}(-s_i)$ with $m_i \geq 1$ and $k_i \geq 1$,
\begin{align}
G^{(n)}\left(v, s_{\Lambda} \right) & = \sum_{\substack{m_i \geq k_i \geq 0 \\ k_i \geq m_i \geq 1 \\ \text{for all } i \in \Lambda }} \Biggl( \prod_{j \in \Lambda} s_j^{m_j} \log^{k_j}(-s_j) \Biggr) G^{(n)}_{\{ m_{l};k_{l}\}_{l\in {\Lambda}}}(v) \,. \label{eq:GnAsymp}
\end{align}
The amplitude $M^{(n)}$ is the term of the asymptotics with $m_i = 0$ and $k_i=0$ for all $i\in \Lambda$. 

In order to formulate the conformal anomaly equation for the amplitude $M^{(n)}$, only $2|\Lambda|$ coefficients in the asymptotic expansion~\p{eq:GnAsymp} are required. They are those with one pair $(m_i,k_i)=(1,1)$ or $(m_i,k_i) =(1,0)$ and all remaining pairs $(m_j,k_j)=(0,0)$. 
Given their importance, let us introduce a notation for them,
\begin{align}
    M^{(n)}\left( v \right) := \ &  G^{(n)}\left( v, s_{\Lambda} \right) |_{s_i =0}\,, \\
       G^{(n)}_{i;\text{pow}}( v) := \ &  \bigl[ G^{(n)}\left( v, s_{\Lambda} \right) \bigr]_{s_i}\,,  \\
            G^{(n)}_{i;\text{coll}}( v)   := \ &  \bigl[ G^{(n)}\left( v, s_{\Lambda} \right) \bigr]_{s_i \log(-s_i)}\,.
\end{align}

Acting on the asymptotic expansion~\p{eq:GnAsymp} with the conformal boost generator~\p{eq:Knsum} (with eqs.~\p{eq:Konshell} and~\p{eq:Koffshell}), and putting the legs $\Lambda$ on shell ($s_i \to 0$), we immediately conclude that the anomaly comes from the $\hat{K}^{(i)}_{d-\Delta_\gamma}$ piece~\p{eq:hatKi} of the conformal boost generator. It picks out the $\sim s_i$ and $\sim s_i \log(-s_i)$ terms of the asymptotic expansion of the amputated correlator~\p{eq:GnAsymp}. 

Putting everything together, we find the following  anomalous conformal Ward identity for the at the conformal fixed point,
\begin{equation}
\boxed{\Biggl(\sum_{i=1}^n p_i^\mu \, \mathbb{K}^{(i)}_{d-\Delta_\gamma} \Biggr) M^{(n)}\left( v \right) = {\cal A}^{(n)\mu} } \label{eq:KMnAnom}
\end{equation}
with the anomaly given by
\begin{equation} \label{eq:KMnAnomaly}
\boxed{{\cal A}^{(n)\mu} \equiv - 4\sum_{i \in \Lambda} p_i^\mu \left( \gamma G^{(n)}_{i;
\text{pow}}(v) + G^{(n)}_{i;\text{coll}}(v) \right) 
}
\end{equation}
This, together with the specific form of the conformal anomaly terms to be derived in the next section, is the main result of this paper.
It generalises a result of \cite{Chicherin:2017bxc} for finite conformal integrals to amplitudes at a conformal fixed point in $d$ dimensions.
Let us discuss the properties of our anomalous CWIs.
\begin{itemize}
\item The anomaly in eq. (\ref{eq:KMnAnomaly}) has a local nature: for each external leg $i\in \Lambda$ that is put on-shell, there is a separate contribution, proportional to the momentum $p_i^{\mu}$.
\item The first anomaly term is proportional to the field anomalous dimension. Therefore constraining the amplitude $M^{(n)}$ up to a certain order in $\eps$ requires the computation of $G^{(n)}_{i;\text{pow}}$ to one order lower only.
\item The second anomaly term, $G^{(n)}_{i;\text{coll}}$, has a collinear origin. We will show in the next subsection that it is described by a convolution of lower-loop amplitudes with a universal collinear kernel, which we determine here up to the two-loop order.
\item The last two points imply that determining the anomaly in perturbation theory is easier compared to computing the amplitudes directly.
This makes the Ward identities useful in practice.
\end{itemize}

\subsection{Collinear anomaly from analysis of regions}
\label{sec:collinear}

In this section we use the method of the analysis of regions of Feynman integrals~\cite{Beneke:1997zp,Smirnov:1999bza,Jantzen:2011nz} to find a formula which allows us to compute efficiently the collinear part of the conformal anomaly, $G^{(n)}_{i,\text{coll}}$ in eq.~\eqref{eq:KMnAnomaly}, and to unveil its origin from the region of the loop integration where some of the loop momenta become collinear with one of the external on-shell momenta.
 We find that $G^{(n)}_{i,\text{coll}}$ is given by the convolution of a universal function and lower-loop, higher-point amplitudes.
This generalises the previous result of ref.~\cite{Chicherin:2017bxc}
to non-integer spacetime dimensions. 

\subsubsection{Review of the collinear anomaly in integer dimensions}

The authors of ref.~\cite{Chicherin:2017bxc} studied finite Feynman integrals in integer dimensions in a number of quantum field theories with classically conformal Lagrangian. They discovered that, whenever one of the external legs is on shell, the conformal invariance of the integral is broken by a contact term localised on the configurations where the loop momentum is collinear to the external on-shell momentum. For scalar $\phi^3$ theory in $d=6$ dimensions, the master formula which gives the contribution to the conformal anomaly from each on-shell corner (i.e.\ a trivalent vertex with an external on-shell leg) is
\begin{align} \label{eq:CollinearAnomaly}
    K^{\mu}_{d-\Delta_{\phi}} \frac{1}{[q^2 +\textup{i}0][(p+q)^2+\textup{i}0]} = 4 \textup{i} \pi^3 p^{\mu} \int_0^1 \mathrm{d}\xi \, \xi (1-\xi) \delta^{(6)}(q+\xi p) \,,
\end{align}
where $p$ is the external on-shell momentum ($p^2 = 0$), $q$ is off-shell ($q^2\neq 0$), and $d-\Delta_{\phi} = 4$. Whenever the on-shell corner is part of a loop, the $\delta$ function localises the adjacent loop integration on the collinear configuration $q = - \xi p$. 

Consider now some (finite) $\ell$-loop Feynman diagram ${\cal F}^{(n+1)}_{d=6}$ in the $su(\textsc{n})$ $\phi^3$ theory. We focus on the corner around the on-shell momentum $p$,
\begin{align}
{\cal F}^{(n+1)}_{d=6}(p^a,p_1^{a_1},\ldots,p_n^{a_n}) = - \textup{i} g d^{abc}\int \frac{\mathrm{d}^6 q}{(2\pi)^6} \frac{1}{q^2 (p+q)^2} {\cal G}^{(n+2)}_{d=6}(q^b,(p-q)^c,p_1^{a_1},\ldots,p_n^{a_n}) \,.  \label{eq:Fd6split}   
\end{align}
Equation~\eqref{eq:CollinearAnomaly} then gives
\begin{align}
 K^{\mu}_{d-\Delta_{\phi}}   {\cal F}^{(n+1)}_{d=6}(p^a,p_1^{a_1},\ldots,p_n^{a_n}) = 4 p^\mu \frac{g d^{abc}}{(4\pi)^3} \int\limits^1_0 \mathrm{d}\xi \, \xi \bar\xi \, {\cal G}^{(n+2)}_{d=6}(\xi p^b, \bar\xi p^c, p_1^{a_1},\ldots,p_n^{a_n}) \,, \label{eq:F6anomaly1} 
\end{align}
where we use the short-hand notation $\bar\xi := 1-\xi$.
According to eqs.~\p{eq:CollinearAnomaly} and~\eqref{eq:F6anomaly1}, the only contribution comes from the region of loop integrations where the collinear light-like momenta $\xi p$ and $(1-\xi) p$ flow through the two propagators in eq.~\p{eq:Fd6split}. In other words, all momenta entering the subdiagram $({\cal F}_{d=6} \backslash  {\cal G}_{d=6})$ are collinear in the relevant region of loop integrations. 
We give a pictorial representation of this relation in figure~\ref{fig:collinear_master}. 

As a result of eq.~\eqref{eq:F6anomaly1}, the anomaly is entirely determined by lower-loop information. This allows for the computation of suitable integrals through the solution of the corresponding anomalous CWIs~\cite{Chicherin:2018rpz,Zoia:2018sin,Chicherin:2018ubl}. {An analogous collinear mechanism is also responsible for the anomalies of the Yangian symmetry of the on-shell fishnet graphs~\cite{Chicherin:2022nqq}.}

\begin{figure}[t!]
\begin{center}
\begin{tikzpicture}[scale=0.4, decoration={markings, mark=at position 0.6 with {\arrow{>}}}]
      
	\node[text width=1.cm] at (-7.5,0.6){\begin{equation*} K^{\mu}_{d-\Delta_{\phi}} \int \mathrm{d}^6 q \end{equation*}};
	\node[text width=5.cm] at (9.7,0.16){\begin{equation*} \sim p^{\mu} \int_0^1 \mathrm{d}\xi \, \xi \bar{\xi}\end{equation*}};

	\draw (2,0) ellipse (2.2cm and 1.7cm);

	\draw (.22,-1) -- (-1.3,0);
	\draw (.22,1) -- (-1.3,0);
	\draw (-2.8,0) -- (-1.3,0);	
	
	\draw (3.77,-1.) -- (5.8,-1.7);
	\draw (3.77,1.) -- (5.8,1.7);
	\draw (4.075,-0.55) -- (5.8,-1.);
	\draw (4.075,0.55) -- (5.8,1.);
	\node[text width=0.2cm] at (4.8,0) {...};
	
	\draw (18.5,0) ellipse (2.2cm and 1.7cm);	

	\draw (16.72,1) -- (15.2,1.5);
	\draw (16.72,-1) -- (15.2,-1.5);
	\node[text width=1.cm] at (15.,-1.62){$\xi p$};	
	\node[text width=1.cm] at (15.,1.6){$\bar{\xi} p$};
	
    \draw (20.27,-1.) -- (22.3,-1.7);
	\draw (20.27,1.) -- (22.3,1.7);
	\draw (20.575,-0.55) -- (22.3,-1.);
	\draw (20.575,0.55) -- (22.3,1.);
	\node[text width=0.2cm] at (21.3,0) {...};

	\draw (.22,-1) -- (-1.3,0);
	\draw (.22,1) -- (-1.3,0);
	\draw (-2.8,0) -- (-1.3,0);	
	
	\node[text width=0.3cm] at (-1.82,0.54) {$p$};
	\node[text width=0.3cm] at (-0.83,-1.07) {$q$};

\end{tikzpicture}
\end{center}
\caption{
Pictorial representation of the collinear anomaly mechanism. The conformal boost generator, acting on a UV-finite diagram in six dimensions with an on-shell external leg ($p^2=0$), produces a contact anomaly localised on the configuration where the loop momentum is collinear to $p^{\mu}$. The contact anomaly freezes the adjacent loop integration, giving a finite anomaly entirely determined by lower loop information. The collinear anomaly receives one such contribution for each of the external on-shell legs.}
\label{fig:collinear_master}
\end{figure}

\subsubsection{Collinear anomaly from logarithmically enhanced terms of the correlator}

Let us now formulate the collinear anomaly mechanism in a way which allows for the extension to non-integer $d$. 
Recall from eq.~\eqref{eq:commutator} that in general we expect two terms in the conformal anomaly equation. In integer dimensions, however, we put $\gamma$ to zero, and we focus on the other term.
On the one hand, we can see from eq.~\eqref{eq:commutator} that the conformal boost generator picks out only the term $p^2 \log (-p^2)$ of the amputated correlator as $p^2\to 0$.
On the other hand, eq.~\eqref{eq:F6anomaly1} gives a precise formula for the conformal anomaly. 
Comparing the two expressions yields
\begin{align}
\left[ {\cal F}^{(n+1)}_{d=6}(p^a,p_1^{a_1},\ldots,p_n^{a_n}) \right]_{p^2 \log(-p^2)} = \frac{g d^{abc}}{(4\pi)^3} \int\limits^1_0 \mathrm{d}\xi \, \xi \bar\xi \, {\cal G}^{(n+2)}_{d=6}(\xi p^b, \bar\xi p^c, p_1^{a_1},\ldots,p_n^{a_n}) \,. \label{eq:F6xlogx} 
\end{align}
This provides a practical recipe to evaluate the term $p^2 \log(-p^2)$ in the asymptotic expansion of ${\cal F}_{d=6}$, based on the knowledge of the lower-loop subdiagram ${\cal G}_{d=6}$.

\begin{figure}
    \centering
    \includegraphics[width=0.25\textwidth]{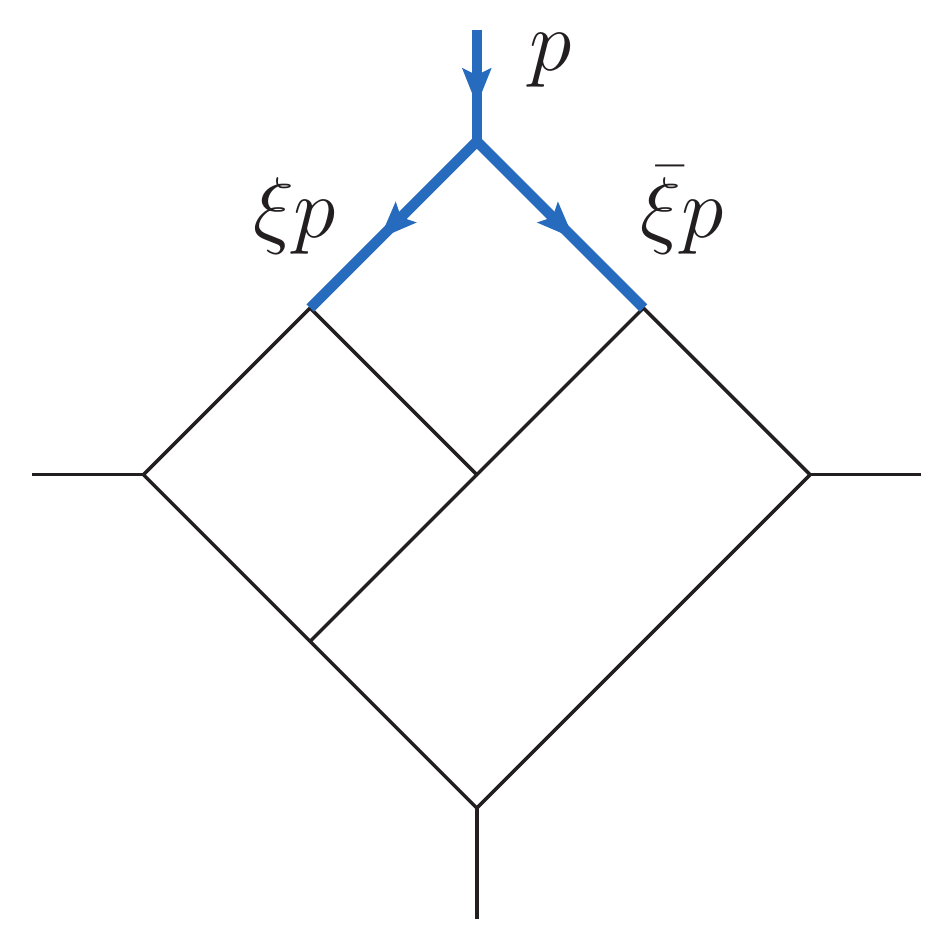}
    \caption{An example of a four-point three-loop Feynman diagram ${\cal F}^{(4)}_{d=6}$ having the tennis-court topology.
       The only region of loop integration which contributes to the asymptotics $p^2 \log(-p^2)$ at $p^2 \to 0$ is the one of collinear momenta, as indicated in the figure.
       }
    \label{fig:collanom4pt}
\end{figure}

Let us now reproduce and understand eq.~\eqref{eq:F6xlogx} in a different way.
We are interested in the kinematic limit $p^2 \to 0$ of a correlation function, and in particular in the term proportional to $p^2 \log (-p^2)$ in that limit.
We use the region expansion analysis~\cite{Beneke:1997zp,Smirnov:1999bza,Jantzen:2011nz} of Feynman diagrams to analyse this limit.

When an external momentum $p^\mu$ is taken to be on-shell, we find
different relevant contributions from the region expansion analysis.
In addition to the so-called hard region, which corresponds to the na\"ive limit, one finds that collinear regions, where one or more loop momenta become collinear to $p^\mu$, are relevant.

For example, for the diagram shown in figure~\ref{fig:collanom4pt}, we find that the only term contributing to $p^2 \log(-p^2)$ in the limit comes from the one-loop collinear region indicated in the figure.
In fact, for (ultraviolet-)finite Feynman diagrams (as those considered in ref.~\cite{Chicherin:2017bxc}), one finds that only one-loop collinear regions can produce logarithmically enhanced terms in the limit. In this way, one recovers eq.~\eqref{eq:F6xlogx}.
In this paper, we do not want to make any assumptions on ultraviolet power counting, and perform an analysis in arbitrary dimension $d$. As we will see in the next subsection, this means that in addition to one-loop collinear regions, also multi-loop collinear regions are relevant to the conformal anomaly.

\subsubsection{Region analysis in non-integer dimensions}

Let us now reproduce the previous results from a region analysis, and generalise the six-dimensional formula~\eqref{eq:CollinearAnomaly} to the $d$-dimensional case. To make the discussion more accessible we begin with explicit examples at one- and two-loop order, before we present the general result.

{\it One-loop example:} We start with the one-loop triangle diagram which gives the one-loop term of the amputated correlator $G^{(3)}$. We put on-shell $p_1$ ($s_1=0$) and we want to compute the $s_1 \log(-s_1)$-term of the asymptotic expansion in the limit $s_1 \to 0$. We express this diagram in terms of the triangle integral from the family defined in eq.~\eqref{eq:3massTriangle} as
\begin{equation}
\vcenter{\hbox{\begin{tikzpicture}
\node at (0,0){\includegraphics[scale=0.13]{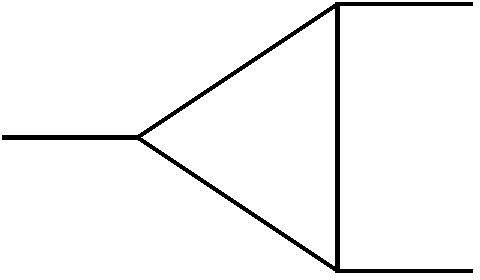}};
\node at (-1.45,0){$p_1^{a}$};
\node at (1.5,-0.65){$p_2^{b}$};
\node at (1.5,0.65){$p_3^{c}$};
\end{tikzpicture}}} = \textup{i} g u \dfrac{\textsc{n}^2-12}{2 \textsc{n}} d^{a b c} I(1,1,1) \,.
\label{eq:triangFD}
\end{equation}
From the expansion in eq.~\eqref{eq:triangleExp}, we see that the $s_1 \log(-s_1)$-term of the triangle integral is given by $\eps$ times the zeroth term from the collinear region. We compute the latter using the method of the expansion by regions in appendix~\ref{app:AnomalyComputation}. Using eq.~\eqref{eq:Icoll0int}, where we replace $\alpha_1$ with $\xi$ to make contact with ref.~\cite{Chicherin:2017bxc}, we find
\begin{align}
\begin{aligned}
    \left[ I(1,1,1) \right]_{s_1 \log(-s_1)} & = \eps I^{\text{coll}}_{[0]}(1,1,1) \\
      & = e^{\eps \gamma_{\text{E}}} \eps \Gamma\left(\eps-1\right) \int_0^1 \mathrm{d}\xi \frac{(\xi \bar{\xi})^{1-\eps}}{\xi (s_2-s_3)-s_2} \,.
\end{aligned}
\end{align}
The denominator of the integrand can be viewed as a tree-level four-point diagram with momenta (all incoming) $\left(\xi p_1, \bar{\xi} p_1, p_2, p_3\right)$ in the $(p_2+\xi p_1)^2$ channel. The $s_1 \log(-s_1)$-term of the asymptotic expansion of the triangle diagram can then be expressed as
\begin{equation}
\label{eq:traing_coll}
\Biggl[ \vcenter{\hbox{\begin{tikzpicture}
\node at (0,0){\includegraphics[scale=0.13]{figures/triangle_collinear.png}};
\node at (-1.45,0){$p_1^{a}$};
\node at (1.5,-0.65){$p_2^{b}$};
\node at (1.5,0.65){$p_3^{c}$};
\end{tikzpicture}}} \Biggr]_{s_1 \log(-s_1)} = \frac{g \, d^{a h_1h_2}}{(4\pi)^{3}}  \int_0^1 \mathrm{d}\xi \, \Omega^{\text{1L}}\left(\xi,\eps\right) 
\vcenter{\hbox{
\begin{tikzpicture}
\node at (0,0){\includegraphics[scale=0.13]{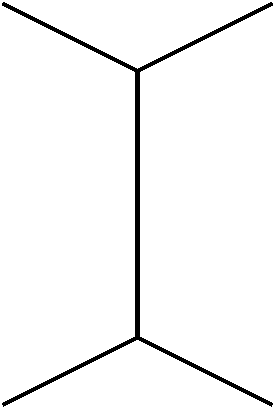}};
\node at (-1.,-1){$\xi p_1^{h_1}$};
\node at (-1.,1){$\bar{\xi} p_1^{h_2}$};
\node at (0.9,-1){$p_2^{b}$};
\node at (0.9,1){$p_3^{c}$};
\end{tikzpicture}}} \,,
\end{equation}
where the one-loop collinear function $\Omega^{\text{1L}}$ is given by
\begin{align} \label{eq:Om1loop}
\begin{aligned}
     \Omega^{\text{1L}}\left(\xi,\eps\right) & = \left[-\eps \Gamma(\eps-1) e^{\eps \gamma_{\text{E}}} \right] \, (\xi \bar{\xi})^{1-\eps}  \\
     & = \xi \bar{\xi} + \ep \, \xi \bar\xi \left[1-\log(\xi \bar\xi)\right] +{\cal O}(\ep^2)  \,.
\end{aligned}
\end{align}
This is in agreement with the six-dimensional eq.~\p{eq:F6xlogx}, and generalises it by also taking into account $\ep$-corrections which are necessary to go to higher orders in $\eps$ at the conformal fixed point.

{\it Two-loop example:}
As a more nontrivial example, we consider a two-loop triangle which is one of several Feynman diagrams contributing to the correlator $G^{(3)}$ at the two-loop order, 
\begin{align}
\label{eq:Jdiagr}
\vcenter{\hbox{\begin{tikzpicture}
\node at (0,0){\includegraphics[scale=0.3]{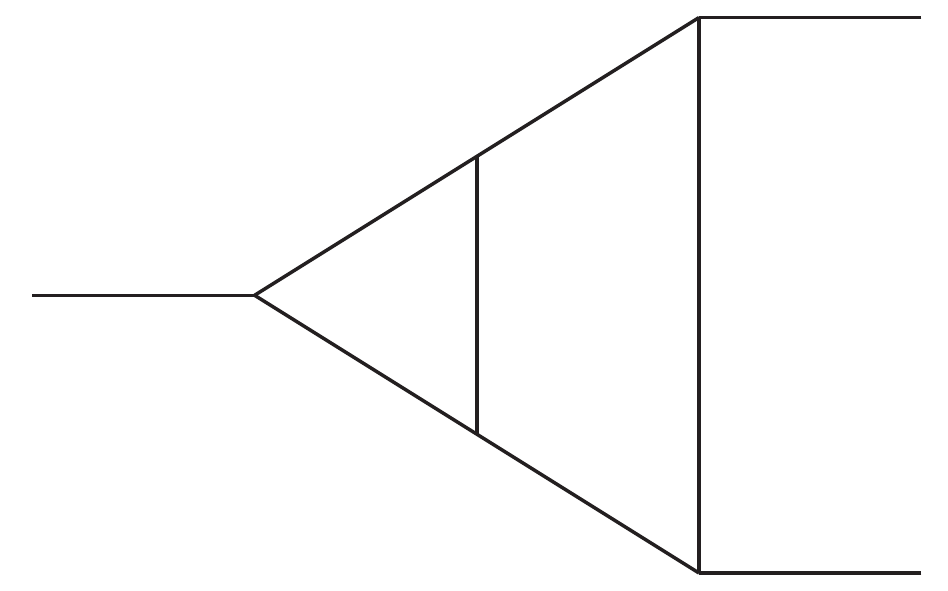}};
\node at (-1.65,0){$p_1^{a}$};
\node at (1.7,-0.8){$p_2^{b}$};
\node at (1.7,0.8){$p_3^{c}$};
\end{tikzpicture}}} 
= \textup{i} g u^2 \left(\dfrac{\textsc{n}^2-12}{2 \textsc{n}} \right)^2 d^{abc} J
\end{align}
where $J$ is the following two-loop scalar Feynman integral,
\begin{align}
J = e^{2\ep \gamma_\text{E}} \int \frac{\mathrm{d}^d k \, \mathrm{d}^d l}{\bigl(\textup{i}\pi^{\frac{d}{2}}\bigr)^2} \frac{1}{k^2 \, l^2 \, (k+l)^2 (k+p_1)^2 (l-p_1)^2 (l+p_2)^2}\,.
\end{align}
The propagators are assigned small imaginary parts as in eq.~\eqref{eq:3massTriangle}.
The asymptotics of $J$ at $p_1^2 \equiv s_1 \to 0$ is similar to that of the one-loop triangle given in eq.~\p{eq:triangleExp}. The new feature is that there are two regions of loop integrations which are responsible for the collinear terms. We refer to them as one-loop-collinear and two-loop-collinear,
\begin{align}
\label{eq:asympJ}
J = \sum_{m\ge 0} (-s_1)^m J^{\text{hard}}_{[m]} + \sum_{m\ge 0} (-s_1)^{1-\eps+m} J^{\text{1L-coll}}_{[m]} + \sum_{m\ge 0} (-s_1)^{1-2\eps+m} J^{\text{2L-coll}}_{[m]}\,.     
\end{align}
Both of them contribute to the $s_1 \log(-s_1)$-term of the asymptotics,
\begin{align}
\left[ J \right]_{s_1 \log(-s_1)} = \ep J^{\text{1L-coll}}_{[0]} + 2 \ep J^{\text{2L-coll}}_{[0]}\,,
\end{align}
which is responsible for the collinear part of the amplitude conformal anomaly. 
The expansion by regions analysis yields the following expressions for the one-loop-collinear and the two-loop-collinear  contributions, 
\begin{align}
\label{eq:2Ltraing_coll1L}
\textup{i} g u^2 \left(\dfrac{\textsc{n}^2-12}{2 \textsc{n}} \right)^2 d^{abc} \, J^{\text{1L-coll}}_{[0]} =
\frac{g \, d^{a h_1h_2}}{(4\pi)^{3}} \int^1_0 \mathrm{d}\xi
\,\Omega^{\text{1L}}(\xi,\ep) 
\vcenter{\hbox{
\begin{tikzpicture}
\node at (0,0){\includegraphics[scale=0.42]{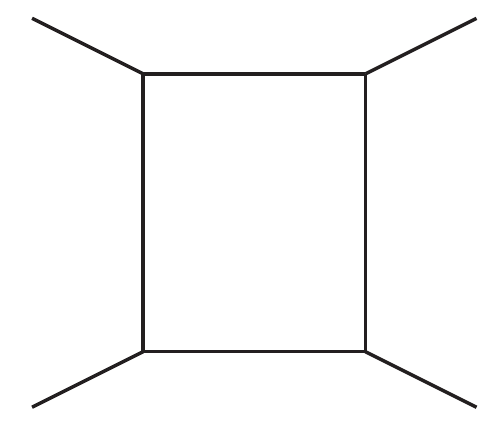}};
\node at (-1.4,-0.9){$\xi p_1^{h_1}$};
\node at (-1.4,0.9){$\bar{\xi} p_1^{h_2}$};
\node at (1.3,-0.9){$p_2^{b}$};
\node at (1.3,0.9){$p_3^{c}$};
\end{tikzpicture}}}
\end{align}
where $\Omega^{\text{1L}}(\xi,\ep)$ is the same as in the one-loop example \p{eq:Om1loop}, and
\begin{align}
\label{eq:2Ltraing_coll2L}
\textup{i} g u^2 \left(\dfrac{\textsc{n}^2-12}{2 \textsc{n}} \right)^2 d^{abc}
J^{\text{2L-coll}}_{[0]}=
\frac{g u \, d^{a h_1h_2}}{(4\pi)^{3}} \int^1_0 \mathrm{d}\xi  \, \Omega^{\text{2L}}_{\rm dbl\;triang}(\xi,\ep)
\vcenter{\hbox{
\begin{tikzpicture}
\node at (0,0){\includegraphics[scale=0.13]{figures/triangle_collinear_anomaly.png}};
\node at (-1.,-1){$\xi p_1^{h_1}$};
\node at (-1.,1){$\bar{\xi} p_1^{h_2}$};
\node at (0.9,-1){$p_2^{b}$};
\node at (0.9,1){$p_3^{c}$};
\end{tikzpicture}}}
\end{align}
with the two-loop collinear function  
\begin{align}
\label{eq:Ome2L triang}
\Omega^{\text{2L}}_{\rm double\;triangle}(\xi,\ep)=  \dfrac{\textsc{n}^2-12}{4 \textsc{n}} \left[ \frac{2\xi\bar\xi}{\ep} + \left( 10\xi\bar\xi - 3\xi\bar\xi \log(\xi\bar\xi) + \xi\log(\xi) + \bar\xi\log(\bar\xi) \right)+ {\cal O}(\ep) \right] \,.
\end{align}
There are two different contributions to the $\ep$-pole in the previous expression. One comes from the UV-divergence of the one-loop triangle subdiagram of \p{eq:Jdiagr}, and the other stems from the collinear divergence in the one-loop box subdiagram of \p{eq:Jdiagr}. The UV-divergence is cancelled out by the corresponding counter-term diagram, which is the one-loop  diagram \p{eq:triangFD} with the counter-term vertex. The collinear function $\Omega$ for the counter-term diagram contains an $\ep$-pole which cancels out the UV-contribution to the pole in eq.~\p{eq:Ome2L triang}. The cancellation of the collinear contribution to the pole is discussed below. 

Equations~\p{eq:traing_coll}, \p{eq:2Ltraing_coll1L}, and \p{eq:2Ltraing_coll2L} are very suggestive. The term $s_1\log(-s_1)$ in the asymptotics of a Feynman diagram comes from subdiagrams with all internal momenta localised on a collinear configuration, which turn into a function $\Omega$ of the collinear splitting parameter $\xi$ and $\eps$.

\begin{figure}
    \centering
    \includegraphics[width=0.25\textwidth]{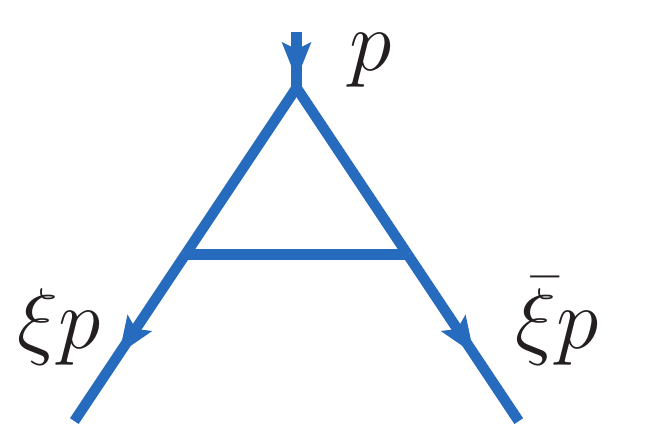}\qquad\qquad
    \includegraphics[width=0.25\textwidth]{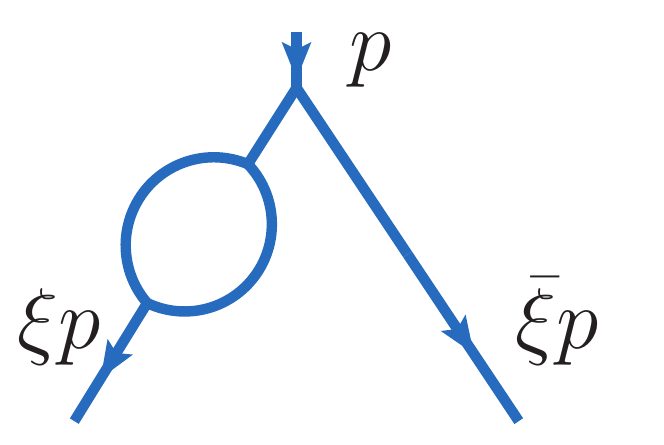}
    \qquad\qquad
    \includegraphics[width=0.25\textwidth]{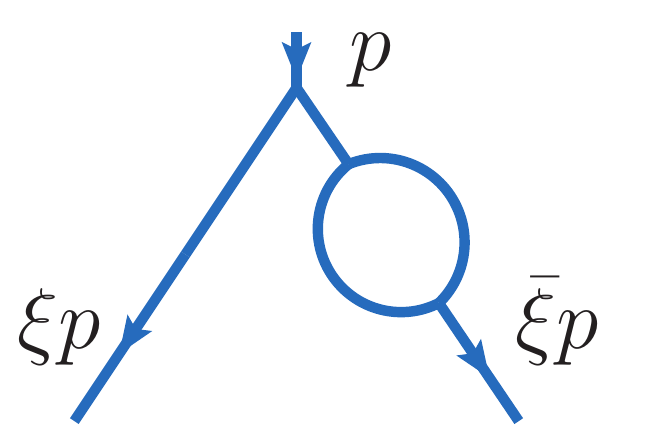}
    \caption{Subdiagrams with five propagators which contribute to the two-loop collinear function $\Omega$. They are to be complemented with the corresponding counter-term diagrams. In the relevant region of loop integrations all internal momenta are collinear with $p$.}
    \label{fig:omega123}
\end{figure}

Besides the Feynman diagram \p{eq:Jdiagr}, there are several other Feynman diagrams (including those with counter-term vertices) contributing to the three-point correlator $G^{(3)}$ at the two-loop order. Their asymptotics at $s_1 \to 0$ has the same form as in \p{eq:asympJ}. Calculating the $\Omega$ function for each of them and taking their sum, we obtain the full two-loop collinear function (see figure~\ref{fig:omega123}),
\begin{align} \label{eq:Om2loop}
\Omega^{\text{2L}}(\xi,\ep) = \, & \frac{\textsc{n}^2-12}{4 \textsc{n}} \frac{\xi \bar\xi}{\ep} + \frac{59 \textsc{n}^2 - 884}{36 \textsc{n}} \xi \bar\xi - \frac{5\textsc{n}^2 - 68}{12 \textsc{n}} \xi \bar\xi \log(\xi \bar\xi) \notag\\
&+ \frac{\textsc{n}^2-12}{4 \textsc{n}} \left[ \xi \log(\xi) + \bar\xi \log(\bar\xi) \right] + {\cal O}(\ep) \,.
\end{align}
The $\ep$-pole in the two-loop $\Omega$ is of collinear origin.

{\it Power counting in the collinear regions:} 
A simple power-counting argument enables us to determine the asymptotics of a given collinear region of loop integrations. Let us consider a connected subdiagram ${\cal H}$ with $E$ external points and involving $N$ propagators. The subdiagram ${\cal H}$ is adjacent to the corner with inflowing momentum $p^\mu$, which approaches the light-cone $p^2\sim 0$ (see the one- and two-loop examples in figures~\ref{fig:collanom4pt} and~\ref{fig:omega123}). We keep in mind that ${\cal H}$ is to be attached to a diagram ${\cal G}$ such that all momenta flowing through ${\cal H}$ can be collinear with $p^\mu$ whereas the momenta flowing through ${\cal G}$ are generic, and $\ell$ independent momenta flowing through ${\cal H}$ are integrated over.\footnote{Let us note that $\ell$ is not equal to the number of loops in ${\cal H}$, i.e.\ the independent nontrivial one-cycles. There are $\ell + 2 -E$ independent nontrivial one-cycles in ${\cal H}$.} The region of loop integrations over $\ell$ momenta of ${\cal H}$, where the momenta flowing through all $N$ propagators are collinear with $p^\mu$, has the asymptotics
\begin{align}
\left(\frac{1}{p^2} \right)^{N} \left( \mathrm{d}^{d} p \right)^{\ell} \sim (p^2)^{E-2-\ell\ep}\,. \label{eq:countp}
\end{align}
Here 
we applied standard graph relations to obtain $3\ell-N = E-2$. Expanding in $\ep$ the right-hand side of eq.~\p{eq:countp}, we see that the $p^2 \log(p^2)$ terms come from the subdiagrams with $E=3$. The subdiagrams in figures~\ref{fig:collanom4pt} and~\ref{fig:omega123} have $E=3$ external points, corresponding to momenta $p,\,\xi p,\,\bar\xi p$. The subdiagrams with $E>3$, where $p^\mu$ is split into $E-1$ collinear fractions, contribute to $(p^2)^{E-2} \log(p^2)$, and so are more suppressed as compared to $p^2 \log(p^2)$ and do not contribute to the conformal anomaly.

UV-finite Feynman diagrams in $d=6$ do not contain two- and three-point subdiagrams with nontrivial one-cycles. We therefore conclude that only subdiagrams with $\ell=1$ of finite Feynman integrals have $E=3$. 
See e.g.\ figure~\ref{fig:collanom4pt}. As a result, in $d=6$ the collinear anomaly is a ``short-range'' effect, namely it receives contribution only from the collinear region of the loop integration adjacent to the on-shell momentum. However, generic Feynman graphs in the $d=6-2\eps$ dimensional model~\p{eq:action} contain multi-loop UV-divergent subgraphs with $E=3$ and $\ell>1$. Figure~\ref{fig:omega123} shows examples with $\ell=2$. Thus, to study the collinear part of the conformal anomaly we have to take into account ``long-range'' effects as well.

Finally, let us note that the counting of eq.~\p{eq:countp} changes for UV-finite Feynman integrals in $d=4$ Yukawa theory. As a result, the ``long-range'' effects impact the collinear anomaly, namely subdiagrams with $\ell>1$ have to be taken into account as well.

{\it General structure:} Based on the previous one-loop and two-loop examples, and the power counting argument, let us summarise the outcome of expansion by region analysis for a generic multi-loop Feynman diagram.
In order to describe the $p^2\log(-p^2)$ contribution to the asymptotics of a Feynman diagram ${\cal F}$ at $p^2 \sim 0$, we split ${\cal F}$ into a subdiagram ${\cal G}$ with $n+2$ inflowing momenta $q,p-q,p_1,\ldots,p_n$, and the complementary subdiagram $({\cal F}\backslash {\cal G})$ with three inflowing momenta $p,-q,-p+q$.\footnote{In the notations of the previous paragraph ${\cal H} = ({\cal F}\backslash {\cal G})$ and $E=3$.} The subdiagram ${\cal G}$ is $\ell'$-loop ($0 \leq \ell'<\ell$), and $(\ell-\ell')$ independent loop momenta flow through $({\cal F}\backslash {\cal G})$. The subdiagram $({\cal F}\backslash {\cal G})$ contains $3(\ell-\ell')-1$ propagators.
The terms $p^2\log(-p^2)$ come from the region of loop integrations of the subdiagram $({\cal F}\backslash {\cal G})$ where all propagators carry momenta collinear with $p$, so its inflowing momenta are also collinear: $p,-\xi p, -\bar\xi p$. This collinear region of loop integrations results in a function $\Omega_{{\cal F} \backslash {\cal G}}(\xi)$, and summing over all possible ways of splitting ${\cal F}$ into ${\cal G}$ and $({\cal F}\backslash{\cal G)}$ as described above we find
\begin{align}
\left[ {\cal F}(p^a,p_1^{a_1},\ldots,p_n^{a_n}) \right]_{p^2 \log(-p^2)} = \frac{g d^{abc}}{(4\pi)^3} \sum_{{\cal G} \subset {\cal F}} \int\limits^1_0 \mathrm{d}\xi \, \Omega_{{\cal F} \backslash {\cal G}}(\xi,\ep,u)\, {\cal G}(\xi p^b, \bar\xi p^c, p_1^{a_1},\ldots,p_n^{a_n}) \,. \label{eq:Fdxlogx}
\end{align}

So far we have considered the asymptotics of individual Feynman diagrams. It is insightful to sum eq.~\p{eq:Fdxlogx} over all Feynman diagrams ${\cal F}$ contributing to the amputated renormalised $(n+1)$-point correlator $G^{(n+1)}$, obtaining
\begin{align}
\left[ G^{(n+1)}(p^a,p_1^{a_1},\ldots,p_n^{a_n}) \right]_{p^2\log(-p^2)} =  \frac{g d^{abc}}{(4\pi)^3} \int^1_0 \mathrm{d}\xi\, \Omega(\xi,\ep,u) \, G^{(n+2)}_{\rm reg}(\xi p^b,\bar\xi p^c,p_1^{a_1},\ldots,p_n^{a_n}) \,. \label{eq:Gn+1CollAnom}
\end{align}
The function $\Omega(\xi,\ep,u)$ is the sum of $\Omega_{{\cal F} \backslash {\cal G}}$ over all three-point collinear subdiagrams.
The ``regularised'' amputated correlator $G^{(n+2)}_{\rm reg}$ is the $(n+2)$-point amputated correlator  $G^{(n+2)}(q_1,q_2,p_1,\ldots,p_n)$ calculated in the collinear configuration $q_1 = \xi p,\, q_2 = \bar\xi p$. Moreover, the Feynman diagrams contributing to $G^{(n+2)}$ which are trivially singular as $q_1$ and $q_2$ are collinear to $p$ are dropped out.\footnote{The singular Feynman diagrams are those involving the propagator $1/(q_1+q_2)^2$.} Let us stress that usually we cannot take the expression for the renormalised  correlator $G^{(n+2)}$ in generic kinematics and substitute the collinear configuration $q_1 = \xi p,\, q_2 = \bar\xi p$ into it. That would result into collinear divergences which should be properly regularised. Thus, there are $\ep$-poles in $G_{\rm reg}$ which are due to collinear divergences. They are compensated by the collinear $\ep$-poles in $\Omega$. The terms of order $u^0$ and $u^1$ of $\Omega$, required to compute the collinear anomaly up to two-loop order, are given by eqs.~\eqref{eq:Om1loop} and~\eqref{eq:Om2loop}, respectively.

In conclusion, we find that the $p^2 \log p^2$-term of the asymptotic expansion of the (renormalised) amputated correlator $G^{(n)}$ is given by the convolution of a universal function, $\Omega$, with a $(n+1)$-point regularised correlator with two of the incoming momenta carrying fractions of the on-shell momentum $p$. Because of the overall factor of $g$ on the right-hand side of eq.~\eqref{eq:Gn+1CollAnom}, and of the additional particle in the correlator under convolution, computing the collinear part of the conformal anomaly for an amplitude at a given loop order requires lower-loop information only.

\subsection{Applications of the conformal Ward identity}
\label{sec:checks}

In this section we give three instructive examples of the computation of the conformal anomaly~\eqref{eq:KMnAnomaly}. First we consider the three-point amplitude with one on-shell leg, to all orders using the method of expansion by regions in section~\ref{sec:exactsol}, and perturbatively at one loop in section~\ref{sec:anomaly3pt1loop}. Then we study the fully on-shell four-point amplitude perturbatively at one loop in section~\ref{app:4pt_amp}.

\subsubsection{Exact solution of the anomalous Ward identity for the three-point amplitude}
\label{sec:exactsol}

In this section we derive the exact, closed-form expressions of the anomaly $\mathcal{A}^{(3)\mu}$ and of the amplitude $M^{(3)}$ in eq.~\eqref{eq:anomCWI}. 
We recall that we put on-shell only the leg with momentum $p_1$. Remarkably, the anomalous CWI~\eqref{eq:anomCWI} constrains both the amplitude and the anomaly up to two kinematic-independent coefficients. The computation of the anomaly is then required only to fix these coefficients.

The starting point are the two scalar anomalous CWIs given by eqs.~\eqref{eq:scalarCWIM3s1} and~\eqref{eq:scalarCWIM3s2}. We denote by $A^{(3)}$ the anomaly on the right-hand side of the latter, namely
\begin{align} \label{eq:A3def}
    A^{(3)} = 4\gamma G^{(3)}_{1;0}(s_2, s_3) + 4 G_{1;1}^{(3)}(s_2,s_3) \,.
\end{align}
It is convenient to define dimensionless functions of the ratio $y=s_2/s_3$ as
\begin{align} \label{eq:M3tildeDef}
&    M^{(3)}(s_2,s_3) = (s_2 s_3)^{\frac{\eps - 3\gamma}{4}} \tilde{M}^{(3)}(y) \,, \\
&    A^{(3)}(s_2,s_3) = -s_2 (s_2 s_3)^{\frac{\eps - 3\gamma}{4}-1} \tilde{A}^{(3)}(y) \,.
\end{align}
The dimensionful factor multiplying $\tilde{M}^{(3)}(y)$ is chosen so as to be invariant under swapping $p_2$ and $p_3$. As a result, $\tilde{M}^{(3)}(y)$ is symmetric under $y\to 1/y$. Substituting these definitions into the scalar Ward identities gives the following system of DEs,
\begin{align} \label{eq:M3system}
\begin{cases} \Bigl[
16 (1-y) y^2\frac{\mathrm{d}^2}{\mathrm{d}y^2} + 
 8 y \left[\eps - \gamma + y (\eps-\gamma-4)\right] \frac{\mathrm{d}}{\mathrm{d}y}  + (1-y) (\eps - 3 \gamma) (\eps + \gamma-4) \Bigr] \tilde{M}^{(3)}(y) = 0 \,, \\[1.0em]
 \Bigl[16 y^2 \frac{\mathrm{d}^2}{\mathrm{d}y^2} + 8 y (\eps - \gamma) \frac{\mathrm{d}}{\mathrm{d}y} + (\eps - 3 \gamma) (\eps + \gamma-4)   \Bigr] \tilde{M}^{(3)}(y) = - 4 y \tilde{A}^{(3)}(y) \,. \\
\end{cases}
\end{align}
We can combine these two equations so as to eliminate the second derivative, obtaining an equation relating the derivative of the amplitude $\tilde{M}^{(3)}(y)$ directly to the anomaly $\tilde{A}^{(3)}(y)$,
\begin{align} \label{eq:dM3toA3}
    \frac{\mathrm{d}}{\mathrm{d}y} \tilde{M}^{(3)}(y) = \frac{(1 - y)}{4 y (\eps - \gamma-2)} \tilde{A}^{(3)}(y) \,.
\end{align}
The second independent equation can then be chosen to be a second-order ODE for either the amplitude alone (the first equation of the system~\eqref{eq:M3system}), or the anomaly alone,
\begin{equation} \label{eq:anomConsistency}
\begin{aligned}
   \biggr\{ & 16 (1-y) y^2 \frac{\mathrm{d}^2}{\mathrm{d}y^2} + 8 y \left[\eps - \gamma+y(\eps - \gamma-8)  \right] \frac{\mathrm{d}}{\mathrm{d}y} + \\
   & y (3 \gamma+4-\eps) (\eps+\gamma-8) + (\eps -  3 \gamma) (\eps + \gamma-4) \biggr\} \tilde{A}^{(3)}(y) = 0 \,,
\end{aligned}
\end{equation}
which can be viewed as a consistency condition for the anomaly.
Remarkably, thus, conformal symmetry puts very strong constraints also on the anomaly. 

Both the consistency condition~\eqref{eq:anomConsistency} for the anomaly and the second-order ODE for the amplitude in the system~\eqref{eq:M3system} can be solved in terms of ordinary hypergeometric functions, as
\begin{equation} \label{eq:A3general}
\begin{aligned}
   \tilde{A}^{(3)}(y) = y^{\frac{3 \gamma-\eps}{4}} \biggl\{ & a_1(\eps,\gamma) \, {}_2F_1\left( 2 +\frac{\gamma-\eps}{2} , 1+\frac{3\gamma-\eps}{2}, 4+\gamma-\eps ; 1-y \right) + \\
   & + a_2(\eps,\gamma) \, (1-y)^{\eps-\gamma-3} \,  {}_2F_1\left(\frac {\eps - \gamma} {2}-1,  \frac {\gamma + \eps}{2} -2 , \eps - \gamma -2 ; 1-y\right) \biggr\} \,,
\end{aligned}
\end{equation}
and
\begin{equation} \label{eq:M3general}
\begin{aligned}
    \tilde{M}^{(3)}(y) = y^{\frac{3\gamma-\eps}{4}} \biggl\{& b_1(\eps,\gamma) \, {}_2F_1\left(1+\frac{\gamma-\eps}{2}, \frac{3\gamma-\eps}{2},2+\gamma-\eps;1-y \right) +  \\
& + b_2(\eps,\gamma) \, (1-y)^{\eps-\gamma-1} \, {}_2F_1\left(\frac{\eps-\gamma}{2}, \frac{\eps+\gamma}{2}-1,\eps-\gamma; 1-y\right) \biggr\} \,,
\end{aligned}
\end{equation}
where $a_i(\eps,\gamma)$ and $b_i(\eps,\gamma)$ are arbitrary kinematic-independent coefficients.
By plugging these solutions into eq.~\eqref{eq:dM3toA3} we can then relate the coefficients of the amplitude to those of the anomaly,
\begin{equation} \label{eq:bi2ai}
\begin{aligned}
  & b_1(\eps,\gamma) = \frac{4(3 - \eps + \gamma)}{(3 \gamma-\eps)(2+\gamma-\eps) (\eps+\gamma-4)} \, a_1(\eps,\gamma)  \,, \\
  & b_2(\eps,\gamma) = \frac{a_2(\eps,\gamma)}{4 (2+\gamma-\eps)(\eps-\gamma-1)} \,.
\end{aligned}
\end{equation}
The term proportional to $b_2(\eps,\gamma)$ on the right-hand side of eq.~\eqref{eq:M3general} is singular at $y=1$, or equivalently at $s_2 = s_3$. 
The amplitude is free of this singularity, then 
\begin{align}
 b_2(\eps,\gamma) \equiv 0\,,
\end{align}
and also $a_2(\eps,\gamma)\equiv 0$.
We thus see that conformal symmetry ---~supplemented with a simple physical argument~--- constrains $M^{(3)}$ up to the kinematic-independent normalisation without requiring any knowledge of the anomaly. The latter fixes the overall factor.

We compute the closed-form expression of the anomaly using the method of the expansion by regions~\cite{Beneke:1997zp,Smirnov:1999bza,Jantzen:2011nz}. We discuss this in appendix~\ref{app:AnomalyComputation}, and give here the resulting expressions for the anomaly coefficients:
\begin{equation}\label{eq:aiSol}
\begin{aligned} 
    & a_1(\eps,\gamma) = \frac{\tilde{c}_{123}}{(\tilde{c}_{12})^3} (3\gamma-\eps) (\gamma + \eps - 4) \frac{e^{\eps \gamma_\text{E}}  \Gamma\left( 1-\gamma \right)\, \Gamma^2\left(2 - \frac{\eps}{2} + \frac{\gamma}{2}\right)}{\Gamma\left(4 - \eps + \gamma\right) \, 
      \Gamma^2\left(2 - \frac{\eps}{2} - \frac{\gamma}{2}\right)} \,, \\
    & a_2(\eps,\gamma) = 0 \,.
\end{aligned}
\end{equation}
Finally, by putting eqs.~\eqref{eq:aiSol}, \eqref{eq:bi2ai}, \eqref{eq:M3general} and~\eqref{eq:M3tildeDef} together we obtain the closed-form expression of the amplitude:
\begin{equation} \label{eq:M3closedForm}
\begin{aligned}
    M^{(3)}(s_2,s_3) = \ & 4 \frac{\tilde{c}_{123}}{(\tilde{c}_{12})^3} \frac{\eps-\gamma-3}{\eps - \gamma-2}
      \frac{e^{\eps \gamma_\text{E}}  \Gamma\left( 1-\gamma \right)\, \Gamma^2\left(2 - \frac{\eps}{2} + \frac{\gamma}{2}\right)}{\Gamma\left(4 - \eps + \gamma\right) \, \Gamma^2\left(2 - \frac{\eps}{2} - \frac{\gamma}{2}\right)} (-s_3)^{\frac{\eps-3\gamma}{2}} \times \\
    & \qquad {}_2F_1\left(1 - \frac{\eps}{2} + \frac{\gamma}{2}, 
 \frac{3 \gamma}{2} -\frac{\eps}{2}, 2 - \eps + \gamma; 1 - \frac{s_2}{s_3}\right) \,.
\end{aligned}
\end{equation}
We validated this result with an independent computation done with the method of the expansion by regions, as discussed in appendix~\ref{app:AnomalyComputation}.
Using the explicit expressions for the normalisation factors $\tilde{c}_{12}$ and $\tilde{c}_{123}$ ---~given in eqs.~\eqref{eq:ctilde12expl} and \eqref{eq:ctilde123} up to order $\eps^2$~--- we can expand the amplitude around $\eps=0$.\footnote{To expand the hypergeometric functions we used the \textsc{Mathematica} package \textsc{HypExp}~\cite{Huber:2005yg,Huber:2007dx}.} Up to order $\eps$ it is given by
\begin{align}
    M^{(3)}(s_2,s_3) = \textup{i} g^* \biggl[ 1+ u^* \frac{\textsc{n}^2-12}{4 \textsc{n}} \frac{3 s_2 - 3 s_3 - s_2 \log(-s_2) + s_3  \log(-s_3)}{s_2-s_3} + \mathcal{O}\left((u^*)^2\right) \biggr] \,, \label{eq:M3oneloop}
\end{align}
which we cross-checked against a direct one-loop perturbative computation.

In this subsection we have shown that conformal symmetry fixes both the three-point amplitude at the conformal fixed point and its conformal anomaly to all orders in $\eps$ up to two kinematic-independent coefficients, which we computed in closed form.

\subsubsection{Conformal Ward identity for the one-loop three-point amplitude}
\label{sec:anomaly3pt1loop}

We now re-calculate the one-loop approximation of the conformal anomaly of the three-point amplitude $M^{(3)}$ (see eq.~\p{eq:anomCWI}). We label the momenta following the notation of section~\ref{sec:collinear}, and consider the three-point correlator $G^{(3)}(p,p_1,p_2)$ at $p^2 \sim 0$. We would like to apply eq.~\p{eq:Gn+1CollAnom}. Since we aim for the one-loop approximation, we need the lowest order approximation of $\Omega$, namely $\Omega^{\text{1L}}$ from eq.~\p{eq:Om1loop}, and the lowest order approximation for $G^{(4)}$, namely the tree-level approximation,
\begin{align}
G^{(4)}_{\rm reg} (q_1^b,q_2^c,p_1^{a_1},p_2^{a_2}) = -\textup{i} g^2 \left[ \frac{d^{a_1 b f} d^{a_2 c f}}{(q_1 + p_1)^2} + \frac{d^{a_1 c f} d^{a_2 b f}}{(q_2 + p_1)^2} \right] \,,
\end{align}
where we discarded the contribution from the third crossing channel since it is singular where $q_1$ and $q_2$ are collinear with $p$. Substituting all these ingredients into 
eq.~\p{eq:F6xlogx}
gives
\begin{align}
\left[ G^{(3)}(p^a,p_1^{a_1},p_2^{a_2}) \right]_{p^2 \log(-p^2)} & = - \textup{i} \frac{g^3 d^{a a_1 a_2}}{(4\pi)^3} \frac{\textsc{n}^2 -12}{2\textsc{n}} \int^1_0 \mathrm{d}\xi \,  \frac{2\xi\bar\xi}{\xi s_1 + \bar\xi s_2} + {\cal O}(g^3\ep) \label{eq:G3xlogx} \\
& = - \textup{i} g u d^{a a_1 a_2}\frac{\textsc{n}^2-12}{2\textsc{n}}\frac{s_1^2-s_2^2 - 2 s_1 s_2 \log(s_1/s_2)}{(s_1-s_2)^3} + {\cal O}(g u \ep)\,. \notag
\end{align}
At the conformal fixed point $u = u^*$ \p{eq:fixed_point},
the $p^2\log(-p^2)$ term in the asymptotics of $G^{(3)}$ is the collinear contribution in the conformal anomaly, which is denoted as $G^{(3)}_{1;1}$. As for the second contribution to the anomaly, $G^{(3)}_{1;0}$ in eq.~\p{eq:anomalyM3}, note that the amputated three-point correlator is a constant at the tree-level,
\begin{align}
G^{(3)}(p^a,p_1^{a_1},p_2^{a_2}) = \textup{i} g d^{aa_1 a_2} + {\cal O}(g^3) \,.
\end{align}
Consequently, the coefficients $G_{m;k}^{(3)}$ with $m\geq 1$ and $k \geq 0$ in the asymptotic expansion~\p{eq:G3exp} at $p^2 \sim 0$ are of order ${\cal O}(g^3)$. This is in agreement with the explicit expression for $G_{1;1}^{(3)}$ in eq.~\p{eq:G3xlogx}. The perturbative expansion of the coefficient $G_{1;0}^{(3)}$ also starts at order ${\cal O}(g^3)$. The contribution $\gamma \, G_{1;0}^{(3)}$ in the conformal anomaly~\p{eq:anomalyM3} at the conformal fixed point is therefore of order ${\cal O}(\ep^2)$, and is thus neglected in the one-loop approximation. Finally, we find the conformal anomaly~\p{eq:anomalyM3},
\begin{align}
    \mathcal{A}^{(3) \mu} = \textup{i} g^* p^{\mu} \biggl\{ u^*  \frac{\textsc{n}^2-12}{\textsc{n}}  \frac{s_1^2-s_2^2 - 2 s_1 s_2 \log(s_1/s_2)}{(s_1-s_2)^3} + \mathcal{O}\left(\eps^2\right) \biggr\} \,,
\end{align}
which agrees with the $K^{\mu}_{d-\Delta_{\phi}}$-variation of the one-loop expression of $M^{(3)}$ given in eq.~\p{eq:M3oneloop}, and with the all-order result obtained in section~\ref{sec:exactsol} (given by eqs.~\eqref{eq:A3general} and~\eqref{eq:aiSol}).

\subsubsection{Conformal Ward identity for the one-loop four-point amplitude}
\label{app:4pt_amp}

As a simple illustration of the CWI beyond three points, we consider the four-point amputated correlator $G^{(4)}$. As independent kinematic variables, we choose the standard bi-particle Mandelstam variables,
\begin{align}
s = (p_1+p_2)^2 ,\quad
t = (p_2 + p_3)^2 \,,
\end{align} 
complemented by $p_i^2$ with $i=1,\ldots,4$. We define the four-point amplitude $M^{(4)}$ by putting on-shell all four legs of the amputated correlator,
\begin{align} \label{eq:M4}
M^{(4)}(s,t) = \left( \prod_{i=1}^{4} \lim_{p_i^2 \to 0} \right) G^{(4)}\left(s,t,p_1^2,p_2^2,p_3^2,p_4^2\right) \,.
\end{align}
Rewriting the conformal boost generator with conformal weight $\Delta = d-\Delta_\gamma$ in the Mandelstam variables as in eqs.~\p{eq:Konshell} and \p{eq:Koffshell}, we find the second-order scalar differential operators in the variables $s,t$ which play the role of the conformal boost generators for the amplitude, 
\begin{align}
& \mathbb{K}^{(1)} = 4 s \pa_s^2 + 4 t \pa_t \pa_s + 4(2+2\gamma-\ep) \pa_s \,,  \notag\\
& \mathbb{K}^{(3)} = \mathbb{K}^{(1)} |_{s \leftrightarrow t} \,,\qquad
 \mathbb{K}^{(2)} = \mathbb{K}^{(1)} + \mathbb{K}^{(3)}  \,. \label{eq:Kop4pt}
\end{align}
Since the momentum $p_4^\mu$ does not appear in the chosen on-shell Mandelstam variables, $\mathbb{K}^{(4)} \equiv 0$. In other words, $p_4^\mu$ is eliminated by the total momentum conservation. Then the anomalous CWI for the amplitude $M^{(4)}$~\eqref{eq:M4} takes the following form,
\begin{align}
\left( \sum_{i=1}^{3} p_i^\mu \, \mathbb{K}^{(i)}\right)  M^{(4)}\left(s,t \right) =  {\cal A}^{(4)\mu} \,. 
\end{align}
The situation of the four-point amplitude with all legs on shell is special. In this case the anomaly in fact vanishes, ${\cal A}^{(4)\mu} = 0$. Indeed, it is easy to check that, for any function $f$,
\begin{align}
\mathbb{K}^{(i)} \, s^{-1+\ep-2\gamma}\, f\left(\frac{s}{t}\right) = 0 \, , \qquad i = 1,2,3\,,   
\end{align}
where the overall factor of $s$ carries the energy dimensions of the amplitude. We verify explicitly up to one-loop order that all nontrivial contributions to the anomaly cancel out.

According to eq.~\p{eq:KMnAnom} there are several contributions to the anomaly ${\cal A}^{(4)\mu}$. They are the terms in the asymptotic expansion of the correlator at $p_i^2 \to 0$,
\begin{align}
G^{(4)} = M^{(4)}\left( s,t \right) + \sum_{i =1}^{4} \left( p_i^2 \, G^{(4)}_{i;\text{pow}}( s,t) + p_i^2 \log\left(-p_i^2\right)\, G^{(4)}_{i;\text{coll}}( s,t ) \right) + \ldots \,. \label{eq:G4exp}
\end{align}
Despite the fact that the coefficients $ G^{(4)}_{i;\text{coll}}$ and $ G^{(4)}_{i;\text{pow}}$ are nontrivial, the anomaly is zero. Indeed, the contributions to the asymptotics \p{eq:G4exp} from each of the four legs are identical, and the anomaly vanishes due to the total momentum conservation. For the sake of illustration, we present here the coefficients $ G^{(4)}_{i;\text{coll}}$ and $ G^{(4)}_{i;\text{pow}}$. Expanding the tree-level correlator in small $p_i^2$ we find   
\begin{align} \label{eq:G4inu}
\frac{\textup{i}}{g^2}G^{(4)}_{i;\text{pow}}( s,t ) =  -\frac{d^{acf}d^{bdf}}{(s+t)^2} + {\cal O}(u)\,, \quad i=1,\ldots,4 \,.
\end{align}
There are no $p^2_i \log(-p^2_i)$ terms in the asymptotic expansion of the tree-level correlator. Such terms appear in the one-loop correlator. They are captured by the collinear anomaly mechanism of section~\ref{sec:collinear}. For example, the one-loop contribution from leg $1$ is
\begin{align} 
G^{(4)}_{1;\text{coll}}( s,t ) = \frac{\textup{i} g d^{a h g}}{(4\pi)^3} \int^1_0 \mathrm{d}\xi \,\xi \bar\xi \, M^{(5)}_{\rm reg}\left(\xi p_1^h ,\bar\xi p_1^g,p_2^b,p_3^c,p_4^d\right) + g^2 {\cal O}(u^2)\,,
\end{align} 
where $ M^{(5)}_{\rm reg}$ is the five-point tree-level amplitude with the Feynman diagrams which diverge in the forward limit dropped. Pictorially we have
\begin{align}
\left[ G^{(4)}_{1;\text{coll}} \right]_{\text{1-loop}} = \frac{\textup{i} g d^{ahg}}{(4\pi)^3}\int^{1}_0 \mathrm{d}\xi \,\xi \bar\xi\left[ 
\begin{array}{c}
\includegraphics[width=2.8cm]{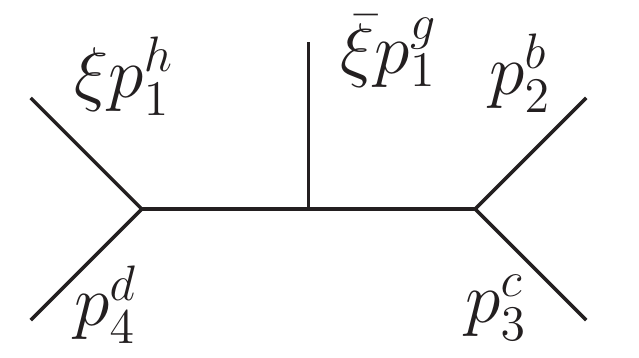} 
\end{array} + 
\begin{array}{c}
\includegraphics[width=2.8cm]{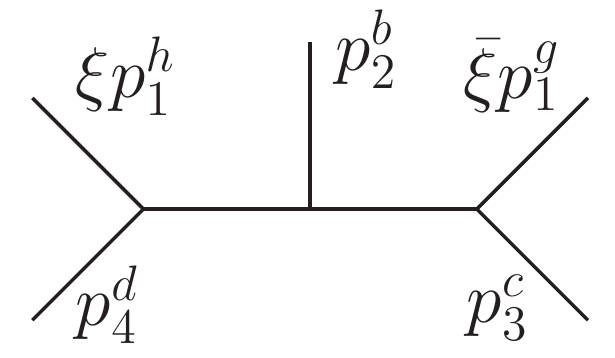} 
\end{array} + \text{cross terms} \right] ,
\end{align}
where the cross-terms are permutations of the legs $2,3,4$, i.e.\ permutations of $p_2^b, p_3^c, p_4^d$. Carrying out the one-fold integration gives
\begin{align} \label{eq:G4icoll}
\frac{\textup{i}}{g^2}G^{(4)}_{i;\text{coll}}( s,t ) = u \left[ - \frac{\textsc{n}^2-12}{4\textsc{n}} \frac{d^{abf}d^{cdf}}{s^2} - d^{a e f} d^{b f g} d^{c g h} d^{d h e}\frac{1}{st} + \text{cross terms} \right] + {\cal O}(u^2) \,.
\end{align}
We see from eqs.~\eqref{eq:G4inu} and~\eqref{eq:G4icoll} that the contributions to the anomaly from the $i$-th on-shell leg are independent of $i$, and hence sum to zero in the anomaly $\mathcal{A}^{(4)}$ because of momentum conservation.
This completes the check.

\section{Discussion and outlook}
\label{sec:conclusions}

In this paper we analysed quantum field theories at a conformal fixed point in dimensional regularisation. 
We derived conformal symmetry constraints for on-shell scattering amplitudes, in the form of anomalous conformal Ward identities.

Previously, studies of implications of conformal symmetry in momentum space were mostly restricted to off-shell quantities.
It turns out that the interplay of conformal symmetry with the on-shell limit is rather subtle. We carefully studied how conformal symmetry acts on LSZ--reduced amplitudes. We found that, in the case where the scattered particles have an anomalous dimension, the conformal symmetry generator does not commute with the on-shell condition $p^2=0$. 
As a consequence of this non-commutativity, an anomalous term appears in the conformal Ward identity. 

One of the main results of this paper is the anomaly formula eq.~\eqref{eq:KMnAnom} for general $n$-particle amplitudes,
generalising to the $d$-dimensional case a previous result due to ref.~\cite{Chicherin:2017bxc}, which is valid for finite conformal loop integrals.
We showed that the anomaly in eq.~\eqref{eq:KMnAnomaly} is comprised of two terms that have different physical origin: while the first term is proportional to the elementary field anomalous dimension, the
 second term can be traced to certain collinear regions of the loop momenta. 
 Importantly, for both terms, at $L$-loops, only $(L-1)$-loop information is required to compute the anomaly. 
 This makes our anomaly formula useful for practical computations.

Furthermore, we showed that for $n$-particle amplitudes the conformal anomaly is a sum of local contributions.
In particular, leveraging the method of regions, we found that the collinear anomaly term can be written as a convolution
of lower-loop amplitudes over a certain collinear kernel, $\Omega$, cf.\ eq.~\eqref{eq:Gn+1CollAnom}, which we computed to two loops in perturbation theory.
This provides an important piece of universal information for using our conformal Ward identities in practice.

There are a number of interesting future directions:
\begin{itemize}
\item Test our $n$-point anomaly formula for one-loop $\phi^3$ amplitudes, and use it to bootstrap two-loop amplitudes. 
\item Understand the collinear kernel $\Omega$ more generally, and explore its relationship to splitting functions considered in the QCD literature.
\item Extend the methods developed in this work to gauge theory scattering amplitudes (which are famously conformally invariant at tree level \cite{Witten:2003nn}).
A natural starting point may be the finite one-loop all-plus and single-minus amplitudes.
It is known that the one-loop all-plus helicity amplitudes are conformally invariant \cite{Henn:2019mvc}, but the single-minus helicity amplitudes are not. Understanding the origin of its non-invariance would be very interesting, and may shed light on the conformal properties of loop amplitudes in gauge theories. 
\end{itemize}

\section*{Acknowledgements}
We are grateful to Vladimir Braun, Einan Gardi, Grisha Korchemsky, Emery Sokatchev and Alexander Manashov for many insightful discussions.
We thank Bl\'aith\'in Power for collaboration at the early stage of this project~\cite{blaithin}. JMH thanks Erik Panzer and Yang Zhang for collaboration on canonical differential equations for Feynman integrals with non-integer powers.
SZ wishes to thank Robin Br\"user for helping with \textsc{Qgraf}.
We have used \textsc{JaxoDraw}~\cite{Binosi:2008ig} to draw Feynman diagrams. 
This project has received funding from the European Union's Horizon 2020 research and innovation programmes \textit{Novel structures in scattering amplitudes} (grant agreement No 725110), and \textit{High precision multi-jet dynamics at the LHC} (grant agreement No 772099). DC is supported by the French National Research Agency in the framework of the \textit{Investissements d’avenir} program (ANR-15-IDEX-02).
DC and SZ gratefully acknowledge the computing resources provided by the Max Planck Institute for Physics.
This research was supported by the Munich Institute for Astro-, Particle and BioPhysics (MIAPbP) which is funded by the Deutsche Forschungsgemeinschaft (DFG, German Research Foundation) under Germany's Excellence Strategy – EXC-2094 – 390783311.

\appendix

\section{Conformal Ward identity for renormalised correlation functions}
\label{app:CWIren}

In this appendix we review the CWIs in the renormalised theory adopting the presentation from \cite{Derkachov:1993uw,Vasilev:2004yr,Braun:2003rp}. We rely on the path-integral representation for the renormalised correlators,
\begin{align}
\vev{\phi(x_1)\ldots \phi(x_n)} = \frac{1}{N}\int \left[ {\cal D} \phi \right] e^{\textup{i} S} \phi(x_1)\ldots \phi(x_n)\,,
\end{align}
where the normalisation $N = \int \left[ {\cal D} \phi \right] e^{\textup{i} S} $ is irrelevant in what follows. The path integration is weighted with the renormalised action of the theory \p{eq:action},
\begin{align} \label{eq:Sren}
S = \int \mathrm{d}^d x\, {\cal L}(x) \,,\qquad {\cal L} := \frac{Z_1}{2} (\partial \phi^a)^2 + \frac{1}{6}g \mu^\ep Z_g Z_1^{\frac{3}{2}} d^{abc} \phi^a \phi^b \phi^c\,,
\end{align}
with renormalisation factors $Z_1=Z_1(u,\ep)$ and $Z_g=Z_g(u,\ep)$.
The renormalisation factors $Z_1$ and $Z_3 = Z_g Z_1^{3/2}$ were calculated in the minimal subtraction scheme up to order $u^2$ in ref.~\cite{Braun:2013tva}. We spell them out here for the convenience of the readers:
\begin{equation} \label{eq:Zfactors}
\begin{aligned}
& Z_1 = 1 - \frac{\textsc{n}^2-4}{2 \textsc{n}} \left[ \frac{u}{6 \eps} - \frac{u^2}{36}  \left( \frac{1}{\eps^2} \frac{ \textsc{n}^2-16}{ \textsc{n}} - \frac{1}{\eps} \frac{ \textsc{n}^2 - 100}{12  \textsc{n}} \right)\right] + \mathcal{O}(u^3) \, , \\
& Z_3 = 1 - \frac{u}{4 \eps} \frac{\textsc{n}^2-12}{\textsc{n}} + \frac{u^2}{16} \left(\frac{1}{\eps^2} \frac{\textsc{n}^2-16}{\textsc{n}} \frac{\textsc{n}^2-12}{\textsc{n}} - \frac{1}{\eps} \frac{\textsc{n}^4-100 \textsc{n}^2 + 960}{6 \textsc{n}^2} \right) + \mathcal{O}(u^3) \,.
\end{aligned}
\end{equation}

The path integral is invariant under the infinitesimal variations of the fields given in eq.~\eqref{eq:confvarphi}, which result in the conformal Ward identities\footnote{The measure of the path integration is not invariant under conformal variations, but the resulting contributions vanish in the dimensional regularisation.}
\begin{align}
& D_{\Delta} \vev{\phi(x_1)\ldots \phi(x_n)} = - \textup{i} \vev{\left(D_{\Delta} S \right) \phi(x_1)\ldots \phi(x_n)} \,, \label{eq:DWard} \\
& K^\mu_{\Delta} \vev{\phi(x_1)\ldots \phi(x_n)} = - \textup{i}\vev{\left(K^\mu_{\Delta} S \right) \phi(x_1)\ldots \phi(x_n)} \,. \label{eq:KWard}
\end{align} 
It remains to choose $\Delta$ such that the right-hand sides of these Ward identities evaluate as a UV-finite correlation function.

The bare action \p{eq:action} in six dimensions is invariant under conformal variations \p{eq:confvarphi} with $\Delta = 2$, which is the classical dimension of the scalar field. Thus, a nonzero variation of the renormalised action in $d$-dimensions \p{eq:Sren} may result only from the $\Delta$-dependent terms of the conformal generators in eqs.~\eqref{eq:Dill} and~\eqref{eq:KX}, which count the dimension. In the following the choose $\Delta = \Delta_\gamma$ \p{eq:Deltag}. Then the dimension of the kinematic term of the action equals to 
$ 2\Delta_\gamma+2 - d = 2\gamma$ and the dimension of the cubic interaction 
$3 \Delta_\gamma -d = 3\gamma-\ep$, namely
\begin{align}
\begin{aligned} \label{eq:DS}
\textup{i} D^\mu_{\Delta_\gamma} S & =  2 \gamma Z_1 \int \mathrm{d}^d x  \frac{1}{2} (\pa \phi^a)^2 +  (3\gamma-\ep) g Z_g Z_1^{\frac{3}{2}} \frac{\mu^\ep}{6} \int \mathrm{d}^d x\, d^{abc} \phi^a \phi^b \phi^c \,, \\ 
-\textup{i} K^\mu_{\Delta_\gamma} S & =  2 \gamma Z_1 \int \mathrm{d}^d x (-2 x^\mu)  \frac{1}{2} (\pa \phi^a)^2 +  (3\gamma-\ep) g Z_g Z_1^{\frac{3}{2}} \frac{\mu^\ep}{6} \int \mathrm{d}^d x (-2 x^\mu) d^{abc} \phi^a \phi^b \phi^c \,.
\end{aligned}
\end{align}

Further, we would like to establish that the given above conformal variations of the action are proportional to the beta-function. We will need the following expressions for derivatives of the renormalisation constants. According to the definition of the anomalous dimension $\gamma$~\p{eq:gamma} and of the $\beta$-function~\p{eq:beta}, we have that
\begin{align}
\gamma(u) = \frac{1}{2} \frac{d u}{d \log \mu}  \, \pa_u \log Z_1  = \frac{\beta(u)}{2 Z_1} \pa_u Z_1 \,. \label{eq:gammaZ1}
\end{align}
Moreover, the bare coupling does not depend on the renormalisation scale, namely $
d g_0/d\mu = 0$. Hence, through to eq.~\p{eq:renormalisation}, it follows that
\begin{align}
\ep g Z_g = - \beta(u) \pa_u(g Z_g) \,. \label{eq:epgZg}
\end{align}
Equations~\p{eq:gammaZ1} and~\p{eq:epgZg} enable us to simplify the coefficients in eqs.~\p{eq:DS},
\begin{align}
& 2\gamma Z_1 = \beta(u) \pa_u Z_1 \,, \\
& (3\gamma(u)-\ep) g Z_g Z_1^{\frac{3}{2}} = \beta(u) g Z_g\, \pa_u Z_1^{\frac{3}{2}} + \beta(u) Z_1^{\frac{3}{2}} \pa_u(g Z_g) = \beta(u) \,\pa_u(g Z_g Z_1^{\frac{3}{2}})  \,. 
\end{align}
Thus we find that the conformal variations of the action are proportional to the $\beta$-function,
\begin{align}
\textup{i} D_{\Delta_\gamma} S & = \beta(u) \int \mathrm{d}^d x \,  \pa_u {\cal L}(x)  = \beta(u) \,\pa_u S \,, \label{eq:DSbeta}
\\
-\textup{i} K^\mu_{\Delta_\gamma} S & = \beta(u) \int \mathrm{d}^d x (-2 x^\mu)\,  \pa_u {\cal L}(x) \,. \label{eq:KSbeta}
\end{align}
It remains to be proven that $\pa_u {\cal L}(x)$ is a UV-finite local operator.

The renormalised correlator is UV-finite, and so is its derivative in the coupling. The latter results in the insertion of the action,
\begin{align}
\pa_u \vev{\phi(x_1) \ldots \phi(x_n)} = \textup{i} \vev{ \phi(x_1) \ldots \phi(x_n) \pa_u S} \,. \label{eq:pacorr}
\end{align}
Consequently, the operator $\pa_u S = \int \mathrm{d}^d x \, \pa_u {\cal L}$ is UV-finite. In perturbation theory, the renormalised action~\p{eq:Sren} is split into the tree-level part, $S_{\rm tree} := S\bigl|_{Z_g,Z_1 \to 1}$, and the counter terms. The role of the counter-terms is to cancel out UV-divergences in the loop corrections. Thus, by adding counter-terms to the operator $\pa_u S_{\rm tree}$ we a obtain UV-finite operator $\pa_u S$, namely
\begin{align}
\int \mathrm{d}^d x \, \left[\pa_u {\cal L}_{\rm tree} \right]_R (x)  = \int \mathrm{d}^d x \, \pa_u {\cal L}(x) \,, \label{eq:LtreeRen}
\end{align}
where explicitly $\pa_u {\cal L}_{\rm tree} = g \mu^\ep d^{a b c} \phi_a \phi_b \phi_c/(12 u)$.
Combining together eqs.~\p{eq:DWard}, \p{eq:DSbeta} and~\p{eq:LtreeRen} we obtain the Ward identity for the dilatation, eq.~\p{eq:WardIdDil}.

Lifting up integration over $x$ in \p{eq:LtreeRen}, we conclude that the renormalised local operator $\left[\pa_u {\cal L}_{\rm tree}\right]_R(x)$ can differ from $\pa_u {\cal L}(x)$ by a total derivative. There is no local operator with a single total derivative in the theory which could mix with the Lagrangian upon renormalisation, e.g.\ $\pa_\mu\left( \phi^a \pa^\mu \phi^a \right) = \pa^2 (\phi^a)^2/2$, therefore
\begin{align}
\left[\pa_u {\cal L}_{\rm tree} \right]_R(x) = \pa_u {\cal L}(x) + \pa^2 F(x) \,,
\end{align}
with a polynomial $F(x)$ in the fields $\phi^a$ and their space-time derivatives. Hence, we can insert a factor of $x^\mu$ into the integration in eq.~\p{eq:LtreeRen} without introducing a contribution from the unknown $F$,
\begin{align}
\int \mathrm{d}^d x \, x^\mu \pa_u {\cal L}(x) = \int \mathrm{d}^d x \, x^\mu \left[\pa_u {\cal L}_{\rm tree} \right]_R (x) \,. \label{eq:xLren}
\end{align}
Combining together eqs.~\p{eq:KWard}, \p{eq:KSbeta} and~\p{eq:xLren} we obtain the Ward identity for the conformal boost, eq.~\p{eq:KCx}.

Finally, let us note that the dilatation Ward identity is equivalent to the renormalisation group equation. Indeed, 
according to eqs.~\p{eq:DWard}, \p{eq:DSbeta} and~\p{eq:pacorr}, we have that
\begin{align}
\textup{i} D_{\Delta_{\gamma}} \vev{\phi(x_1)\ldots \phi(x_n)} = -\textup{i} \beta(u) \vev{\left(\pa_u S \right) \phi(x_1)\ldots \phi(x_n)} = - \beta(u) \pa_u  \vev{\phi(x_1)\ldots \phi(x_n)}\,. \label{eq:Dcorr}
\end{align}
Furthermore, the renormalised correlator carries the canonical dimension, namely
\begin{align}
\mu \pa_\mu \vev{\phi(x_1) \ldots \phi(x_n)} = \sum_{j=1}^n \left( x_j^\mu \pa_{x_j^\mu} + \Delta_{\phi} \right) \vev{\phi(x_1) \ldots \phi(x_n)} \,.     \label{eq:pamuCorr}
\end{align}
Putting eqs.~\p{eq:Dcorr} and~\p{eq:pamuCorr} together gives the renormalisation group equation,
\begin{align}
\left[ \mu \pa_\mu + \beta(u) \pa_u + n \gamma \right] \vev{\phi(x_1) \ldots \phi(x_n)}= 0  \,.   
\end{align}

\section{Implications of conformal symmetry for amplitudes away from the conformal fixed point}
\label{app:implications_away}

In this appendix we show how the knowledge of a (renormalised) $n$-point correlator at the conformal fixed point $u^*$ constrains strongly its form away from the conformal fixed point. Completely analogous observations hold for the scattering amplitudes.
Let us assume we know the correlator $C$ at the conformal fixed point,
\begin{align} \label{eq:CconfExp}
C \bigl|_{u=u^*} = \sum_{k\ge 0} \eps^k C^{*}_k\,.
\end{align}
The dependence on the kinematics is understood.
Away from the conformal fixed point, the renormalised correlator has the generic form
\begin{align} \label{eq:CpertExp}
C = C_0 + \sum_{\ell\ge1} u^{\ell} \sum_{m\ge 0} C_{\ell}^{[m]} \eps^m \,.
\end{align}
By equating eq.~\eqref{eq:CconfExp} with eq.~\eqref{eq:CpertExp} evaluated at $u=u^*$,
\begin{align}
u^{*} = \sum_{m\ge 1} u_m \eps^m \,,
\end{align}
and solving for the $C_{\ell}^{[0]}$'s we obtain
\begin{equation}
\begin{aligned}
& C_0 = C^*_0 \,, \\
& C_{1}^{[0]} = \frac{1}{u_1} C^{*}_1 \,, \\
& C_{2}^{[0]} = \frac{1}{u_1^2} C^{*}_2 - \frac{u_2}{u_1^3} C^{*}_1 - \frac{1}{u_1} C_{1}^{[1]}  \,,
\end{aligned}
\end{equation}
and so on. Up to one loop conformal symmetry fixes entirely the finite part of the renormalised correlator. This follows from the fact that the tree-level correlator does not depend on $\eps$. The finite part of the two-loop correlator is instead determined by conformal symmetry together with the order-$\eps$ part of the one-loop correlator. In general, the finite part of a renormalised correlator at a given loop order is determined by the correlator at the conformal fixed point together with higher-$\eps$ but lower-loop information.

\section{Conformal symmetry and momentum conservation}
\label{app:momcons}

Momentum-space correlations functions and scattering amplitudes are always accompanied by an overall $\delta$ function imposing momentum conservation. In this appendix we motivate why, when studying the conformal properties of these objects, it is possible to effectively neglect such $\delta$ function.

Let us consider a $n$-point correlation function $\mathcal{C}^{(n)}$ in momentum space. We define the reduced correlator $C^{(n)}$ by stripping off the overall momentum-conservation $\delta$ function,
\begin{align}
\mathcal{C}^{(n)}(p_1,\ldots,p_n) = \delta^{(d)}(P) \, C^{(n)}(p_1,\ldots,p_n)\,,
\end{align}
where $P = p_1 + \ldots + p_n$. The commutator between the conformal boost generator $K^{\mu}_{\Delta}$ given in eq.~\eqref{eq:Kmu} and the momentum-conservation $\delta$ function is
\begin{align} \label{eq:Kdelta}
\left[ K^{\mu}_{\Delta} \,,  \delta^{(d)}(P) \right] C^{(n)}(p_1,\ldots,p_n) = 2 \left( [C^{(n)}] + (n-1) d - n \Delta \right)  (\partial_{P^{\mu}} \delta) C^{(n)}(p_1,\ldots,p_n)\,,
\end{align} 
where $[C^{(n)}]$ is the dimension of $C^{(n)}$ in units of energy. The dimension of the \emph{full} reduced correlator $[C^{(n)}]$ (i.e.\ not truncated to some perturbative order) is exactly such that the right-hand side of eq.~\eqref{eq:Kdelta} vanishes. Therefore we can focus on the action of the conformal generator on the reduced correlator,
\begin{align}
K^{\mu}_{\Delta} \delta^{(d)}\left(P\right) C^{(n)}(p_1,\ldots,p_n) =  \delta^{(d)}\left(P\right) K^{\mu}_{\Delta} C^{(n)}(p_1,\ldots,p_n)  \,.
\end{align}
Effectively this means we can neglect the momentum-conservation $\delta$ function and, for this reason, we refer to the reduced correlator as correlator throughout this paper.

It is important to stress that $K^{\mu}_{\Delta}$ and $\delta^{(d)}\left(P\right)$ do \emph{not} commute if applied only to certain terms of the perturbative expansion of $C^{(n)}$. For instance, $C^{(3)}$ has dimension $-6 +\eps$, whereas its tree-level term, 
\begin{align}
C^{(3)}_0 = \frac{1}{p_1^2 s_2 s_3} \,,
\end{align}
has dimension $-6$. For this reason,
\begin{align}
    \left[ K^{\mu}_{\Delta}\,,   \delta^{(d)}(P) \right] C^{(3)}_0 = \mathcal{O}(\epsilon)\,.
\end{align}
This is the case whenever we truncate the perturbative expansion to a certain order: differences due to different ways of implementing momentum conservation kick in at higher orders in $\eps$, as expressed by eq.~\eqref{eq:Kdelta}.


\section{Exact results for two- and three-point correlators}
\label{sec:correlators}

\begin{figure}
\centering
\begin{tabular}{ccccc}
  \raisebox{-.5\height}{\includegraphics[scale=0.16]{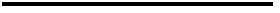}} & 
  \raisebox{-.5\height}{\includegraphics[scale=0.16]{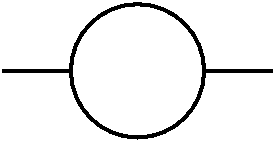}} & 
  \raisebox{-.5\height}{\includegraphics[scale=0.16]{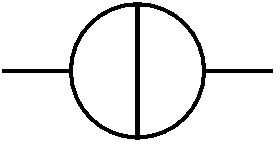}} & 
  \raisebox{-.5\height}{\includegraphics[scale=0.16]{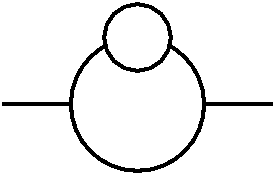}} & 
  \raisebox{-.5\height}{\includegraphics[scale=0.16]{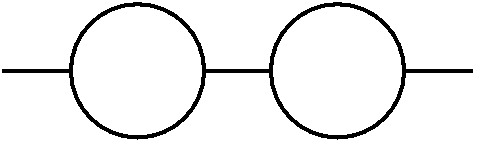}} \\ 
\end{tabular}
\caption{Representative Feynman diagrams contributing to the two-point correlator $C^{(2)}$ up to two-loop order. 
}
\label{fig:diags_2pt}
\end{figure}

In this section we discuss how conformal symmetry constrains the renormalised two- and three-point correlators at the conformal fixed point. We start off in position space, where the correlators have particularly simple expressions, and then Fourier-transform them to momentum space. While the Fourier transform of the two-point correlator is trivial, that of the three-point one requires the computation of a one-loop triangle integral with $\eps$-dependent powers of the propagators. We then compute perturbatively the correlators up to two loop orders using Feynman diagrams, and show that the perturbative result matches the conformal prediction at the conformal fixed point.

We begin by defining the notation.
The main characters of this section are the two- and three-point correlators at the conformal fixed point \p{eq:fixed_point},
\begin{align} \label{eq:correlators_def_x}
\begin{aligned}
    & \mathcal{C}^{(2)}_{ab}(x_1,x_2) = \langle \phi^a(x_1) \phi^b(x_2) \rangle\bigl|_{u=u^*} \,, \\
    & \mathcal{C}^{(3)}_{abc}(x_1,x_2,x_3) = \langle \phi^a(x_1) \phi^b(x_2) \phi^c(x_3) \rangle \bigl|_{u=u^*} \,.
\end{aligned}
\end{align}
Both correlators have a single and overall $su(\textsc{n})$ factor, which it is convenient to strip off as 
\begin{align} \label{eq:correlators_colour}
\begin{aligned}
    & \mathcal{C}^{(2)}_{ab}(x_1,x_2) = \delta^{ab} C^{(2)}(x_1,x_2) \,, \\
    & \mathcal{C}^{(3)}_{abc}(x_1,x_2,x_3) = d^{abc} C^{(3)}(x_1,x_2,x_3) \,.
\end{aligned}
\end{align}
We define the momentum-space correlators by Fourier transforming the position-space ones as in eq.~\eqref{eq:correlators_Fourier}.\footnote{To avoid the proliferation of symbols we slightly abuse the notation, and use $C^{(n)}$ for the correlators in both momentum and position space. The arguments make the distinction.}
It is long known that conformal symmetry fixes them up to the overall normalisation (see e.g.\ ref.~\cite{Polyakov:1970xd}). In position space their expressions are extremely simple, 
\begin{align}
\label{eq:C2confX}
    & C^{(2)}(x_1,x_2) = \frac{c_{12}(u^*)}{\left(x_{12}^2 - \textup{i} 0\right)^{\Delta_{\gamma}}} \,, \\
    & C^{(3)}(x_1,x_2,x_3) = \frac{c_{123}(u^*)}{\left(x_{12}^2- \textup{i} 0\right)^{\frac{\Delta_{\gamma}}{2}} \left(x_{23}^2- \textup{i} 0\right)^{\frac{\Delta_{\gamma}}{2}} \left(x_{31}^2- \textup{i} 0\right)^{\frac{\Delta_{\gamma}}{2}}} \,,
\end{align}
where $x_{ij}=x_i - x_j$. We recall that we are working at the conformal fixed point $u=u^*$, hence $\Delta_{\gamma}$ and the normalisation factors $c_{12}$ and $c_{123}$ are functions of $\eps$. The anomalous dimension $\gamma$ at the conformal fixed point is given by
\begin{align} \label{eq:gammaeps}
\begin{aligned}
\gamma\left(u^*\right) & = \sum_{k\ge1} \gamma_k \, \eps^k \\
& =-\eps \, \frac{\textsc{n}^2-4}{3 (\textsc{n}^2-20)} - 
 4 \eps^2 \frac{(\textsc{n}^2-4) (\textsc{n}^4 - 94 \textsc{n}^2 + 840)}{27 (\textsc{n}^2-20)^3} + \mathcal{O}\left(\eps^3\right) \,.
 \end{aligned}
\end{align}
In the next subsections we will consider these correlators in momentum space, and compare them against the result of a perturbative computation up to two-loop order.

\subsection{Fourier transform of the conformal correlators}
\label{sec:C2}

To warm up, we begin with the two-point correlator given by eq.~\eqref{eq:C2confX}. Since it is a function of a single scalar invariant only, the Fourier transform has no mystery. It is given by
\begin{align} \label{eq:C2conf}
    C^{(2)}\left(p^2\right) = \tilde{c}_{12}(u^*) \, \left(-p^2 - \textup{i} 0 \right)^{-1+\gamma} \,,
\end{align}
where we absorbed the factors coming from the Fourier transform into the normalisation,
\begin{align} \label{eq:ctilde12}
    \tilde{c}_{12}(u^*) = - \textup{i} \, e^{ \textup{i}\pi (\gamma-\eps) } \frac{ 4^{1-\gamma} \pi^{\frac{d}{2}} \Gamma\left(1-\gamma \right) }{\Gamma(2-\ep+\gamma)} \, c_{12}(u^*) \,.
\end{align}
We choose the minus sign for $p^2$ in eq.~\eqref{eq:C2conf} so that the series expansion around $\eps=0$ can be directly compared against the perturbative results containing Feynman integrals, which we compute in the Euclidean region, namely $p^2<0$. 

\medskip

The Fourier transform of the three-point conformal correlator~\eqref{eq:C3confX} is substantially more involved, and has indeed been subject of deep investigation~\cite{Barnes:2010jp,Coriano:2013jba,Bzowski:2013sza,Bzowski:2015yxv,Isono:2019ihz,Bautista:2019qxj,Gillioz:2019lgs,Bzowski:2020lip,Bzowski:2019kwd,Bzowski:2020kfw}. 

As anticipated in section~\ref{sec:conformalVSperturbative}, we represent the Fourier transform as a Feynman integral,
\begin{align} \label{eq:C3I}
    C^{(3)}\left(p_1,p_2,p_3\right) = \tilde{c}_{123}(u^*) \, I\left(2-\frac{\eps+\gamma}{2}, 2-\frac{\eps+\gamma}{2}, 2- \frac{\eps+\gamma}{2} \right) \,,
\end{align}
where
\begin{align}
    \tilde{c}_{123}(u^*) = c_{123}(u^*) \, \left( 2^{d - \Delta_{\gamma}}   \pi^{\frac{d}{2}} e^{\textup{i} \frac{\pi}{2} (\gamma-\ep)} \frac{
  \Gamma\left(\frac{d-\Delta_{\gamma}}{2}\right)}{\Gamma\left(\frac{\Delta_{\gamma}}{2}\right)} \right)^3 \left( \frac{e^{-\eps \gamma_\text{E}}}{(4\pi)^{\frac{d}{2}}} \right) \,,
\end{align}
and $I(e_1,e_2,e_3)$ is a Feynman integral of the one-loop three-mass triangle family~\eqref{eq:3massTriangle}
depicted in figure~\ref{fig:triangle1L}.\footnote{Note that we could equally have considered  the amputated correlator directly. It corresponds to the Feynman integral $I$ with $e_1 = e_2 = e_3 = 1+(\gamma-\ep)/2$, and with a change in the normalisation factor.}
The difference with the standard loop-integral computations is that the powers of the propagators are not integers. We compute the integral in eq.~\eqref{eq:C3I} using the method of the differential equations~\cite{KOTIKOV1991158,Bern:1993kr,Remiddi:1997ny,Gehrmann:1999as} in the canonical form~\cite{Henn:2013pwa}. For this purpose, we consider the family of integrals of the form
\begin{align} \label{eq:Iab}
    I\left(a_1 + \eps b_1, a_2 + \eps b_2, a_3 + \eps b_3 \right) \,,
\end{align}
for constant integer $a_i$ and arbitrary $b_i$. We call the latter \emph{shift parameters}, and denote them cumulatively by $b = (b_1,b_2,b_3)$. They can be functions of $\eps$, provided they have a Taylor expansion around $\ep = 0$. Eventually we will in fact be interested in setting
\begin{align}
b_i = b^* = -\frac{1}{2}\left(1+\frac{\gamma}{\eps}\right) \,,
\end{align}
for $i=1,2,3$, to obtain the integral in eq.~\eqref{eq:C3I}. Since $\gamma$ is of order $\eps$, $b^*$ admits a Taylor expansion around $\eps=0$. We can therefore treat the shift parameters $b$ as additional variables, compute the integrals of the form~\eqref{eq:Iab} as a Laurent expansions around $\eps=0$ up to a certain order, and then substitute $b_i = b^*$ in the result. 

The integrals of the form~\eqref{eq:Iab} for integer $a_i$ and fixed $b_i$ admit a finite dimensional basis. In other words, any integral of the form~\eqref{eq:Iab} can be expressed as a linear combination of a finite number of basis integrals, which are typically called master integrals in the literature. In this case there are four, and we denote them cumulatively by $\vec{g} = (g_1, g_2, g_3, g_4)^T$. We discuss their computation in a Laurent expansion around $\eps=0$ using the method of the canonical differential equations in appendix~\ref{app:DEs}. Using integration-by-parts identities (IBPs)~\cite{Tkachov:1981wb,Chetyrkin:1981qh,Laporta:2000dsw} we can then express the integral relevant for the three-point correlator in terms of the master integrals,
\begin{align} \label{eq:I222red}
I\left(2-\frac{\eps+\gamma}{2}, 2-\frac{\eps+\gamma}{2}, 2- \frac{\eps+\gamma}{2} \right) = \sum_{j=1}^4 r_j(s,\eps) \,  g_j\left(s,\eps\right) \,,
\end{align}
where we denote cumulatively by $s = (s_1,s_2,s_3)$ the independent kinematic variables, with $s_i = p_i^2$. 

We find that the algebraic coefficients $r_i$ in eq.~\eqref{eq:I222red} are given by
\begin{align} \label{eq:Cri}
\begin{aligned}
& r_1(s,\eps) =  2 \frac{\gamma-\eps}{
 \eps (\gamma+\eps-2)^3} \frac{s_1^2 - \left(s_2 - s_3\right)^2}{s_1 s_2 s_3 \lambda(s)} \,, \\ 
& r_2(s,\eps) = r_1(s,\eps)\bigl|_{s_1\to s_2, s_2 \to s_3, s_3 \to s_1} \,, \\
& r_3(s,\eps) =  r_1(s,\eps)\bigl|_{s_1\to s_3, s_2 \to s_1, s_3 \to s_2} \,, \\
& r_4(s,\eps) = \frac{16 (1 - 2 \eps) s_1 s_2 s_3 +2 (\gamma-\eps)  (s_1 + s_2 + s_3) \lambda(s)}{\eps (\eps + \gamma-2)^3 s_1 s_2 s_3 \lambda(s)^{\frac{3}{2}}} \,,
\end{aligned}
\end{align}
with the K\"{a}ll\'{e}n function
\begin{align} \label{eq:lambdap}
    \lambda(s) = s_1^2 + s_2^2 + s_3^2 - 2 s_1 s_2 - 2 s_2 s_3 - 2 s_3 s_1 \,.
\end{align}

Let us now perform a consistency check and use eq.~\eqref{eq:I222red} to verify that the momentum-space expression we obtained for the three-point correlator is indeed conformally invariant,
\begin{align}
    K^{\mu}_{\Delta_{\phi}+\gamma} \, I\left(2-\frac{\eps+\gamma}{2}, 2-\frac{\eps+\gamma}{2}, 2- \frac{\eps+\gamma}{2} \right) = 0 \,,
\end{align}
where we recall that $u=u^*$. The conformal boost generator is expressed as a differential operator in the scalar invariants in eq.~\eqref{eq:Kexp}.
Clearly we know how to differentiate the rational coefficients $r_i(s,\eps)$, and the system of DEs satisfied by the master integrals $\vec{g}$ (see eq.~\eqref{eq:canDEs} in appendix~\ref{app:DEs}) tell us how to express the derivatives of the master integrals $\vec{g}$ as linear combinations of master integrals themselves. We can therefore straightforwardly express the conformal variation of the three-point correlator as a combination of master integrals. Since the latter are linearly independent, their coefficients vanish identically.
Having completed this consistency check, let us now return to the result we obtained.

By substituting the expressions of the master integrals computed in Appendix~\ref{app:DEs} into eq.~\eqref{eq:I222red}, and multiplying by the normalisation factor $\tilde{c}_{123}$, we obtain the analytic expression of the three-point conformal correlator in momentum space.
Here we give the first few terms in the perturbative expansion in $\eps$,
\begin{align} \label{eq:C3confExpl}
\begin{aligned}
 C^{(3)}&(s)  = \frac{\tilde{c}_{123}\left(u^*\right)}{s_1 s_2 s_3} \biggl\{ -1 + \eps \biggl[ - \frac{3}{2} \left(1 + \gamma_1\right) +  (3 \gamma_1 - 1) \frac{s_1 s_2 s_3}{\lambda(s)^{\frac{3}{2}}} f^{(2)}(s) \\
 & + 
 \frac{1}{2\lambda(s)} \left( \left( s_1 (s_2 + s_3 - s_1) + \gamma_1  \left( s_1^2 - 2 \left(s_2 - s_3\right)^2 + s_1 \left(s_2 + s_3\right)\right) \right) \log\left(-s_1\right) + \text{cyc}\right)\biggr] \\
 & + \mathcal{O}\left(\eps^2\right) \biggr\} \,,
\end{aligned}
\end{align}
where $h(s_1,s_2,s_3) + \text{cyc} := h(s_1,s_2,s_3) + h(s_2,s_3,s_1) + h(s_3,s_1,s_2)$
for a generic function $h$,
$\gamma_1$ is defined by eq.~\eqref{eq:gammaeps}, and $f^{(2)}(s)$ is a transcendental function,
\begin{align} \label{eq:f2}
 f^{(2)} = 2 \, \text{Li}_2\left( \tau_2 \right) + 2 \, \text{Li}_2\left( \tau_3 \right) + \log\left( \frac{\tau_3}{\tau_2} \right) \log \left( \frac{1-\tau_3}{1-\tau_2} \right) + \log\left(-\tau_2 \right) \log\left( -\tau_3 \right) + \frac{\pi^2}{3} 
 \,,
\end{align}
with the arguments involving the square root of the K\"{a}ll\'{e}n function~\p{eq:lambdap} defined by
\begin{align}
\tau_2 = \frac{-2 \, s_2}{s_1 - s_2 - s_3 - \sqrt{\lambda(s)}} \, , \qquad \tau_3 = \frac{-2 \, s_3}{s_1 - s_2 - s_3 - \sqrt{\lambda(s)}} \,.
\end{align}
It is worth noting that $f^{(2)}(s)$ is the finite part of the four-dimensional triangle integral (normalised by $\sqrt{\lambda(s)}$), and that it corresponds to the well-known Bloch-Wigner dilogarithm in the variables $z$ and $\bar{z}$, related to $s$ through $s_2 = s_1 z \bar{z}$ and $s_3 = s_1 (1-z)(1-\bar{z})$~\cite{Chavez:2012kn}.

We truncate the result to order $\eps^2$ because this is what is necessary in order to compare against the two-loop perturbative computation which we discuss in section~\ref{sec:perturbative}. Only at two-loop order, in fact,  certain crucial features of the CWIs for the on-shell amplitudes become relevant (see section~\ref{sec:CWI_amplitudes}). However, the computation of the master integrals for the non-integer-power triangle family can be straightforwardly extended to any fixed order in $\eps$ (provided enough computing power for performing the required linear algebra is available). At higher orders more complicated transcendental functions make their appearance. In general, the class of functions corresponding to the alphabet in eq.~\eqref{eq:alphabet} is that of the two-dimensional harmonic polylogarithms~\cite{Gehrmann:2001jv}. Conformal symmetry captures entirely this complexity, to all orders. In the next section we will compare this simplicity against a ``traditional'' perturbative computation using Feynman diagrams, and marvel at how the many pieces fall into place at the conformal fixed point to reproduce the conformal result.

\subsection{Two-loop perturbative computation of the three-point correlator}
\label{sec:perturbative}

In this section we present the perturbative computation of the two- and three-point correlators up to two-loop orders. 
We generate the Feynman diagrams contributing to the bare correlators using \textsc{Qgraf}~\cite{Nogueira:1991ex}. Representative Feynman diagrams are shown in figures~\ref{fig:diags_2pt} and~\ref{fig:diags_3pt}. We express the Feynman diagrams in terms of scalar Feynman integrals, and rewrite the latter in terms of independent master integrals by solving the IBP relations~\cite{Tkachov:1981wb,Chetyrkin:1981qh,Laporta:2000dsw} using \textsc{LiteRed}~\cite{Lee:2012cn,Lee:2013mka}. The two-point one- and two-loop master integrals have long been known analytically because of their simplicity.
The one- and two-loop three-particle integrals were computed analytically in refs.~\cite{Boos:1987bg,Usyukina:1992jd,Usyukina:1992wz,Davydychev:1992xr,Usyukina:1994iw,Davydychev:1999mq,Birthwright:2004kk,Chavez:2012kn}. 
In order to have a uniform setup between the perturbative computation and the Fourier transform of the three-point correlator, we re-computed these Feynman integrals using the method of the differential equations~\cite{KOTIKOV1991158,Bern:1993kr,Remiddi:1997ny,Gehrmann:1999as} in the canonical form~\cite{Henn:2013pwa}. 
We constructed the canonical bases of master integrals using the \textsc{Mathematica} packages \textsc{DlogBasis}~\cite{Wasser:2018qvj,Henn:2020lye} and \textsc{INITIAL}~\cite{Dlapa:2020cwj}, and the heuristic rules discussed in refs.~\cite{Henn:2013tua,Henn:2014qga}.
We fixed the boundary values by imposing that all master integrals are finite on the hypersurface where $\lambda(s) = 0$, and using the well-known closed-form expressions for the bubble-type integrals. We checked the boundary values by evaluating the integrals numerically using \textsc{pySecDec}~\cite{Borowka:2017idc}.

The renormalised correlators are obtained by multiplying the bare ones by a factor of $Z_1^{-\frac{1}{2}}$ for each external particle, and replacing the bare coupling with the renormalised one according to eq.~\eqref{eq:renormalisation}. We work in the modified minimal subtraction ($\overline{\text{MS}}$) scheme, and we set the renormalisation scale $\mu$ to
\begin{align}
\mu^2 = \frac{e^{\gamma_\text{E}}}{4 \pi} \,,
\end{align}
this way absorbing the factors of the Euler–Mascheroni constant $\gamma_\text{E}$ and of $\pi$ coming from the Feynman integrals.

For the two-point renormalised correlator we obtain
\begin{align} \label{eq:C2pert}
\begin{aligned}
    & C^{(2)}_{\text{pert}}(p^2; u, \eps) = \frac{\textup{i}}{p^2} \biggl\{ 1 \\ 
    & \ + u \, \frac{\textsc{n}^2-4}{432 \textsc{n}} \biggl[36 \, \log(-p^2)-96 + \eps \biggl( 3 \pi^2 -208  + 96 \log(-p^2) - 18 \log^2(-p^2)\biggr) + \mathcal{O}\left(\eps^2\right) \biggr] \\
    & \ + u^2 \, \frac{\textsc{n}^2-4}{1728 \textsc{n}^2} \biggl[ 3764 - 81 \textsc{n}^2 + 4 (17 \textsc{n}^2-548) \log(-p^2) - 12 (\textsc{n}^2-28) \log^2(-p^2) + \mathcal{O}\left(\eps\right)  \biggr] \\
    & \ + \mathcal{O}\left(u^3\right) \biggr\} \,.
\end{aligned}
\end{align}
We use the subscript ``$\text{pert}$'' to distinguish the correlators computed perturbatively from those at the conformal fixed point determined by conformal symmetry. Indeed, setting $u=u^*$ in eq.~\eqref{eq:C2pert}, and comparing against the Fourier transform of the conformal correlator in eq.~\eqref{eq:C2conf} allows us to determine the overall normalisation constant up to order $(u^*)^2$ (or equivalently order $\eps^2$). In this way we find
\begin{align} \label{eq:ctilde12expl}
\tilde{c}_{12}(u^*) = -\textup{i} \biggl(1 - 2 u^* \frac{\textsc{n}^2-4}{\textsc{n}} - (u^*)^2 (\textsc{n}^2-4) \frac{396 - 127 \textsc{n}^2 + 3 (\textsc{n}^2-20) \pi^2}{1728 \textsc{n}^2 } + \mathcal{O}\left((u^*)^3\right)\biggr) \,.
\end{align}

For the three-point renormalised correlator we obtain
\begin{align} \label{eq:C3pert}
\begin{aligned}
    & C^{(3)}_{\text{pert}}(s; u, \eps) = \frac{g}{s_1 s_2 s_3} \biggl\{ 1 \\
    & \ + \frac{u}{12 \textsc{n} \lambda(s)} \biggl[(\textsc{n}^2-76) \lambda(s) + \bigl(r_a(s) \log(-s_1) + \text{cyc} \bigr) -6 (\textsc{n}^2-12) \frac{s_1 s_2 s_3}{\sqrt{\lambda(s)}}  f^{(2)}(s)  \\
    & \ + \eps \biggl( \frac{\lambda(s)}{3} (11 \textsc{n}^2-548+6 \pi^2)  - 6(\textsc{n}^2-12) \frac{s_1 s_2 s_3}{\sqrt{\lambda(s)}}  f^{(3)}(s) \\
    & \ + \left( r_b(s) \log(-s_1) + r_c(s) \log^2(-s_1) + \text{cyc} \right) \biggr) + \mathcal{O}\left(\eps^2\right) \biggr] \\
    & \ + \mathcal{O}\left(u^2\right) \biggr\} \,,
\end{aligned}
\end{align}
where
\begin{align}
\begin{aligned}
& r_a(s) = s_1 (s_2 + s_3) (\textsc{n}^2-28) - 2 s_1^2 (\textsc{n}^2-16) + (s_2 - s_3)^2 (\textsc{n}^2-4) \,,\\
& r_b(s) = \frac{1}{3} \biggl[s_1^2 (292 - 19 \textsc{n}^2) + 8 (s_2 - s_3)^2 (\textsc{n}^2-4) + 
    s_1 (s_2 + s_3) (11 \textsc{n}^2-260) \biggr]\,, \\
& r_c(s) = (\textsc{n}^2-16) s_1^2 - \frac{\textsc{n}^2-28}{2} s_1 (s_2 + s_3)   - 
 \frac{\textsc{n}^2-4}{2} (s_2 - s_3)^2 \,,
\end{aligned}
\end{align}
$f^{(2)}$ is defined in eq.~\eqref{eq:f2}, and $f^{(3)}$ is a weight-three transcendental function.
We do not spell out the order-$u^2$ term for the sake of conciseness.
We then compare the perturbative result at the conformal fixed point against the Fourier transform of the conformal correlator in eq.~\eqref{eq:C3confExpl}. We find agreement, and fix the normalisation constant up to order $(u^*)^2$ as
\begin{align} \label{eq:ctilde123}
\begin{aligned}
    \tilde{c}_{123}(u^*) = - g^* \biggl[ & 1 + u^* \frac{\textsc{n}^2-40}{3 \textsc{n}} + \frac{(u^*)^2}{\textsc{n}^2} \biggl( \frac{46 \textsc{n}^4 - 5501 \textsc{n}^2 + 79156}{432} \\
    & + \frac{(\textsc{n}^2-28) (\textsc{n}^2-20)}{192} \pi^2 + 2 (\textsc{n}^2-10) \zeta_3\biggr) + \mathcal{O}\left((u^*)^3\right) \biggr] \,.
\end{aligned}
\end{align}
Let us take a moment to appreciate how non-trivial this is. The perturbative computation involves dozens of Feynman diagrams, which evaluate to complicated transcendental functions. 
All these pieces fit together perfectly to reproduce, at the conformal fixed point, the conformal correlators. What is more, the incredibly compact position-space expressions for the conformal correlators actually capture the entire perturbative expansion of the correlators at the conformal fixed point, which here we have probed up to two loops.

\section{Details of the exact calculation of the triangle correlator and its conformal anomaly} 
\label{app:calculation_3pt}

\subsection{Canonical differential equation for the triangle integrals with \texorpdfstring{$\epsilon$}{eps}-dependent exponents}
\label{app:DEs}

In this appendix we discuss the computation of the master integrals of the one-loop three-mass triangle family with shifted propagator powers, $ I\left(a_1 + \eps b_1, a_2 + \eps b_2, a_3 + \eps b_3 \right)$, defined in eq.~\eqref{eq:3massTriangle}. We adopt the method of the differential equations~\cite{KOTIKOV1991158,Bern:1993kr,Remiddi:1997ny,Gehrmann:1999as} in the canonical form~\cite{Henn:2013pwa}. We emphasize that, in contrast with the traditional applications of this method, here we consider a family of integrals with non-integer, $\eps$-dependent powers of the propagators. We recall that the indices $a_i$ are integers, whereas we treat the shift parameters $b_i$ as additional variables. 

We generate the analytic integration-by-parts identities (IBPs)~\cite{Tkachov:1981wb,Chetyrkin:1981qh,Laporta:2000dsw} and Lorentz-invariance identities~\cite{Gehrmann:1999as} manually, and ``seed'' them only with respect to the integer parameters $a_i$.
We solve the resulting system of equations with the Laporta algorithm~\cite{Laporta:2000dsw} using the dense linear solver implemented in the finite-field framework \textsc{FiniteFlow}~\cite{Peraro:2016wsq,Peraro:2019svx}. There are four master integrals, which the Laporta algorithm chooses as
\begin{align}
\begin{aligned}
    \biggl\{ & I(1+\eps b_1, 1+\eps b_2, \eps b_3) \,, I(\eps b_1, 1+\eps b_2, 1+\eps b_3) \,, \\
    & I(1+\eps b_1, \eps b_2, 1+\eps b_3) \,,
    I(1+\eps b_1, 1+\eps b_2, 1+\eps b_3)\biggr\}\,.
\end{aligned}
\end{align}
This is in complete analogy with the usual integer-power case. Indeed, by setting the shift parameters $b_i$ to zero we get the bubbles in the three possible channels, and the triangle. In contrast with the integer-power case, however, this family contains no bubble integral for $b_i \neq 0$. All integrals have an inevitable triangle component.

We construct a custom basis of master integrals, $\vec{g} = (g_1, g_2, g_3, g_4)$, which satisfy a system of DEs in the canonical form~\cite{Henn:2013pwa},
\begin{align} \label{eq:canDEs}
    \mathrm{d} \vec{g}(s,b,\eps) = \eps \, \mathrm{d}\tilde{A}(s,b) \cdot \vec{g}(s,b,\eps) \,,
\end{align}
where we differentiate with respect to the independent kinematic variables $s = (s_1,s_2,s_3)$, with $s_i = p_i^2$.
The matrix $\tilde{A}$ has the form
\begin{align} \label{eq:Atilde}
    \tilde{A}(s,b) = \sum_{i=1}^{6} A_i(b) \log W_i(s) \,,
\end{align}
where the $A_i$'s are matrices which depend linearly on the shift parameters $b$, and
the arguments of the logarithms are algebraic functions of the kinematic variables called letters~\cite{Chavez:2012kn},
\begin{align} \label{eq:alphabet}
\begin{alignedat}{4}
    & W_1 = s_1\,, && W_2 = s_2 \,, && W_3 = s_3 \,, \\
    & W_4 = \frac{s_1 - s_2 - s_3 - \sqrt{\lambda(s)}}{s_1 - s_2 - s_3 + \sqrt{\lambda(s)}} \,, \quad \quad 
    && W_5 = \frac{s_2 - s_1 - s_3 - \sqrt{\lambda(s)}}{s_2 -s_1 - s_3 + \sqrt{\lambda(s)}}\,, \quad \quad
 && W_6 = \sqrt{\lambda(s)} \,.
\end{alignedat}
\end{align}
 Three master integrals are the non-integer-power equivalent of the bubble integrals in $d=2-2 \eps$ dimensions (normalised by their scale), which are known to satisfy canonical DEs. 
The fourth integral is the non-integer-power triangle in $d=4-2\eps$ dimension. We fix its normalisation by requiring that the DEs take the canonical form, and that the matrices $A_i(b)$ in eq.~\eqref{eq:Atilde} depend at most linearly on the shift parameters $b$. 
We expressed these integrals in terms of integrals in $d=6-2 \eps$ dimensions through the dimensional recurrence relations~\cite{Tarasov:1996br} implemented in \textsc{LiteRed}~\cite{Lee:2012cn,Lee:2013mka}.
Explicitly, they are given by
\begin{align}
\begin{aligned}
g_1 & =
\eps \, s_1 \biggr[2 b_3 \eps (1 + b_1 \eps) I \left(2 +b_1 \eps, 1 + b_2 \eps, 1 + b_3 \eps \right) \\ 
& + 2 b_3 \eps (1 + b_2 \eps) I\left(1 + b_1 \eps, 2 + b_2 \eps, 1 + b_3 \eps \right) + b_3 \eps (1 + b_3 \eps) I\left(1 + b_1 \eps, 1 + b_2 \eps, 2 + b_3 \eps \right) \\
& + (1 + b_1 \eps) (2 + b_1 \eps) I\left(3 + b_1\eps, 1 + b_2\eps, b_3 \eps \right) + (1 + b_2\eps) (2 + b_2 \eps) I\left(1 + b_1 \eps, 3 + b_2 \eps, b_3 \eps \right) \\
& + 2(1 + b_1 \eps)(1 + b_2 \eps) I\left(2 + b_1 \eps, 2 + b_2 \eps, b_3 \eps \right) \biggr]\,,
\end{aligned}
\end{align}
\begin{align}
\begin{aligned}
g_2 & = \eps \, s_2 \biggr[ 
 2 b_1\eps(1 + b_2\eps)I\left(1 + b_1\eps, 2 + b_2 \eps, 1 + b_3 \eps \right) \\ 
& + 2 b_1\eps(1 + b_3\eps)I\left(1 + b_1\eps, 1 + b_2\eps, 2 + b_3\eps \right) + b_1\eps(1 + b_1\eps)I\left(2 + b_1 \eps, 1 + b_2\eps, 1 + b_3\eps\right) \\
& + (1 + b_2 \eps)(2 + b_2\eps)I\left(b_1\eps, 3 + b_2\eps, 1 + b_3\eps \right)
+ (1 + b_3 \eps)(2 + b_3 \eps)I\left(b_1 \eps, 1 +b_2 \eps, 3 +b_3 \eps \right) \\
& + 2(1 + b_2\eps)(1 + b_3 \eps) I\left(b_1\eps, 2 + b_2\eps, 2 + b_3\eps \right) \biggr]\,,
\end{aligned}
\end{align}
\begin{align}
\begin{aligned}
g_3 & = \eps \, s_3 \biggr[ 
2 b_2\eps(1 + b_3\eps)I\left(1 + b_1\eps, 1 + b_2\eps, 2 + b_3\eps \right) \\
& + 2b_2\eps(1 + b_1\eps)I\left(2 + b_1\eps, 1 + b_2\eps, 1 + b_3\eps \right) + b_2\eps(1 + b_2\eps)I\left(1 + b_1\eps, 2 + b_2\eps, 1 + b_3\eps \right) \\
& +(1 + b_3\eps)(2 + b_3\eps) I\left(1 + b_1\eps, b_2\eps, 3 + b_3\eps \right) + (1 + b_1\eps)(2 + b_1\eps) I\left(3 + b_1\eps, b_2\eps, 1 + b_3\eps \right) \\
& + 2(1 + b_1\eps)(1 + b_3\eps)I\left(2 + b_1\eps, b_2\eps, 2 + b_3\eps \right) \biggr]\,,
\end{aligned}
\end{align}
\begin{align}
    \begin{aligned}
g_4 & = \eps^2 (2 + b_1 + b_2 + b_3) \sqrt{\lambda(s)} \biggr[(1 +b_1\eps) I\left(2 + b_1\eps, 1 + b_2 \eps, 1 + b_3 \eps \right) + \\
& + (1 + b_2 \eps) I\left(1 + b_1\eps, 2 + b_2\eps, 1 + b_3\eps \right) + (1 + b_3 \eps)I\left(1 + b_1\eps, 1 + b_2 \eps, 2 + b_3 \eps \right)  \biggr]\,,
    \end{aligned}
\end{align}
where $\lambda(s)$ is defined in eq.~\eqref{eq:lambdap}. For this choice of basis, the kinematic-independent matrices $A_i$ in eq.~\eqref{eq:Atilde} are
\begin{align}
    A_1(b) = \begin{pmatrix} -1 - b_1 - b_2 & 0 & 0 & 0 \\ 
    -\frac{b1}{2} & - \frac{b_1}{2} & \frac{b_1}{2} & 0 \\
    -\frac{b_2}{2} & \frac{b_2}{2} & -\frac{b_2}{2} & 0 \\
    0 & 0 & 0 & -1-\frac{b_1+b_2}{2} \end{pmatrix} \,,
\end{align}
\begin{align}
 A_2(b) = \begin{pmatrix} -\frac{b_3}{2} & -\frac{b_3}{2} & \frac{b_3}{2} & 0 \\ 
 0 & -1 - b_2 - b_3 & 0 & 0 \\
 \frac{b_2}{2} & -\frac{b_2}{2} & -\frac{b_2}{2} & 0 \\
 0 & 0 & 0 & -1 - \frac{b_2+b_3}{2} \end{pmatrix} \,,
 \end{align}
\begin{align}
 A_3(b) = \begin{pmatrix} 
 -\frac{b_3}{2} & \frac{b_3}{2} & -\frac{b_3}{2} & 0 \\
 \frac{b_1}{2} & -\frac{b_1}{2} & -\frac{b_1}{2} & 0 \\ 
 0 & 0 & -1 - b_1 - b_3 & 0 \\
 0 & 0 & 0 & -1-\frac{b_1+b_3}{2} \end{pmatrix} \,,
\end{align}
\begin{align}
 A_4(b) = \begin{pmatrix} 0 & 0 & 0 & -\frac{b_3}{2} \\
 0& 0& 0& 0 \\
 0& 0& 0& \frac{b_2}{2} \\
 -1-\frac{b_2+b_3}{2} & \frac{b_3-b_2}{2} & 1+\frac{b_2+b_3}{2} & 0
 \end{pmatrix} \,,
\end{align}
\begin{align}
 A_5(b) = \begin{pmatrix}
  0& 0& 0& 0 \\ 
  0& 0& 0& -\frac{b_1}{2} \\ 
  0& 0& 0& \frac{b_2}{2} \\
  \frac{b_1 - b_2}{2} & -1-\frac{b_1+b_2}{2} & 1+\frac{b_1+b_2}{2} & 0 \end{pmatrix} \,,
\end{align}
\begin{align}
 A_6(b) = \begin{pmatrix}
 0& 0& 0& 0 \\ 0& 0& 0& 0 \\ 0& 0& 0& 0 \\ 0& 0& 0& 2
\end{pmatrix} \,.
\end{align}

\begin{figure}[t]
\centering
\includegraphics[scale=0.2]{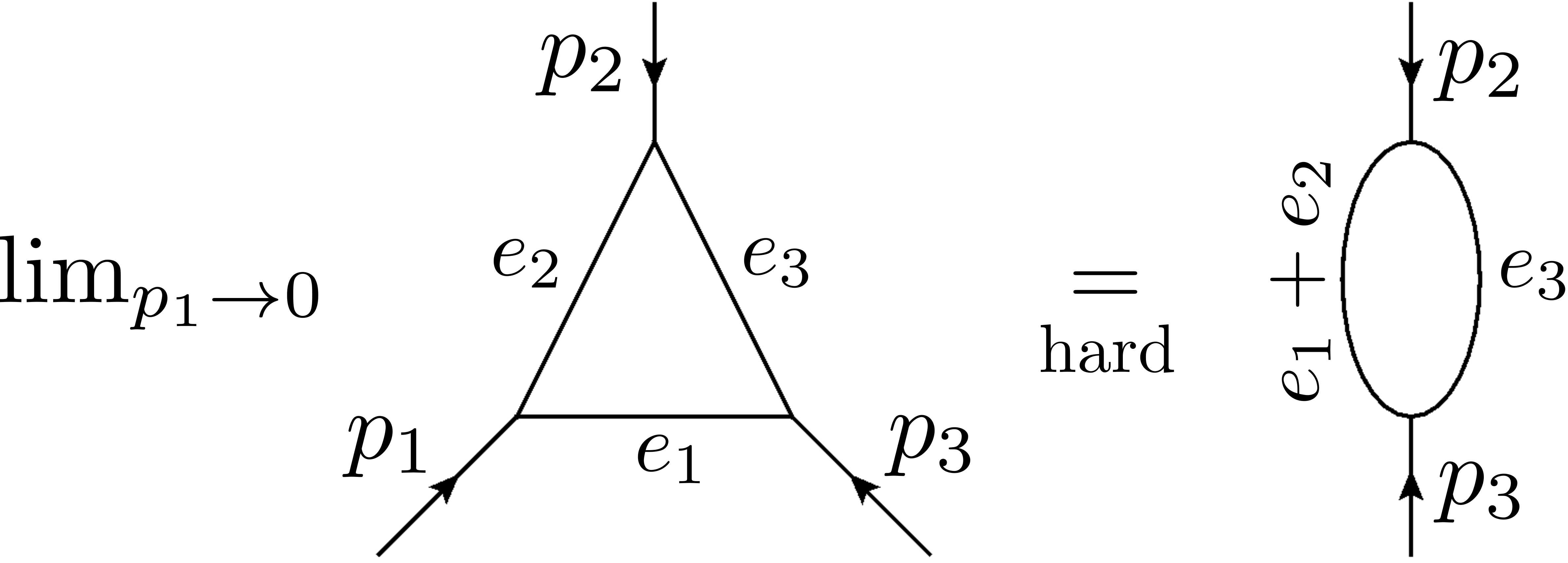}
\caption{The hard region of a triangle integral in the soft limit $p_1 \to 0$ is given by a bubble integral. The arrows denote the momentum flow, while $e_1$, $e_2$ and $e_3$ are generic propagator powers.}
\label{fig:soft_limit_triangle}
\end{figure}

The DEs~\eqref{eq:canDEs} determine the master integrals up to an integration constant. In order to uniquely fix the solution we need the values of the integrals at an arbitrary point. We choose a point in the Euclidean region, where $s_i<0$ for any $i=1,2,3$,
\begin{align} \label{eq:p0}
    s^{(0)} \equiv \left\{s_1 = -1, \, s_2 = -1, \, s_3 = -1 \right\} \,, \qquad \qquad \sqrt{\lambda\left(s^{(0)}\right)} = \textup{i} \sqrt{3}\,.
\end{align}
In the non-integer-power case it is sufficient to impose that all integrals are finite on the hypersurface where  $\lambda(s) = 0$ to relate the value of the triangle integral to that of the bubbles, which are then easily computed in closed form. As we already pointed out, there is no bubble integral for generic values of the shift parameters $b$. In order to fix the boundary values we therefore complement the finiteness at $\lambda(s)=0$ with the analysis of the soft limits $p_i \to 0$ with the method of the expansion by regions~\cite{Beneke:1997zp,Smirnov:1999bza,Jantzen:2011nz}. 
First, we integrate the canonical DEs from $s^{(0)}$ to a point $s^*$ such that $\lambda(s^*) = 0$.\footnote{We made use of the \texttt{Mathematica} package \texttt{HPL}~\cite{Maitre:2005uu} to manipulate the harmonic polylogarithms appearing in the solution of the DEs.} Requiring that the integrals are finite at $s^*$ gives constraints among the boundary values, order by order in $\eps$. Next, we study the soft limits of the integrals. The asymptotic expansion of the master integrals can be computed systematically through the canonical DEs (see e.g.\ ref.~\cite{wasow1965asymptotic} for a detailed discussion, and ref.~\cite{Caron-Huot:2020vlo} for an explicit application to Feynman integrals). For instance, the integral $g_1$ develops three regions in the limit $p_1\to 0$,
\begin{align}
g_1 \underset{p_1 \to 0}{\sim} g_1^{\text{hard}}(s_2,s_3,b,\eps) + g_1^{\text{soft},1}(s_2,s_3,b,\eps) \, s_1^{-b_1-b_2} + g_1^{\text{soft},2}(s_2,s_3,b,\eps) \, s_1^{-2(1+b_1+b_2)}  \,,
\end{align}
up to infinitesimal terms. The hard region corresponds to setting $p_1 = 0$ at the integrand level, which results in a bubble integral as shown in figure~\ref{fig:soft_limit_triangle}. We can thus compute closed-form expressions for the hard region. 
Matching the closed-form expressions for the hard regions in the soft limits against the asymptotic solution of the canonical DEs order by order in $\eps$ gives further constraints on the boundary values. Putting these together with those originating from the finiteness at $\lambda(s^*)=0$ fixes entirely the values of the master integrals at $s^{(0)}$.
 In order to compare against the perturbative computation up to two loops, we need to determine the conformal correlator $C^{(3)}$ up to order $\eps^2$. This requires the boundary values of the master integrals up to order $\eps^3$. We obtain
\begin{align} \label{eq:bdryvals}
\begin{aligned}
 & g_1(s^{(0)}) = \frac{2 + b_1 + b_2}{(1+b_1)(1+b_2)} \biggl(1 -\eps^2 \frac{\pi^2}{12} \biggr) - \frac{\zeta_3}{3} \biggl\{  8  (4 + b_1 + b_2 + b_3) (b_3-2) -32 \frac{1 + b_2}{1 + b_1} \\
& + \frac{2+b_1+b_2}{(1+b_1)(1+b_2)} \left[55 + 60 b_2 + 6 b_2^2 + 6 b_1^2 (1 + b_2) + 
 b_1 (28 + 34 b_2 + 6 b_2^2)\right] \biggr\} \eps^3 +\mathcal{O}\left(\eps^4\right)  \,,\\
 & g_2(s^{(0)}) = g_1(p^{(0)})\biggl|_{b_1\rightarrow b_2, b_2\rightarrow b_3, b_3\rightarrow b_1} \,, \\
 & g_3(s^{(0)}) = g_1(p^{(0)})\biggl|_{b_1\rightarrow b_3, b_2\rightarrow b_1, b_3\rightarrow b_2}  \,, \\
 & g_4(s^{(0)}) = \textup{i} (2 + b_1 + b_2 + b_3) \biggl\{ \alpha_2  \eps^2 +  \left(\frac{17}{45}  \pi^3 + \frac{1}{5}  \alpha_3 \right) \eps^3 \biggr\} +\mathcal{O}\left(\eps^4\right) \,,
\end{aligned}
\end{align}
where $\alpha_2$ and $\alpha_3$ are transcendental constants,
\begin{align}
\begin{aligned}
    & \alpha_2 = 4 \, \text{Im}\left[\text{Li}_2\left(e^{\textup{i} \frac{\pi}{3}}\right)\right] \,, \\
    & \alpha_3 = \pi \log^2(3) - 48 \, \text{Im}\left[\text{Li}_3\left(\frac{\textup{i}}{\sqrt{3}}\right)\right] \,.
\end{aligned}
\end{align}
 We validated the values in eq.~\eqref{eq:bdryvals} by evaluating numerically some finite integrals of the family. We worked out their Feynman parameterisation, expanded it around $\eps=0$, and integrated each term numerically at $s^{(0)}$ using \textsc{pySecDec}~\cite{Borowka:2017idc}. 

Once the boundary values are determined, the canonical DEs~\eqref{eq:canDEs} can be solved straightforwardly in terms of Chen's iterated integrals~\cite{Chen:1977oja,Brown:2013qva} or two-dimensional harmonic polylogarithms~\cite{Gehrmann:2001jv} (see e.g.\ the lecture notes~\cite{Henn:2014qga}). Up to the order required for our purposes (namely $\eps^3$), logarithms and classical polylogarithms are all we need. For the sake of illustration, we give explicitly the expressions of the canonical master integrals up to order $\eps^2$,
\begin{align} \label{eq:gexpl}
\begin{aligned}
g_1&(s,b,\eps) =  \frac{1}{(1 + b_1) (1 + b_2)} \biggl\{ (2 + b_1 + b_2) - \eps \, \biggl[ (1 + b_1 + b_2) (2 + b_1 + b_2) \log\left(-s_1\right) \\
& + b_3 (1 + b_1) \log\left(-s_2\right)  + b_3 (1 + b_2)  \log\left(-s_3\right)
\biggr] + \eps^2 \biggl[- \frac{\pi^2}{12} (2 + b_1 + b_2) \\
&  + \frac{1}{2} (1 + b_1 + b_2)^2 (2 + b_1 + b_2) \log^2\left(-s_1\right) + \frac{1}{2} (1 + b_1) b_3 (1 + b_2 + b_3) \log^2\left(-s_2\right) \\
& +  \frac{1}{2} (1 + b_2) b_3 (1 + b_1 + b_3) \log^2\left(-s_3\right) + (1 + b_1) (1 + b_1 + b_2) b_3 \log\left(-s_1\right) \log\left(-s_2\right) \\
& -(1 + b_1) (1 + b_2) b_3 \log\left(-s_2\right) \log\left(-s_3\right) + (1 + b_2) (1 + b_1 + b_2) b_3 \log\left(-s_3\right) \log\left(-s_1\right) \biggr] \biggr\} \\ 
&  + \mathcal{O}\left(\eps^3 \right)\,, \\
g_2&(s,b,\eps) = g_1(p,b,\eps) \bigl|_{p_1\to p_2, p_2 \to p_3, p_3 \to p_1, b_1 \to b_2, b_2 \to b_3 , b_3 \to b_1}  \,, \\
g_3&(s,b,\eps) = g_1(p,b,\eps) \bigl|_{p_1\to p_3, p_2 \to p_1, p_3 \to p_2, b_1 \to b_3, b_2 \to b_1, b_3 \to b_2}  \,, \\
g_4&(s,b,\eps) = \eps^2 (2 + b_1 + b_2 + b_3) f^{(2)}(s) + \mathcal{O}\left(\eps^3 \right) \,,
\end{aligned}
\end{align}
where $f^{(2)}(s)$ is the Bloch-Wigner dilogarithm defined in eq.~\eqref{eq:f2}.

\subsection{Asymptotic expansion of the three-point correlator and closed-form expression of the anomaly}
\label{app:AnomalyComputation}

In this appendix we present the computation of the closed-form expression of the anomaly in the CWIs~\eqref{eq:anomCWI}. To do so, we compute the asymptotic expansion in the small $s_1$ limit of the amputated correlator $G^{(3)}$ using the method of the expansion by regions~\cite{Beneke:1997zp,Smirnov:1999bza,Jantzen:2011nz}.

We begin by expressing the amputated correlator in terms of an integral of the three-mass triangle family~\eqref{eq:3massTriangle}, through eqs.~\eqref{eq:C3I} and~\eqref{eq:GnDef}, as
\begin{align} \label{eq:G3I}
    G^{(3)} = \frac{\tilde{c}_{123}}{(\tilde{c}_{12})^3} (-s_1 s_2 s_3)^{1-\gamma}  I\left(2-\frac{\eps+\gamma}{2},2-\frac{\eps+\gamma}{2},2-\frac{\eps+\gamma}{2}\right) \,.
\end{align}
The triangle integrals~\eqref{eq:3massTriangle} have the following asymptotic expansion in the light-like limit $s_1\to 0$,
\begin{align} \label{eq:Iregions}
I(e_1,e_2,e_3) = \sum_{m=0}^{\infty} (-s_1)^m I^{\text{hard}}_{[m]}(e_1,e_2,e_3) + \sum_{m=0}^{\infty} (-s_1)^{\frac{d}{2}-e_1-e_2+m} I^{\text{coll}}_{[m]}(e_1,e_2,e_3)  \,.
\end{align}
This can be shown by analysing the DEs of the basis integrals for this family (see appendix~\ref{app:DEs}), or by using the \texttt{Mathematica} package \texttt{asy2.m}~\cite{Pak:2010pt,Jantzen:2012mw}.
Substituting eq.~\eqref{eq:Iregions} into eq.~\eqref{eq:G3I} gives the asymptotic expansion of $G^{(3)}$, which we anticipated in eq.~\eqref{eq:G3exp1}, with
\begin{align} \label{eq:coll2hard}
\begin{aligned}
    & G^{(3),\text{hard}}_{[m]} = \frac{\tilde{c}_{123}}{(\tilde{c}_{12})^3} (s_2 s_3)^{1-\gamma} I^{\text{coll}}_{[m]}\left(2-\frac{\eps+\gamma}{2},2-\frac{\eps+\gamma}{2},2-\frac{\eps+\gamma}{2}\right)\,, \\
    & G^{(3),\text{coll}}_{[m]} = \frac{\tilde{c}_{123}}{(\tilde{c}_{12})^3} (s_2 s_3)^{1-\gamma} I^{\text{hard}}_{[m]}\left(2-\frac{\eps+\gamma}{2},2-\frac{\eps+\gamma}{2},2-\frac{\eps+\gamma}{2}\right)\,,
\end{aligned}
\end{align}
for $m\ge 0$. We emphasise that collinear and hard regions are exchanged between the amputated correlator and the triangle integrals. This follows from the factor of $(-s_1)^{1-\gamma}$ in eq.~\eqref{eq:G3I}, which turns the integer powers $(-s_1)^{m}$ in the expansion of the integrals~\eqref{eq:Iregions} into $(-s_1)^{1-\gamma+m}$-terms of the amputated correlator, and vice versa the terms of order $(-s_1)^{-1+\gamma+m}$ of the integrals into integer powers of the amputated correlator.

The $m=0$ term from the collinear region gives the amplitude $M^{(3)}$,
\begin{align} \label{eq:M3I}
    M^{(3)}=\frac{\tilde{c}_{123}}{(\tilde{c}_{12})^3} (s_2 s_3)^{1-\gamma} I^{\text{coll}}_{[0]}\left(2-\frac{\eps+\gamma}{2},2-\frac{\eps+\gamma}{2},2-\frac{\eps+\gamma}{2}\right) \,.
\end{align}
Similarly, acting with $\hat{K}^{(1)}_{d-\Delta_{\gamma}}$ on the asymptotic expansion of $G^{(3)}$~\eqref{eq:G3exp1} and letting $s_1=0$ gives the anomaly in the CWI~\eqref{eq:A3def},
\begin{align} \label{eq:A3I}
\begin{aligned}
A^{(3)} & = \lim_{s_1\to 0} \hat{K}^{(1)}_{d-\Delta_{\gamma}} G^{(3)} = \\
& = -4 \gamma \frac{\tilde{c}_{123}}{(\tilde{c}_{12})^3} (s_2 s_3)^{1-\gamma}  I^{\text{coll}}_{[1]}\left(2-\frac{\eps+\gamma}{2},2-\frac{\eps+\gamma}{2},2-\frac{\eps+\gamma}{2}\right)\,.
\end{aligned}
\end{align}
The amplitude and the conformal anomaly are thus determined by the first two terms coming from the collinear region the triangle integral, which ---~because of eq.~\eqref{eq:coll2hard}~--- constitute the first two terms from the hard region of the amputated correlator. We now proceed to compute them.

We start from the Feynman parameterisation of the triangle integrals,
\begin{align}
I(e_1,e_2,e_3) = \Omega_0(e_1,e_2,e_3) \int_{[0,\infty)^3} \left(\prod_{j=1}^3 d\alpha_j \alpha_j^{e_j-1} \right) \delta\left(\alpha_1+\alpha_2-1 \right) \frac{\mathcal{U}^{e_{123}-d}}{\mathcal{F}^{e_{123}-\frac{d}{2}}} \,,
\end{align}
where
\begin{align}
& \Omega_0(e_1,e_2,e_3) = \frac{e^{\eps \gamma_{\text{E}}} \Gamma\left(e_{123}-\frac{d}{2}\right)}{\Gamma(e_1) \Gamma(e_2) \Gamma(e_3)} \,, \\
& \mathcal{U} = \alpha_1 + \alpha_2 + \alpha_3 \,, \\
& \mathcal{F} = -s_1 \alpha_1 \alpha_2 -s_2 \alpha_2 \alpha_3 -s_3 \alpha_3 \alpha_1 \,, \\
& e_{123}=e_1+e_2+e_3 \,.
\end{align}
To compute the terms from the collinear region, we rescale the Feynman parameter $\alpha_3$ as $s_1 \alpha_3$ (as prescribed by \texttt{asy2.m}), expand around $s_1 = 0$ under the integral sign, and integrate term by term.
We obtain the following closed-form expressions for the first two terms,
\begin{align} \label{eq:Ilightlike0}
\begin{aligned}
I^{\text{coll}}_{[0]}(e_1,e_2,e_3) = \, & e^{\eps \gamma_{\text{E}}} \frac{\Gamma\left(e_1+e_2-\frac{d}{2} \right) \Gamma\left(\frac{d}{2}-e_1 \right) \Gamma\left(\frac{d}{2}-e_2 \right)}{\Gamma\left(e_1 \right) \Gamma\left(e_2 \right) \Gamma\left(d-e_1-e_2 \right)} \left(-s_3\right)^{-e_3} \times \\
& \qquad {}_2F_1\left(e_3, \frac{d}{2}-e_1, d-e_1-e_2 ;1-\frac{s_2}{s_3}\right) \,,
\end{aligned}
\end{align}
and
\begin{align} \label{eq:Ilightlike1}
\begin{aligned}
& I^{\text{coll}}_{[1]}(e_1,e_2,e_3) = e_3 (e_{123}-d) e^{\eps \gamma_{\text{E}}} \frac{\Gamma\left(e_1+e_2-\frac{d}{2}-1 \right) \Gamma\left(\frac{d}{2}+1-e_1 \right) \Gamma\left(\frac{d}{2}+1-e_2 \right)}{\Gamma\left(e_1 \right) \Gamma\left(e_2 \right)\Gamma\left(d+2-e_1-e_2 \right)}\times \\
& \ \  \left(-s_3\right)^{-1-e_3} y^{\frac{d}{2}-e_2-e_3}  {}_2F_1\left(\frac{d}{2}+1-e_2, 1+d-e_{123}, d+2-e_1-e_2 ;1-\frac{s_2}{s_3}\right) \,.
\end{aligned}
\end{align}
Let us also spell out the first term from the collinear region prior to carrying out the integration which produces the hypergeometric function,
\begin{align} \label{eq:Icoll0int}
    I^{\text{coll}}_{[0]}(e_1,e_2,e_3) = \frac{e^{\eps \gamma_{\text{E}}} \Gamma\left(e_1+e_2-\frac{d}{2}\right)}{\Gamma(e_1) \Gamma(e_2)} \int_0^1 d\alpha_1 \frac{\alpha_1^{\frac{d}{2}-e_2-1} (1-\alpha_1)^{\frac{d}{2}-e_1-1}}{\left[\alpha_1(s_2-s_3)-s_2\right]^{e_3}} \,.
\end{align}

Substituting eqs.~\eqref{eq:Ilightlike0} and~\eqref{eq:Ilightlike1} into eqs.~\eqref{eq:M3I} and~\eqref{eq:A3I} finally gives closed-form expressions for the amplitude and the anomaly. The resulting expression for the anomaly matches the solution of the anomaly consistency condition given by eq.~\eqref{eq:A3general}, fixing the free coefficients to the values given in eq.~\eqref{eq:aiSol}. The expression for the amplitude agrees with the solution of the CWIs given in eq.~\eqref{eq:M3closedForm}. We arrived at the same results also starting from the canonical DEs for the basis integrals of the triangle family with shifted propagator powers discussed in appendix~\ref{app:DEs}. We solved the DEs asymptotically in the limit $s_1 \to 0$, and substituted the resulting expansions into the correlator through eq.~\eqref{eq:C3I}. We found perfect agreement between the two approaches.

\subsection{The conformal boost generator controls the on-shell expansion of the correlator}
\label{app:K1G3}
In this appendix we prove that the entire asymptotic expansion \p{eq:G3exp} of the amputated correlator $G^{(3)}$ in the on-shell limit $s_1\to 0$ is determined by just two terms, $G^{(3)}_{1;0}$ and $G^{(3)}_{1;1}$. Moreover, using the anomalous CWIs, we show that the off-shell amputated correlator $G^{(3)}$ can be expressed in terms of the amplitude $M^{(3)}$ and $G^{(3)}_{1;1}$ through an operator series which effectively restores the off-shellness.

We begin by plugging the asymptotic expansion of $G^{(3)}$ given in eq.~\eqref{eq:G3exp} into the Ward identity~\eqref{eq:scalarCWIs2},
\begin{equation} \label{eq:KG3exp}
\begin{aligned}
    & \mathcal{K} M^{(3)} + \sum_{m\ge 1} \sum_{k\ge 0} s_1^m \, \mathbf{L}^k \mathcal{K} G^{(3)}_{m;k}  -  \sum_{m\ge 0}\sum_{k\ge 0} s_1^m \, \mathbf{L}^k  \biggl[(\gamma+m)(m+1) G^{(3)}_{m+1;k} + \\ 
    & \ +
     (2m+1+\gamma)(k+1)  G^{(3)}_{m+1;k+1} + (k+1)(k+2)  G^{(3)}_{m+1;k+2} \biggr] = 0 \,,
\end{aligned}
\end{equation}
where we used the short-hand notations
\begin{align}
\mathcal{K} = \frac{1}{4} \hat{K}^{(2)}_{d-\Delta_{\gamma}} \,, \qquad \qquad \qquad \mathbf{L} = \log(-s_1) \,.
\end{align}
Only the terms with $m=0$ survive in the $s_1 \to 0$ limit. Those with $m=0$ and $k=0$ produce the anomaly equation,
\begin{align} \label{eq:KM3v2}
\mathcal{K} M^{(3)} = \gamma G_{1;0}^{(3)} + G^{(3)}_{1;1} \,.
\end{align}
Those with $m=0$ and $k>0$ must cancel out. This implies the recursion relation
\begin{align}
    \gamma G^{(3)}_{1;k} + (1+\gamma)(k+1)  G^{(3)}_{1;k+1} + (k+1)(k+2)  G^{(3)}_{1;k+2}  = 0
\end{align}
for all $k>1$, which can be solved as
\begin{align}
    G^{(3)}_{1;k} = c_1 \frac{(-1)^k}{k!} + c_2 \frac{(-\gamma)^{k-1}}{k!} \,,
\end{align}
where $c_1$ and $c_2$ are independent of $k$. We recall that $G^{(3)}_{1;k}$ is the coefficient of $p_1^2 \log^k(-p_1^2)$ in the asymptotic expansion of the amputated correlator $G^{(3)}$. As such, the lowest perturbative
order of $G_{1;k}^{(3)}$ should increase along with $k$. This implies $c_{1}=0$ and hence that
\begin{align} \label{eq:G1k}
    G^{(3)}_{1;k} = \frac{(-\gamma)^{k-1}}{k!} G_{1;1}^{(3)} \,.
\end{align}
Similarly, by demanding that all remaining terms in eq.~\eqref{eq:KG3exp} cancel out we find
\begin{align}
& \begin{aligned} \label{eq:Gm0}
    G_{m;0}^{(3)} = \ & \frac{\Gamma(\gamma+1)}{m!\Gamma(\gamma+m)} \left[ \mathcal{K} \right]^{m-1} G^{(3)}_{1;0} + \frac{\Gamma(\gamma)}{m! \Gamma(\gamma+m)} \left[ \mathcal{K} \right]^{m-1} G^{(3)}_{1;1} + \\
    & - \frac{\Gamma(\gamma-m)}{(m-1)! \Gamma(\gamma-1) \gamma}  \left[ -\mathcal{K} \right]^{m-1} G^{(3)}_{1;1} \,,
\end{aligned} \\
\label{eq:Gmk}
& G_{m;k} = \frac{(-\gamma)^{k-1}}{(m-1)! k!} \frac{\Gamma(\gamma-m)}{\Gamma(\gamma-1)} \left[- \mathcal{K} \right]^{m-1} G^{(3)}_{1;1}  \,, 
\end{align}
for $m\ge 1$ and $k\ge 1$. We therefore see that all terms of the asymptotic expansion of the amputated correlator $G^{(3)}$ are determined by just two: $G_{1;0}^{(3)}$ and $G^{(3)}_{1;1}$. 

Furthermore, using the recursive expressions for the coefficients $G^{(3)}_{m;k}$ given in eqs.~\eqref{eq:G1k}, \eqref{eq:Gm0} and~\eqref{eq:Gmk}, and the anomaly equation~\eqref{eq:KM3v2}, we can resum the asymptotic expansion in the concise form
\begin{align}
    G^{(3)} = {}_0F_1\left(\gamma; s_1 \mathcal{K} \right) M^{(3)} - \frac{1}{\gamma} \left(-s_1\right)^{1-\gamma} {}_0F_1\left(2-\gamma; s_1 \mathcal{K}\right) G^{(3)}_{1;1} \,, \label{G3asSeries}
\end{align}
where the hypergeometric functions ${}_0F_1$ are understood as power series of the operator $s_1 {\cal K}$. This series representation for $G^{(3)}$ exactly reproduces the expansion into hard and collinear regions in eq.~\eqref{eq:G3exp1}, which we prove in appendix~\ref{app:AnomalyComputation}. Furthermore, by acting with the conformal boost ${\cal K}$ on the right-hand side of eq.~\p{G3asSeries}, one can immediately verify that the off-shell correlator $G^{(3)}$ is conformally invariant, provided that the amplitude $M^{(3)}$ satisfies the anomaly equation~\p{eq:KM3v2}.
In conclusion, thanks to conformal symmetry, the amplitude $M^{(3)}$ and the coefficient $G^{(3)}_{1;1}$ are sufficient to restore the off-shell amputated correlator $G^{(3)}$.

In section~\ref{sec:multi-point} we briefly discuss how to find an analogue of eq.~\p{G3asSeries} for $n$-point correlators with several legs approaching the light cone. In this generic case, a concise expression like \p{G3asSeries} is not known. Still, the conformal boost invariance severely restricts the form the asymptotic expansion~\p{eq:GnAsymp}. Solving the contiguous relations following from the conformal boost invariance of $G^{(n)}$ requires the boundary conditions, which are $2^{|\Lambda|}$ coefficients among $G^{(n)}_{\{ m_{l};k_{l}\}_{l\in \Lambda}}(v)$ of the asymptotic expansion \p{eq:GnAsymp}. All coefficients of the expansion are obtained from the boundary coefficients by acting on them with polynomials in the operators $\mathbb{K}^{(i)},  A^{(ij)}, B^{(ij)}$ from eqs.~\p{eq:Konshell} and~\p{eq:Koffshell}.

\bibliographystyle{JHEP}
\bibliography{conformal.bib}

\end{document}